\documentclass[aps,pra,amsmath,notitlepage,nofootinbib,twocolumn,superscriptaddress,noeprint]{revtex4-2}
\pdfoutput=1
\usepackage[utf8]{inputenc}
\usepackage[english]{babel}

\usepackage{textcomp}
\usepackage[T1]{fontenc}
\usepackage{microtype}

\usepackage{graphicx}
\graphicspath{{Figures/}{./}}

\usepackage{amsmath,amsfonts,amssymb,amsthm,bm,bbm,bbold}
\usepackage{mathtools}
\usepackage{physics}
\usepackage{makecell}
\usepackage[normalem]{ulem}

\usepackage{xcolor}
\usepackage[breaklinks=true,colorlinks=true,linkcolor=teal,urlcolor=teal,citecolor=teal]{hyperref}

\newcommand{\id}{\mathbb{1}}

\newcommand{\I}{\mathrm{i}}
\newcommand{\E}{\mathrm{e}}

\newcommand{\Gperp}{\Gamma_{\!\!\perp}}
\newcommand{\sigplus}{\sigma_{\!+}}
\newcommand{\sigmin}{\sigma_{\!-}}

\let\Re\relax
\let\Im\relax

\DeclareMathOperator{\Re}{Re}
\DeclareMathOperator{\Im}{Im}

\renewcommand{\vec}{\boldsymbol}
\let\originalleft\left
\let\originalright\right
\renewcommand{\left}{\mathopen{}\mathclose\bgroup\originalleft}
\renewcommand{\right}{\aftergroup\egroup\originalright}
\renewcommand{\right}{\aftergroup\egroup\originalright}

\begin{document}
\title{Fundamental limits of pulsed quantum light spectroscopy: Dipole moment estimation}

\author{Francesco Albarelli}
\email{francesco.albarelli@gmail.com}
\affiliation{Dipartimento di Fisica ``Aldo Pontremoli'', Università degli Studi di Milano, via Celoria 16, 20133 Milan, Italy}
\affiliation{Istituto Nazionale di Fisica Nucleare, Sezione di Milano, via Celoria 16, 20133 Milan, Italy}

\author{Evangelia Bisketzi}
\affiliation{Department of Physics, University of Warwick, Coventry, CV4 7AL, United Kingdom}

\author{Aiman Khan}
\affiliation{Department of Physics, University of Warwick, Coventry, CV4 7AL, United Kingdom}

\author{Animesh Datta}
\email{animesh.datta@warwick.ac.uk}
\affiliation{Department of Physics, University of Warwick, Coventry, CV4 7AL, United Kingdom}

\begin{abstract}
We study the fundamental limits of the precision of estimating parameters of a quantum matter system when it is probed by a travelling pulse of quantum light.
In particular, we focus on the estimation of the interaction strength between the pulse and a two-level atom, equivalent to the estimation of the dipole moment.
Our analysis of single-photon pulses highlights the interplay between the information gained from the absorption of the photon by the atom as measured in absorption spectroscopy, and the perturbation to the temporal mode of the photon due to spontaneous emission.
Beyond the single-photon regime, we introduce an approximate model to study more general states of light in the limit of short pulses, where spontaneous emission can be neglected.
We also show that for a vast class of entangled biphoton states, quantum entanglement between the signal mode interacting with the atom and the idler mode provides no fundamental advantage and the same precision can be obtained with a separable state.
We conclude by studying the estimation of the electric dipole moment of a sodium atom using quantum light.
Our work initiates a quantum information theoretic methodology for developing the theory and practice of quantum light spectroscopy.
\end{abstract}

\date{\today}

\maketitle

\section{Introduction}

Spectroscopy seeks to estimate one or more parameters appearing in the model of a matter system by measuring the light that has interacted with it.
Recent technological developments have made it possible to use quantum light in spectroscopy~\cite{Mukamel2020}, e.g., few-photon Fock, squeezed or entangled states that exhibit nonclassical spatial or temporal correlations~\cite{Walmsley2015}.
This resulted in sensing with sensitivity better than the classical shot noise limit~\cite{polzik1992spectroscopy,kalachev2007biphoton,Kalashnikov2014,Dorfman2021b}, in obtaining different scaling of the spectroscopic signals~\cite{Dorfman2016} with incident light intensity, as well as in new spectroscopic techniques~\cite{yabushita2004spectroscopy,saleh1998entangled,Raymer2013,Ishizaki2020,Fujihashi2021}.
Despite many proposals to perform spectroscopy with pulses of quantum light, a rigorous and quantitative assessment of the best attainable precision, and of the potential advantage of using entangled light, remains absent.

\begin{figure}
    \includegraphics[width=.95\columnwidth]{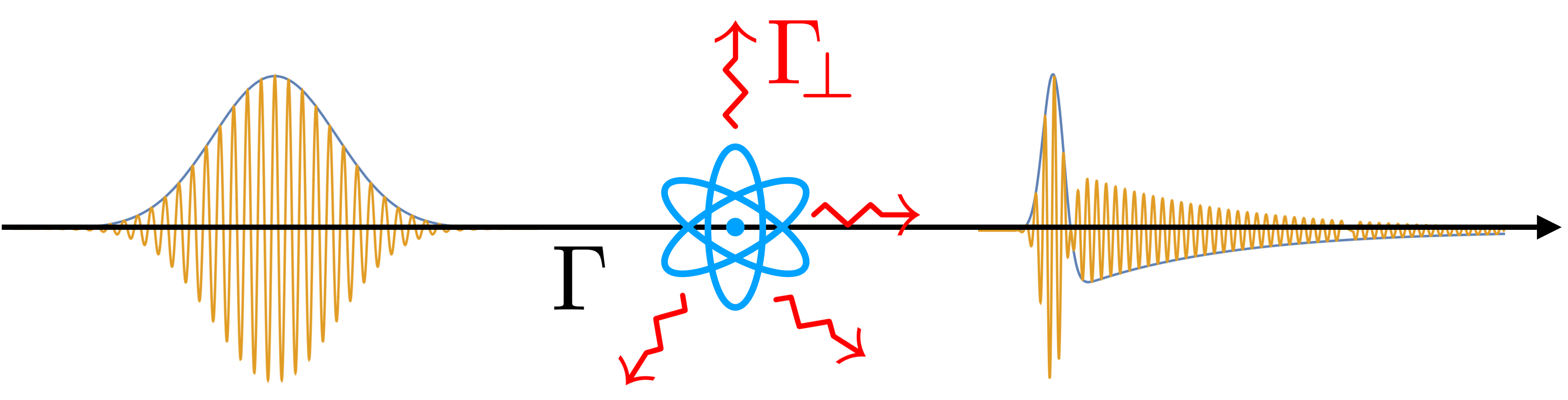}
    \caption{Illustration (not to scale) of the excitation of an atom by a quantum pulse of light (with Gaussian temporal envelope).
    $\Gamma$ represents the interaction strength with the pulse, while $\Gperp$ describes emission into other (inaccessible) orthogonal field modes.
    As illustrated, the shape of the wavepacket is changed by the interaction with the atom.
    }
    \label{fig:scheme_interaction}
\end{figure}

In this paper, we start to uncover the fundamental limits of quantum light spectroscopy by employing the tools of quantum estimation theory that underlie quantum metrology.
Our aim is to understand the extent to which the in-principle enhancements of quantum metrology~\cite{Giovannetti2011,Polino2020} are relevant under the particular circumstances of practical quantum light spectroscopy experiments~\cite{SrimathKandada2021,Mukamel2020}.
Specifically, we focus on a paradigmatic scenario that can be considered a minimal example of quantum spectroscopy: 
A pulse of quantum light is used to probe a single two-level atom, as illustrated in Fig.~\ref{fig:scheme_interaction},
with the objective of estimating the light-atom coupling parameter $\Gamma$, proportional to the square of the atom's electric dipole moment (EDM).

Our work chooses this simplest of matter systems to establish a quantum information theoretic methodology for analysing pulsed quantum light spectroscopy.
We focus on pulses of quantum light as our endeavour is to understand the challenges and methods peculiar to this scenario; we do not aim to compare the performance or practicality of pulsed light with schemes based on continuous waves.

Calculating the fundamental bounds set by quantum mechanics to the precision of estimating $\Gamma$ requires a full description of the quantum state of the light and all mathematically valid detection techniques.
The former is made challenging by the change in the modal structure of pulsed light after the light-atom interaction, and the latter by the infinitude of possibilities.
Nevertheless, for single-photon pulses for instance, we clearly identify two sources of information about the parameter $\Gamma$: A ``classical'' one, related to absorption spectroscopy~\cite{Whittaker2017}, and a ``quantum'' one, related to fluorescence lifetime estimation~\cite{Mitchell2022} and fluorescence spectroscopy~\cite{Schlawin2017a}.
That our methodology can elucidate phenomena typically studied in disparate frameworks speaks to its strength.

Our work thus stands in contrast to previous ones where quantum estimation theory has been applied to estimating the light-matter coupling parameter in cavity-based setups~\cite{Genoni2012b,Bernad2019}, wherein only one discrete mode of the light field is involved.
It also stands in contrast to the analysis of spectroscopic signals that rely on evaluating the expectation values of particular observables~\cite{Schlawin2017a}, different for various spectroscopic setups.
Finally, our methodology relies on studying the exact dynamics of a pulse interacting with a \emph{single} atom.
This is in contrast to ``conventional'' approaches~\cite{Schlawin2017,Mukamel2020} that relate the induced polarizations in ensembles to the measured signals or treat the matter ``effectively''~\cite{Dinani2016,Birchall2019,Biele2021}.

Our main results are as follows:
\begin{enumerate}
\item We derive the fundamental precision bounds for estimating $\Gamma$ with generic single-photon and entangled biphoton states (in the slowly-varying envelope approximation, assumed throughout the paper), in Eqs.~\eqref{eq:QFI_1ph_normalized} and~\eqref{eq:QFI_biph_normalized} respectively.

\item We identify practically reasonable measurements that attain the fundamental precision for single-photon pulses in Sec.~\ref{subsubsec:1phOptMeas}.

\item We show that there is, in general, no quantitative relationship between the precision of estimating $\Gamma$ and the excitation probability using a certain pulse (see e.g., Fig.~\ref{fig:QFIshapes}).
However, the two quantities are related for Fock states in the short-pulse regime, as in Sec.~\ref{subsec:linearabsorption}.

\item For single-photon pulses, we show that the probability of losing the photon from the original travelling pulse wavepacket contains a fraction of the total information about the parameter that would be available at the end of the experiment in the ideal case, i.e., where all the spontaneously emitted light can be measured optimally.
The fraction becomes one half in the limit of short pulses (see e.g., Fig.~\ref{fig:QFIshapes}(c)).

\item In Sec.~\ref{subsec:QFIapproxJC} we show that in the limit of short pulses and detection of the light shortly after the interaction (in units of the atom lifetime) the variance of estimating $\Gamma$ decreases as the inverse of the pulse duration and as $1/n$ if probed by $n$-photon Fock states.

\item In Sec.~\ref{subsec:advantage_biphoton} we show that for entangled biphoton probes with real-valued temporal envelopes, entanglement is not a fundamental resource to enhance the estimation precision, since there always exists an unentangled single-photon probe that performs at least as well.
\end{enumerate}
All the results in this paper, apart from 1. above, are obtained under the assumption of zero detuning between the pulse carrier frequency and the transition frequency of the atom.

The paper is structured as follows.
In Sec.~\ref{sec:theory} we present our theoretical framework: The model of light-matter interaction and a summary of local quantum estimation theory.
In Sec.~\ref{sec:1photon} we present an extensive analysis of the limits of the precision of estimating $\Gamma$ using single-photon pulses.
In Sec.~\ref{sec:short_pulses} we approximate the problem to a much simpler one in the regime of short pulses and present a general solution within this approximation.
In Sec.~\ref{sec:biphoton} we extend the analysis to entangled biphoton pulses, investigating the potential advantage afforded by such states.
In Sec.~\ref{sec:sodium} we apply our general results to the EDM estimation of a sodium atom.
We conclude in Sec.~\ref{sec:discussion} with a brief discussion.

\section{Theoretical framework}
\label{sec:theory}

We begin with a theoretical model of light-atom interaction, followed by a description of the quantum states of a travelling pulse of light.
We then provide a brief introduction to quantum estimation theory necessary to quantify the fundamental and attainable precision in parameter estimation.

\subsection{Model}
\label{subsec:model}

\subsubsection{Atom, field, and their interaction}

We consider a single two-level atom (the ``atom'' or $\mathrm{A}$ subsystem) fixed in space modelled by its free Hamiltonians $H^{\mathrm{A}}.$
The ground and excited states of the atom are denoted by $\ket{g}$ and $\ket{e}$ respectively.
Setting the ground state energy to zero,
\begin{equation}
H^{\mathrm{A}} = \hbar \omega_0 \dyad{e}{e},
\end{equation}
where $\omega_0$ is the transition frequency.

We next consider a travelling pulse of quantized radiation field (the ``pulse'' or $\mathrm{P}$ subsystem), which must be described by a continuum of frequencies.
As is customary in spectroscopic setups, we assume the field to have a well-defined direction of propagation.
This leads to the free field Hamiltonian~\cite{Blow1990a}
 \begin{equation}
    \label{eq:HF}
    H^\mathrm{P} = \hbar \int_0^\infty \! d\omega \, \omega \, a^\dag (\omega) a(\omega),
 \end{equation}
with the bosonic operators $[a(\omega),a^\dag(\omega)]=\delta(\omega-\omega')$ labelled by a continuous frequency $\omega$.
Invoking the slowly-varying envelope approximation, which assumes the central frequency $\bar{\omega}$ of the electric field to be much larger than the spectral width~\cite{Schlawin2017} and is usually valid in the optical regime, we obtain the standard expression
for the positive-frequency part of the electric field operator in the interaction picture with respect to $H^\mathrm{P}$
\begin{equation}\label{eq:Electric_field_1D_continuous}
\mathbf{E}(t) = \I \hbar \bm{\epsilon} \mathcal{A}(\bar{\omega})  \int_{-\infty}^{\infty} \frac{d\omega}{\sqrt{2\pi}} \; a(\omega) \E^{-i \omega t},
\end{equation}
where $\bm{\epsilon}$ is a unit polarization vector, $\mathcal{A}(\bar{\omega}) = \sqrt{\bar{\omega}/(2 \epsilon_{0}c A \hbar)}$ and $A$ is the transverse quantisation area.

The interaction between the travelling pulse and the atom is schematically illustrated in Fig.~\ref{fig:scheme_interaction}.
It is modelled by an interaction term $H_{\mathrm{I}}.$ 
Supplementing the slowly-varying envelope with the dipole approximation so that the spatial extent of the atom is assumed to be much smaller than the wavelength associated with $\bar{\omega}$, it is possible to show that $A = 1/u(\vec{r})^2$, where $u(\vec{r})$ is the transverse spatial mode function of the paraxial beam at the atomic position $\vec{r}$~(for a detailed derivation, see for instance Ref.~\cite{Ko2022}).
Making next the rotating wave approximation, the interaction Hamiltonian in the interaction picture generated by the unitary 
transformation $e^{-\I (H^\mathrm{A}+H^\mathrm{P})t}$ takes the standard form~\cite{Scully1997}
\begin{align}
    \label{eq:interaction_hamiltonian_initial_definition}
H_{\mathrm{I}}^{\mathrm{AP}}(t) &= \mathbf{d}(t)\cdot \mathbf{E}^{\dagger}(t)+\mathbf{d}^{\dagger}(t)\cdot \mathbf{E}(t), \\
& = -\I \hbar \sqrt{\Gamma} \left( \sigma_{\!+} a(t) - \sigma_{\!-} a^\dag(t)  \right)
\label{eq:interaction_hamiltonian_Gamma}
\end{align}
where  $\mathbf{d}(t)=\bm{\mu}_{eg} \sigmin e^{-\I \omega_{0} t}$ is the positive frequency part of the dipole operator, $\bm{\mu}_{eg} = - q_{e} \bra{e}\mathbf{r}\ket{g}$ is the relevant dipole matrix element ($q_{e} $ is the charge of the electron), and $\sigmin = \dyad{g}{e}=\sigplus^{\dagger}$.
In Eq.~\eqref{eq:interaction_hamiltonian_Gamma} we have introduced the so-called ``quantum white-noise'' operators\footnote{Formally, the operators $a(t)$ should be treated using quantum stochastic calculus~\cite{gardiner2004quantum}, but this is unnecessary for our purposes.
We refer the reader to some physics-oriented introductions in the context of light-matter interaction with pulses of light~\cite{Baragiola2014a,Fischer2018a,Ko2022}.}
\begin{equation}
    \label{eq:white_noise_a}
    a(t) = \int_{-\infty}^{\infty} \frac{d\omega}{\sqrt{2\pi}} \; a(\omega) \E^{-\I (\omega-\omega_0) t}
\end{equation}
satisfying $[a(t),a^\dag(t')] = \delta(t-t')$ and the constant $\Gamma=(\bm{\mu}_{eg}\cdot \bm{\epsilon})^2 \mathcal{A}(\bar{\omega})^2$ proportional to the square of the dipole moment.

In addition to the continuum of bosonic modes $a(\omega)$ that describes the pulse degrees of freedom, an atom in free space interacts with an infinitude of other modes of the electromagnetic field (capturing all the other spatial and polarization degrees of freedom beyond those of the pulse).
We account for this by introducing a coupling to an additional continuum of bosonic modes $b(\omega)$ with a free Hamiltonian analogous to Eq.~\eqref{eq:HF}, leading to the interaction-picture Hamiltonian
\begin{equation}
    \label{eq:Hint_PE}
	H_{\mathrm{I}}^\mathrm{APE}(t) =  -\I \hbar \sigma_{\!+} \left( \sqrt{\Gamma} a(t) + \sqrt{\Gperp } b(t) \right) + \mathrm{h.c.},
\end{equation}
where the additional set of white noise operators $b(t)$ satisfying $[b(t),b^\dag(t')] = \delta(t-t')$ 
represents a collective ``environment'' ($\mathrm{E}$) subsystem coupled to the atom, and $\Gperp$ is the coupling strength with such environment.
For completeness, we show in Appendix~\ref{app:effectiveGammmaPerp} that this is equivalent to a more realistic model where the environment consists of a discrete set of infinitely many families of white noise operators, representing all the degrees of the electromagnetic field beyond those described by $a(\omega)$.

Our approach is to treat A, P and E, as distinct subsystems, of which P is the only one that can be measured experimentally.
This will change slightly in Sec.~\ref{sec:biphoton} for entangled biphoton states: the signal (S) subsystem plays the role of P, but an additional idler (I) subsystem that does not interact with either A or E, is also assumed to be measurable.
The atom-environment interaction in Eq.~\eqref{eq:Hint_PE} seeks to capture an experimental scenario where light emitted into the environment is irreversibly lost.
Mathematically, this means tracing out the subsystem E; the resulting reduced dynamics of the atom-pulse state is governed by a master equation in Lindblad form
\begin{equation}
    \frac{d \rho^{\mathrm{AP}}(t)}{dt} =  -\frac{\I}{\hbar} [ H_\mathrm{I}^\mathrm{AP}(t), \rho^{\mathrm{AP}}(t)] + \Gperp \mathcal{D}[ \sigmin] \rho^{\mathrm{AP}}(t),
    \label{eq:LindbladPulseAtom}
\end{equation}
where we have introduced the superoperator $\mathcal{D}[ A ] \rho= A \rho A^\dag - \frac{1}{2} \left( \rho A^\dag A  +  A^\dag A \rho \right) $.
While a master equation treatment is very useful numerically, for single-photon pulses it will be easier to solve the full unitary dynamics.
We take the latter approach in Sec.~\ref{sec:1photon}.

Although the Hamiltonian in Eq.~\eqref{eq:Hint_PE} is obtained in the white noise limit---a Markov approximation, 
this is not enough to have a reduced dynamics in Lindblad form for a generic initial state of the E subsystem.
The Lindblad form is guaranteed because the initial state is the vacuum and thus temporally uncorrelated.
On the contrary, nonclassical initial states of the field, such as Fock or squeezed states that we will choose for the travelling pulse, have temporal correlations and induce a non-Markovian reduced dynamics of the atom~\cite{Dabrowska2021a}.
However, the reduced dynamics of the atom can generally be described by hierarchies of master equations~\cite{Baragiola2012,Baragiola2014a,Gross2022} or by using a virtual cavity with a suitable time-dependent coupling~\cite{Kiilerich2019,Kiilerich2020}.

The interaction Hamiltonian in Eq.~\eqref{eq:Hint_PE} can be used to describe light-matter interactions in different scenarios by assigning different relative magnitudes to  the quantities $\Gperp$ and $ \Gamma$.
In a free-space configuration with the atom probed by paraxial light, there is usually a strong coupling with the environment and a weak coupling with the pulse, resulting in $\Gperp \gg \Gamma$~\cite{Silberfarb2003}.
Nonetheless, by matching the pulse spatial and polarization degrees of freedom to the dipole pattern of the atom, one could in principle obtain a perfect coupling $\Gperp = 0$ even in free space~\cite{Wang2011b}.
However, the scenario $\Gperp = 0$ is mostly employed to study two-level atoms in one-dimensional waveguides~\cite{Konyk2016,Roulet2016a}. 

In this paper, we will not grapple with such model-dependent details and take the Hamiltonian in Eq.~\eqref{eq:Hint_PE} as our starting point until Sec.~\ref{sec:sodium}.
There we will apply our methods to estimating the dipole moment of the Sodium $D_2$ transition using travelling pulses of quantum light.
Furthermore, we will always assume that the atom is initially in the ground state, because we want to model light absorption and the corresponding excitation induced by the pulse.
However, if we kept the same Hamiltonian but started the dynamics with the atom in the excited state and both P and E in the vacuum, the overall decay rate would be $\Gamma_{\mathrm{tot}} = \Gamma + \Gperp $~\cite{Silberfarb2004,Wang2011b,Baragiola2014a,Ko2022}, and in free-space this corresponds to the standard rate obtained from Wigner-Weisskopf theory $\Gamma_{\mathrm{tot}} = \frac{ |\bm{\mu}_{eg}|^2 \omega_0^3}{3 \pi \epsilon_0 \hbar c^3 }$.

\subsubsection{Quantum states of the travelling pulse}

To describe a pulse of light travelling in a well-defined direction quantum mechanically, we introduce the continuous 
Fock states~\cite{Blow1990a}
\begin{equation} \label{eq:fock_state_definition_MT}
	\ket{n_{\xi}} = \frac{1}{\sqrt{n!}} \left( \int_0^\infty d\omega \; \tilde{\xi}(\omega)a^{\dagger}(\omega)\right)^{n}\ket{0},
\end{equation}
where $\tilde{\xi}(\omega)$ is the single-photon spectral amplitude, a square-integrable normalised function describing the wavepacket.
As mentioned before, we assume the light to be sufficiently narrowband around the carrier frequency $\bar{\omega}$, which we further assume to equal the atomic transition frequency.
Thus, $\bar{\omega}=\omega_0.$
In this regime, we can extend the integral in Eq.~\eqref{eq:fock_state_definition_MT} to the whole real line and introduce the temporal amplitude $\xi(t)= \int_{-\infty}^\infty \frac{d \omega}{\sqrt{2\pi}} \tilde{\xi}(\omega) e^{-\I (\omega-\omega_0) t}$, i.e., an envelope that modulates oscillations at the carrier frequency, as depicted in Fig.~\ref{fig:scheme_interaction}.
Then we can introduce the photon-wavepacket creation operator (satisfying $[A_{\xi},A_{\xi}^\dag] = 1$)
\begin{equation}
    \label{eq:temporalmode_annihil}
    A^\dag_{\xi} = \int_{-\infty}^{\infty} dt \; \xi(t) a^\dag(t) = \int_{-\infty}^\infty d\omega \; \tilde{\xi}(\omega) a^{\dagger}(\omega) 
\end{equation}
where $a(t)$ are defined in Eq.~\eqref{eq:white_noise_a}.
In general $\xi(t)$ can be considered as one element of a complete orthonormal basis of functions $\xi_j(t)$, known as temporal modes~\cite{Brecht2015,Raymer2020b} (e.g., Hermite-Gauss polynomials if $\xi(t)$ is Gaussian).
The Hilbert space of the P subsystem is thus the tensor product of the Fock spaces associated to each temporal mode and the most general pure state therein is written as $\sum_j \sum_{n_j} c_{n,j} \ket{n_{\xi_j}}$~\cite{Fabre2020}.

Thus, even if the description of a travelling pulse needs an underlying continuous degree of freedom (i.e., $\omega$ or $t$), the initial state describing the incoming pulse is effectively single-mode using the photon-wavepacket operators.
The Fock states in Eq.~\eqref{eq:fock_state_definition_MT} can be reexpressed as $\ket{n_\xi} = \frac{1}{\sqrt{n!}} A_{\xi}^{\dag n} \ket{0}$ and descriptions of other states such as coherent states
\begin{equation}
    \label{eq:coherent}
	\ket{\alpha_\xi}= e^{ \alpha A_{\xi}^\dag - \alpha^* A_{\xi}    }\ket{0} = e^{-|\alpha|^{2}/2} \sum_{n=0}^\infty \frac{\alpha^n}{\sqrt{n!}} \ket{n_\xi},
\end{equation}
with average photon number $\int_{-\infty}^{\infty} dt \langle \alpha_{\xi} | a^\dag(t) a(t) |  \alpha_{\xi} \rangle  =  |\alpha|^{2}$ and squeezed vacuum states~\cite{Raymer2020b,Gross2022}
\begin{equation}
    \label{eq:squeezed}
	\ket{s_\xi}= e^{  \frac{1}{2} \left( r A_{\xi}^{\dag \, 2} - r^* A_{\xi}^2 \right) }\ket{0} = \frac{1}{\sqrt{c}} \sum_{n=0}^\infty \left( \frac{ s }{ 2 c } \right)^n \frac{\sqrt{(2 n)!}}{n!} \ket{ 2n_\xi},
\end{equation}
with average photon number $\int_{-\infty}^{\infty} dt \langle s_{\xi} | a^\dag(t) a(t) |  s_{\xi} \rangle  = \sinh^2 |r|$ (where $r = |r| e^{i \phi}$, $c = \cosh r$ and $s = e^{i \phi} \sinh r$) follow.

The preceding is not the most general scenario, since generic quantum states of light are supported on multiple temporal modes~\cite{Rohde2007}, e.g., realistic single-beam squeezed states obtained from parametric downconversion~\cite{Blow1990a,Christ2011}.
We leave the analysis of this more general case for future studies.
Note, however, that the entangled biphoton states in Sec.~\ref{sec:biphoton} necessarily involve multiple temporal modes.

Finally, having defined $\xi(t)$ and $a(t)$ as Fourier transforms centered around $\omega_0$, for a detuning $\Delta = \omega_0 - \bar{\omega} \neq 0$ the temporal amplitude $\xi(t)$ acquires a linear temporal phase\footnote{Applying a temporal phase means modifying the temporal amplitude as $\xi(t) e^{\I \phi(t)}$, preserving the distribution $|\xi(t)|^2$.} equivalent to a shift of the spectral amplitude~\cite{Karpinski2021}.
Since a complex-valued temporal amplitude might effectively make the pulse not resonant, a real $\xi(t)$ is to some extent a ``natural assumption on resonance''~\cite{Rag2017}, a choice we will also make in the following.
A detailed study of the effect of detuning on pulsed quantum light spectroscopy of a two-level atom will be presented in following publications~\cite{Khan2022,Khan2022a}.

\subsection{Local quantum estimation theory}\label{subsec:estimation}

Estimation theory quantifies the precision in estimating the value of a parameter $\Gamma$ from experimental observations $x$ distributed according to a probability distribution $p_\Gamma(x)$ belonging to a parametric family.
Quantum mechanically, $p_\Gamma(x)=\mathrm{Tr}\left( \rho_{\Gamma} \mathrm{\Pi}_{x} \right)$--the probability distribution of the collected data is obtained from the Born rule.
Here $\rho_{\Gamma}$ is a quantum state depending on the parameter $\Gamma$, $\Pi_{x}$ is an element of a positive operator-valued measure (POVM), which mathematically describes a quantum measurement~\cite{Heinosaari2011a}, and $x$ labels the possible experimental outcomes. 
For example, in a photon counting measurement $x$ is the number of detected photons.
In this paper, $\rho_{\Gamma}$ may correspond either to the joint state of the atom and light or to the reduced state of the light only.

Denoting by $\tilde{\Gamma}$ an unbiased estimator of $\Gamma$, its variance satisfies the Cramér-Rao bound (CRB)~\cite{VanTrees2013}
\begin{equation}
\label{eq:CCRB}
\operatorname{Var}[\tilde{\Gamma}]
\geq  \frac{1}{M \mathcal{C}(\rho_{\Gamma}, \mathrm{\Pi}_{x}) },
\end{equation}
where $M$ is the number of repetitions of the experiment and $\mathcal{C}(\rho_{\Gamma}, \mathrm{\Pi}_{x})$ is the classical Fisher information (CFI)\footnote{Since the CFI depends only on the classical probability distribution $p_\Gamma(x),$ we will often use $\mathcal{C}(p)$ to denote the CFI of a particular $\Gamma$-dependent distribution $p$, dropping the $\Gamma$ dependence too.} defined as
\begin{equation} \label{eq:CFIM}
	\mathcal{C}(\rho_{\Gamma}, \mathrm{\Pi}_{x}) =
	\sum\limits_x \frac{1}{p_\Gamma( x )}\left( \frac{\partial p_\Gamma(x)}{\partial \Gamma}\right)^{2},
\end{equation}
where the summation becomes an integral for continuous distributions.
Since the inequality~\eqref{eq:CCRB} can be saturated in the limit $M \to \infty$~\cite{VanTrees2013}, 
e.g., by the maximum likelihood estimator, the CRB captures the maximum precision that can be extracted by the collecting data from the distribution $p_\Gamma(x)$.

To identify the fundamental quantum limit on the variance of the estimator, the CFI must be maximised over all possible POVMs, obtaining~\cite{helstrom1976quantum,Holevo2011b,Braunstein1994,Paris2009}
\begin{equation} 
\label{eq:maxCCRB}
\max _{\{\Pi_x\}} ~ 
\mathcal{C}(\rho_{\Gamma}, \mathrm{\Pi}_{x}) = \mathcal{Q}(\rho_{\Gamma}),
\end{equation}
where we introduced the quantum Fisher information (QFI), defined as
\begin{equation}\label{eq:QFIM}
\mathcal{Q}(\rho_{\Gamma})=\mathrm{Tr} \left[ \rho_{\Gamma} L_{\Gamma} ^{2} \right],
\end{equation}
where $L_{\Gamma}$ is the symmetric logarithmic derivative (SLD), an Hermitian operator satisfying the Lyapunov equation 
\begin{equation}\label{eq:Lyap_def}
\frac{\partial \rho_{\Gamma}}{\partial \Gamma} =  \frac{\rho_{\Gamma}L_{\Gamma} + L_{\Gamma} \rho_{\Gamma}}{2}.
\end{equation}
The bounds on the estimation precision are thus
\begin{equation} \label{eq:QCRB}
\operatorname{Var}[\tilde{\Gamma}] \geq \frac{1}{M \mathcal{C}(\rho_{\Gamma}, \mathrm{\Pi}_{x}) }  \geq \frac{1}{M \mathcal{Q}(\rho_{\Gamma}) }.
\end{equation}
The latter inequality is known as the quantum CRB on the variance.
We assume that $M$ can be made sufficiently large, so that we can meaningfully focus on the CFI and QFI as the relevant figures of merit to quantify the estimation precision.
This setting is known as local estimation, since the CFI and QFI are defined locally around the true value of the parameter ($\rho_\Gamma$ and $\partial_{\Gamma} \rho_\Gamma$ are evaluated at the true value of $\Gamma$ in all the equations above).
We stress that, while $M$ copies of are required, they need not be measured collectively to attain the fundamental quantum bound for single-parameter estimation, since the optimization in Eq.~\eqref{eq:maxCCRB} is over measurements on a single copy of the system.\footnote{While the optimal measurement may depend on the unknown parameter, performing first a rough estimate using a sublinear amount of copies is enough to attain the bound~\cite{Barndorff-Nielsen2000}.}
For small $M$, non-local approaches such as Bayesian or minimax estimation are more suitable.

Note that the CFI and QFI are dimensional quantities if the parameter has physical dimensions.
To ease the comparison for different parameter values, we will focus on the dimensionless QFI $\Gamma^2 \mathcal{Q}(\rho_{\Gamma})$ that captures the estimation precision relative to the true parameter value.
It sets a fundamental upper bound on the relative estimation precision $\Gamma^2/\mathrm{Var}[\tilde{\Gamma}]$, formally the squared inverse coefficient of variation of the unbiased estimator.
These quantities are sometimes referred to as the quantum and classical signal-to-noise ratio (SNR)~\cite{Paris2009}; 
we do not adopt this terminology to avoid confusion with homonymous experimental quantities.

While the QFI defined in Eq.~\eqref{eq:QFIM} does not have a simple closed-form expression in general and must be evaluated by diagonalizing the density matrix~\cite{Paris2009,Liu2014a}, in some cases more explicit formulas can be obtained.
For a pure state $\ket{\Psi_\Gamma}$
\begin{equation}
    \label{eq:qfi_pure}
    \mathcal{Q}(\ket{\Psi_{\Gamma}})= 4 \left( \braket{\partial_\Gamma \Psi_{\Gamma}}{\partial_\Gamma \Psi_{\Gamma}} - \left| \braket{\partial_{\Gamma}\Psi_{\Gamma}}{\Psi_{\Gamma}} \right|^2 \right).
\end{equation}
Another case that will be relevant is the rank-2 mixed state
$\rho_\Gamma = \ket*{\widetilde{\psi}_e}\bra*{ \widetilde{\psi}_e} + \ket*{ \widetilde{\psi}_g} \bra*{ \widetilde{\psi}_g}$,
obtained from tracing out the atomic degrees of freedom from a pure state of the form $\ket{\psi_\Gamma} = \ket{e}|\widetilde{\psi}_e\rangle + \ket{g} | \widetilde{\psi}_g \rangle$.
The two vectors $\ket*{\widetilde{\psi}_e}$ and $\ket*{\widetilde{\psi}_g}$ describing the quantum states of the field are neither normalized nor mutually orthogonal, and generally infinite-dimensional.
In this paper we will use the tilde to denote unnormalized state vectors.
In this scenario, the QFI can be evaluated explicitly without rewriting $\rho_\Gamma$ on an orthonormal basis by solving Eq.~\eqref{eq:Lyap_def} using non-orthogonal bases~\cite{Genoni2019,Bisketzi2019,Fiderer2021a}, as recently shown in the context of superresolution imaging.
We use this technique in Sec.~\ref{sec:short_pulses}.

Another useful property of the QFI is the extended convexity~\cite{Alipour2015,Ng2016}
\begin{equation}
    \label{eq:ExtendedConvexityQFI}
    \mathcal{Q}\biggl( \sum_m p_{m,\Gamma}  \rho_{m,\Gamma}  \biggr) \leq \mathcal{C}\left( p_{m,\Gamma} \right) + \sum_i p_{i,\Gamma }\mathcal{Q}\left( \rho_{i,\Gamma} \right),
\end{equation}
where $p_{m,\Gamma}$ is a (potentially parameter-dependent) probability distribution and $\rho_{m,\Gamma}$ are normalized quantum states.
In words, the QFI of a generic mixture is upper bounded by the CFI of the mixing probability plus the average QFI of the states; this reduces to standard convexity when the mixing probability does not depend on the parameter.
This equation can be understood as a consequence of the monotonicity of the QFI under completely positive, trace-preserving maps~\cite{Shitara2016}, since the right-hand side of Eq.~\eqref{eq:ExtendedConvexityQFI} is the QFI of the state $\sum_m p_{m,\Gamma} \rho_{m,\Gamma} \otimes |m \rangle \langle m |  $ 
while the left-hand side is obtained via its partial trace, potentially losing information.
In the context of probabilistic quantum metrology, a state in this form can be obtained by making a selection measurement on an initial state and storing the outcome in an ancillary system that acts as a classical register~\cite{Combes2014}.
If the states $\rho_{m,\Gamma}$ have support in mutually orthogonal subspaces (at least in the neighbourhood of the true parameter value), then the information contained in the classical register $\{ | m \rangle \langle m | \} $ is formally redundant, since they are perfectly distinguishable, and Eq.~\eqref{eq:ExtendedConvexityQFI} is saturated with equality~\cite{Demkowicz-Dobrzanski2009}.

\section{Single-Photon Pulses}
\label{sec:1photon}

We now begin the presentation of our results on quantum light spectroscopy of a two-level atom using single-photon pulses.
This is a simple, yet conceptually rewarding and practically relevant scenario of  quantum light spectroscopy.
The results of this section can be applied to arbitrary pulse shapes, but we present analytical expressions for a few paradigmatic ones studied in the literature~\cite{Wang2011b}, often focusing on a rectangular pulse for clarity.
This is more than a mere theoretical exercise, since the realization of nontrivial single-photon wavepackets is a well-developed experimental field~\cite{Farrera2016,Pursley2018,Morin2019,Karpinski2021,Lipka2021b}.

We first present the general, analytical expression for the QFI for single-photon pulses.
Then we evaluate the QFI for various pulse shapes when the atom is perfectly coupled to the incoming pulse, i.e, $\Gperp =0$.
Finally, we present results for an atom that can also emit spontaneously into an environment, focusing in particular on the free-space case where $\Gperp \gg \Gamma $, where we elaborate on the relation to single-photon absorption spectroscopy.

\subsection{General expressions}

\subsubsection{Unitary evolution of atom, pulse and environment}
\label{subsubsec:1photonEvo}

We start by assuming the atom to be in the ground state.
Then, the global atom-pulse-environment state never contains more than one excitation due to the form of the interaction Hamiltonian in Eq.~\eqref{eq:Hint_PE}.
This state is given by (omitting the explicit time dependence for brevity)
\begin{equation}
    \label{eq:1ph_pulse_full_state}
    \ket{\Psi^\mathrm{APE}} = \psi_e \ket{e} \ket{0^\mathrm{P}} \ket{0^\mathrm{E}} + \ket{g} \left( \ket*{\widetilde{\psi}_{g}^\mathrm{P}} \ket{0^{\mathrm{E}}} + \ket{0^\mathrm{P}} \ket*{ \widetilde{\psi}_{g}^\mathrm{E} } \right),
\end{equation}
where 
$ \psi_e$ is the time-dependent amplitude of excitation of the atom, and
\begin{eqnarray}
\ket*{\widetilde{\psi}_{g}^\mathrm{P}(t)} &=& \int_{-\infty}^\infty d\tau \widetilde{\psi}_g^\mathrm{P}(t,\tau) a^\dag (\tau)  \ket{0^\mathrm{P}},  \\
\ket{\widetilde{\psi}_g^\mathrm{E}(t)} &=& \int_{-\infty}^\infty d\tau \widetilde{\psi}_{g}^\mathrm{E}(t,\tau) b^\dag (\tau)  \ket{0^\mathrm{E}},
\end{eqnarray}
are unnormalized single-photon states in the pulse and environment modes respectively.
For clarity we have explicitly separated the vacuum in the pulse and environment modes.

Solving the Schrödinger equation $ \I  \hbar \, d \ket{\Psi^\mathrm{APE}(t)} /dt =  H_{\mathrm{I}}^\mathrm{APE}(t) \ket{\Psi^\mathrm{APE}(t)} $ 
for the interaction-picture Hamiltonian in Eq.~\eqref{eq:Hint_PE} assuming the initial state $\ket{\Psi^\mathrm{APE} (t_0)} = \ket{g} \left( \int_{-\infty}^\infty d\tau \xi(\tau) a^\dag (\tau)  \ket{0^\mathrm{P}}  \right) \ket{0^\mathrm{E}} $ gives~\cite{Konyk2016} (see also Ref.~\cite[Appendix D]{Ko2022})
\begin{align}
    \psi_e(t) &= - \sqrt{ \Gamma} \int_{t_0}^t \! dt' \, e^{-\frac{1}{2}(\Gamma + \Gperp)(t-t')} \xi(t')  \label{eq:1ph_pulse_psie} \\ 
    \ket*{\widetilde{\psi}_{g}^\mathrm{P}(t)} &= \int_{t_0}^\infty \! d\tau \left( \xi(\tau) + \sqrt{\Gamma} \Theta(t-\tau) \psi_e(\tau) \right) a^\dag(\tau) \ket{0^\mathrm{P}} \label{eq:1ph_pulse_unnorm_state}\\
    \ket*{\widetilde{\psi}_{g}^\mathrm{E}(t)} &= \sqrt{\Gperp} \int_{t_0}^t \! d\tau \, \psi_e(\tau) b^\dag (\tau) \ket{0^\mathrm{E}}, \label{eq:1ph_env_state}
\end{align}
where $\Theta(x)$ is the Heaviside step function.
Note that for $\xi(t) \in \mathbb{R}$ the amplitudes of the evolved wavefunctions remain real, as will be the case for the explicit calculations in the following subsections.
We also assume that $t_0$ is well before the arrival of the pulse so that we may set $t_0=-\infty$ in the integrals.
For $\Gamma >0$ there is a nonzero excitation probability $p_e(t) = | \psi_e(t) |^2$ which tends to zero for large times: $ \lim_{t\to \infty} p_e(t)  = 0$.
This happens even when $\Gperp=0$, meaning that the atom spontaneously emits into the pulse;
in this case the final state $\ket{\psi_{g}^{\mathrm{P},\infty}} \equiv \lim_{t \to \infty} \ket*{\widetilde{\psi}_{g,P}(t)}$ is a normalized one-photon wavepacket with temporal amplitude $\xi(\tau) + \sqrt{\Gamma} \psi_e(\tau)$.
Note also that the global state defined in Eqs.~\eqref{eq:1ph_pulse_full_state}--\eqref{eq:1ph_env_state} is normalized, as we show explicitly in Appendix~\ref{app:normalization1ph}.

Assuming that only the P subsystem is accessible for detection, we trace out both the A and E subsystems, obtaining an incoherent mixture of the vacuum and the modified single photon wavepacket
\begin{equation}
    \label{eq:single_photon_PE_mixed}
    \rho^{\mathrm{P}} = \left( |\psi_e|^2 + \braket*{ \widetilde{\psi}_{g}^\mathrm{E}}{ \widetilde{\psi}_{g}^\mathrm{E} } \right) \ket{0^\mathrm{P}}\bra{0^\mathrm{P}} + \ket*{\widetilde{\psi}_{g}^\mathrm{P}}\bra*{\widetilde{\psi}_{g}^\mathrm{P}},
\end{equation}
where again we have suppressed the explicit time dependence.
In the long-time limit $t\to \infty$ the atom decays to the ground state and becomes disentangled from the light, but for $\Gperp > 0$ the initial photon of the pulse is partly lost to the environment.

\subsubsection{Single-photon QFI: classical and quantum contributions}

The state in Eq.~\eqref{eq:single_photon_PE_mixed} has the form $\rho_{\Gamma}= p_\Gamma \dyad{0}{0} + (1-p_\Gamma )\dyad{\psi_\Gamma}{\psi_\Gamma} $, where we have highlighted the dependence on the parameter of interest $\Gamma$ and written it in terms of a normalized single-photon state $\braket{\psi_\Gamma}{\psi_\Gamma} = 1$.
For clarity,
\begin{equation}
p_\Gamma =  |\psi_e|^2 + \braket*{ \widetilde{\psi}_{g}^\mathrm{E}}{ \widetilde{\psi}_{g}^\mathrm{E} }, ~~~
\ket{\psi_\Gamma} = \ket*{\widetilde{\psi}_{g}^\mathrm{P}} / \sqrt{\braket*{\widetilde{\psi}_{g}^\mathrm{P}}{\widetilde{\psi}_{g}^\mathrm{P}}}.
\end{equation}
The QFI of $\rho_{\Gamma}$ is then the CFI of the two-outcome probability distribution $\{ p_\Gamma,1-p_\Gamma \} $ plus the QFI of the pure single-photon state rescaled by the corresponding probability:
\begin{align}
    \mathcal{Q}(\rho_{\Gamma}) &= \frac{ (\partial_\Gamma p_\Gamma)^2 }{p_\Gamma (1-p_\Gamma)} + (1-p_\Gamma) \mathcal{Q}\left( \ket{\psi_\Gamma} \right) \\ 
    & \equiv ~ \mathcal{C}(p_\Gamma) + \tilde{\mathcal{Q}}(\ket{\psi_\Gamma}).
    \label{eq:QFI_1ph_normalized}
\end{align}
This QFI equals the right-hand side of~\eqref{eq:ExtendedConvexityQFI}, saturating the extended convexity bound, since the two pure states in the mixture are orthogonal and the vacuum contains no information on $\Gamma$.

There are two contributions to the fundamental limit on the precision of estimating $\Gamma$:

(i) The probability $p_\Gamma$ of losing a photon from the pulse due to absorption by the atom, giving the CFI $\mathcal{C}\left( p_\Gamma \right) $ which we call the classical contribution to the total QFI $\mathcal{Q}(\rho_{\Gamma})$, and 

(ii) The perturbation to the temporal shape of the single-photon wavepacket due to sponatenous emission, giving the QFI of the pure single-photon state (rescaled by the corresponding probability of not losing a photon) $\tilde{\mathcal{Q}}(\ket{\psi_\Gamma}) =(1-p_\Gamma) \mathcal{Q}\left( \ket{\psi_\Gamma} \right) = 4 (1-p_\Gamma) \left( \braket{\partial_\Gamma \psi_\Gamma}{\partial_\Gamma \psi_\Gamma}  -  \left\vert \braket{\partial_\Gamma \psi_\Gamma}{ \psi_\Gamma} \right\vert^2 \right)$, which we call the quantum contribution.

In Appendix~\ref{app:1photonQFI} we report more explicit expressions for $\xi(t) \in \mathbb{R}$ in terms of the unnormalized single photon state $\ket*{\widetilde{\psi}_{g}^\mathrm{P}}.$ 
These are more convenient in the calculations of the following sections.

\subsubsection{Optimal measurements}
\label{subsubsec:1phOptMeas}

We now discuss the means of attaining (i) and (ii) above.

The classical term (i) in the QFI~\eqref{eq:QFI_1ph_normalized} is attained by any measurement that perfectly distinguishes the vacuum from the single-photon component.
In principle, this is always possible and corresponds to a POVM with an element $\Pi_0 = |0^\mathrm{P} \rangle\langle 0^\mathrm{P}|$, completed on the single photon subspace by a POVM with outcomes $s$ (continuous or discrete) $\Pi_{1,s} = \int d \tau d \tau' \Pi_{1,s} ( \tau, \tau') a^\dag(\tau) |0^\mathrm{P} \rangle\langle 0^\mathrm{P}| a(\tau')$. 
This yields the joint probabilities $p_0 = \Tr[  \rho_\Gamma \Pi_0 ] \equiv p_\Gamma$ and 
$p_{s,1} = \Tr[  \rho_\Gamma \Pi_{1,s} ] =  (1-p_\Gamma) \langle \psi_\Gamma | \Pi_{1,s} | \psi_\Gamma \rangle$, corresponding to the marginal and conditional probabilities $p_1 = \int \! ds \, p_{s,1} = 1-p_\Gamma$ and $p_{s|1}= p_{s,1} / p_1  =\langle \psi_\Gamma | \Pi_{1,s} | \psi_\Gamma \rangle$.
The chain rule~\cite{Zamir1998} gives the overall CFI $\mathcal{C}\left(p_\Gamma\right) + (1-p_\Gamma) \mathcal{C}\left( p_{s|1} \right)$\footnote{The chain rule can be applied by considering two random variables: the photon number~(either 0 or 1) and the outcome of the single-photon POVM $\Pi_{1,s}$, conditional on having the outcome 1 in first random variable.}.
The first term of this expression is exactly the classical contribution (i) in Eq.~\eqref{eq:QFI_1ph_normalized}, for any choice of the single-photon POVM $\Pi_{1,s}$.
Physically, this is the information obtained by measuring the photon loss, as in single-photon absorption spectroscopy.
Indeed, the CFI $\mathcal{C}\left(p_\Gamma\right)$ represents all the information available when the measurement detects the presence of a photon but is insensitive to the shape of the wavepacket, i.e., a trivial single outcome POVM $\Pi_1(\tau,\tau') = \delta(\tau-\tau')$ (formally the identity in the single-photon subspace such that $\langle \psi_\Gamma | \Pi_{1} | \psi_\Gamma \rangle = 1 $ for any $\ket{\psi_\Gamma}$).

The quantum term (ii) can be attained by choosing an appropriate single-photon POVM.
A projection onto the output state $\ket{\psi_{\Gamma}}$ itself (more precisely the projection on a state $\ket{\psi_{\Gamma'}}$ in the limit $\Gamma'\to \Gamma$) saturates the QFI.
This is an optimal measurement in local pure-state quantum estimation~\cite{Macri2016}.
For completeness we show this explicitly in Appendix~\ref{app:OptProjSameState}.
There are, however, infinitely many POVMs that saturate the QFI for pure states and they generally differ in how robust they are to imperfections in their practical implementation.
The most robust POVM for which the CFI is least degraded by a worst-case small perturbation can be found exactly~\cite{Kurdzialek2022}, and chosen in absence of other practical constraints.

Such formally optimal measurements may however be experimentally impractical.
Consequently, we present two more measurements that are optimal for attaining (ii) when $\xi(t) \in \mathbb{R}$, an assumption that we will make for the explicit results in the next sections.
The first is the POVM 
\begin{equation}
\Pi_{1,s}(\tau,\tau') = \delta(s-\tau) \delta(\tau-\tau')
\label{eq:tcspc}
\end{equation}
yielding the conditional probability density $p_{s|1}=\left| \psi_\Gamma(s) \right|^2$.
For $\psi_\Gamma(s) \in \mathbb{R} $ and $\partial_\Gamma \psi_\Gamma(s) \in \mathbb{R}$ 
we have $\int \! ds \, \partial_\Gamma \psi_\Gamma(s) \psi(s) = 0$ and the CFI of the probability distribution $p_{s|1} = \psi_\Gamma(x)^2$ equals the pure-state QFI, since $\int ds \left[ \partial_\Gamma \psi_\Gamma(s)^2\right]^2 / \psi(s)^2 =4 \int d\tau \left[\partial_\Gamma \psi_\Gamma(s) \right]^2$.

The POVM in Eq.~\eqref{eq:tcspc} operationally corresponds to measurement of photon arrival times, which can be accomplished using time-correlated single-photon counting~(TCSPC)~\cite{becker2004fluorescence}.
While TCSPC is ordinarily used in lifetime measurements using single-photon fluorescence spectroscopy, where the atom is assumed to be already excited and the excitation process itself is not modeled~(TCSPC is, in fact, optimal for single exponential lifetime detection~\cite{Mitchell2022}), our proposal is somewhat different.
The quantum term (ii) is saturated by the TCSPC measurement of the probability distribution $p_{s|1}$, which amounts to the measurement of the modulus-squared of the time-dependent envelope of the conditional single-photon state.
However, to measure also the absorption probability and saturate the classical term (i), one should also know precisely the number $N_{\mathrm{inc}}$ of photons incident on the two-level system, so that the probability can be estimated as $p_0 = 1- N_{\mathrm{TCSPC}}/N_{\mathrm{inc}}$ for large $N_{\mathrm{inc}}$, where $N_{\mathrm{TCSPC}}$ is the total number of TCSPC counts.
This is different from standard fluorescence detection, where one assumes the measured single-photon wavepacket to be a decaying exponential, and $N_{\mathrm{TCSPC}}$ only acts as a normalization factor.

The second optimal measurement is to detect the light in a discrete orthonormal basis of temporal modes that includes the original pulse temporal mode, corresponding to a rank-1 projective POVM $\Pi_{1,s} = | \xi_s \rangle \langle \xi_s |$, where $\ket{\xi_s} = \int_{-\infty}^\infty d\tau  \xi_s(\tau) a^\dag (\tau) \ket{0_P}$.
The optimality of temporal mode-resolved photodetection here has been studied in detail in Ref.~\cite{Khan2022a} and will be presented in a follow-up paper~\cite{Khan2022}. 
Practically, such mode resolved photon counting measurements can be achieved using quantum pulse gating~(QPG) techniques~\citep{Brecht2015,eckstein2011quantum,donohue2018quantum,de2021effects,ansari2021achieving,garikapati2022programmable} for ultrafast pulses, where an incoherent train of pulses interacts with a sufficiently shaped gating pulse in a sum-frequency~(SF) interaction inside a nonlinear crystal.
The shape of the gating pulse determines the mode the incoming pulse is effectively projected onto, presenting at the output as a higher frequency signal than the incoming pulse.
Therefore, with the right toolbox of gating pulses, the optimal measurement for $\Gamma$-estimation is accessible, in principle.

Finally, we consider a suboptimal measurement that will be relevant subsequently: 
Detecting photons only in the original unperturbed temporal mode of the pulse.
For single photons, this corresponds to a two-outcome POVM with elements $\Pi_1= \ket{\xi} \bra{ \xi}$ and $\Pi_0 = \id - \Pi_1,$ where $ \ket{\xi} = \int_{-\infty}^\infty d\tau  \xi (\tau) a^\dag (\tau) \ket{0^\mathrm{P}}$ is the initial state of the single-photon pulse.
The probability of such a detector clicking for the state in Eq.~\eqref{eq:single_photon_PE_mixed} is thus $p_{\mathrm{orig}} = \Tr \left[  \Pi_1 \rho_\Gamma (t) \right] = |\braket{\xi}{\psi_{g}^\mathrm{P}(t)}|^2 $.
Clearly this measurement is suboptimal because it does not 
discriminate the vacuum from temporal modes orthogonal to $\xi$, which become populated due to the spontaneous emission of the atom.
The performance of this measurement in attaining the overall QFI $\mathcal{Q}(\rho_{\Gamma})$ in Eq.~\eqref{eq:QFI_1ph_normalized} is addressed in the next section for some specific examples; we will see that this measurement is optimal or close to optimal when the effect of spontaneous emission is negligible.

\subsection{Perfect atom-pulse coupling}
\label{subsec:1phPerfCouplRes}

\begin{figure}
    \includegraphics{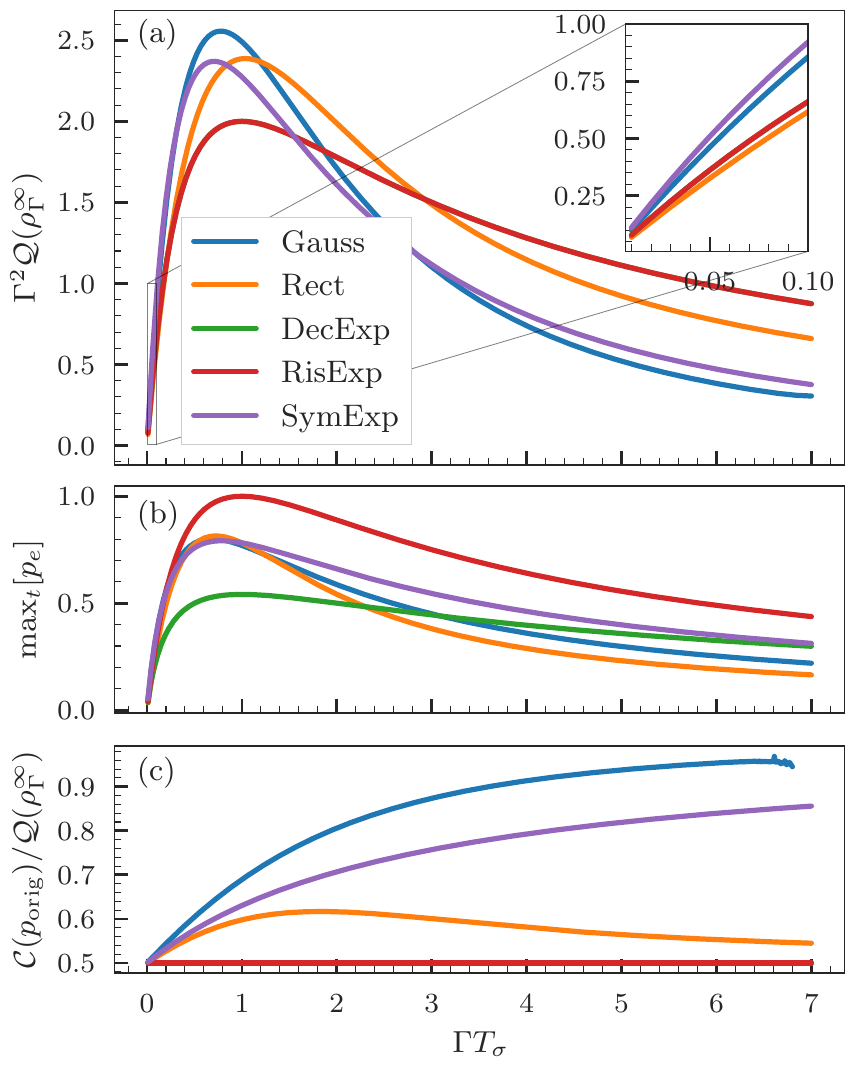}
    \caption{
    Panel (a): Dimensionless QFI $\Gamma^2 \mathcal{Q}(\rho_\Gamma^\infty)$ of the atom-pulse coupling parameter $\Gamma$ for the asymptotic single-photon wavepacket (with perfect coupling, $\Gperp=0$), as a function of the dimensionless quantity $\Gamma T_\sigma$, where $T_\sigma$ is the standard deviation of pulse temporal distribution.
    The pulse shapes are (from top to bottom in the legend): Gaussian, rectangular, decaying and rising exponential (with overlapping lines) and symmetric exponential.
    Panel (b): Maximal excitation probability of the two-level atom.
    Panel (c): Ratio between the FI $\mathcal{C}(p_\mathrm{orig})$ for photodetection in the original pulse mode and the asymptotic QFI, decaying and rising exponential are overlapping. 
    }
    \label{fig:QFIshapes}
\end{figure}

To further clarify our understanding of quantum light spectroscopy using single-photon pulses, we now focus on perfect atom-pulse coupling by setting $\Gperp = 0$.
We study the effect of different pulse shapes by considering a few paradigmatic real-valued temporal amplitudes $\xi(t)$.

\subsubsection{Asymptotically long time}

We start by studying the asymptotic case $t \to \infty$, when the final state of the pulse contains exactly one photon, and is pure and disentangled from the atom, 
i.e., $p_{\Gamma} = 0$.
Then all the information about the parameter $\Gamma$ is encoded in the temporal shape of the wavepacket, which is perturbed due to the interaction.

In Fig.~\ref{fig:QFIshapes}(a) we show the QFI of the asymptotic single-photon state as a function of pulse duration, for various pulse shapes.
We define the pulse duration $T_\sigma$ as the standard deviation of the initial single-photon temporal distribution $\xi(t)^2$ to aid the comparison of different shapes; we will later use a different convention for rectangular pulses, whose duration is unambigously defined. 
The mathematical descriptions of the pulse shapes considered in Fig.~\ref{fig:QFIshapes} are provided in Appendix~\ref{app:1phShapes}, together with the available analytical expressions for the quantities of interest.
Note that the dimensionless QFI depends solely on the dimensionless combination $\Gamma T_\sigma$.

Fig.~\ref{fig:QFIshapes}(a) shows that the various pulse shapes display the same qualitative behaviour.
The QFI increases approximately linearly as the pulse duration increases from $T_\sigma=0$, as shown in the inset.
It reaches a maximum for a value around the fluorescence lifetime $\Gamma T_\sigma = 1.$
Overall, there is a mild dependence on the particular shape.
This behaviour is similar to that of the maximum excitation probability of the atom~\cite{Wang2011b,Rag2017}, shown in Fig.~\ref{fig:QFIshapes}(b).
One might naively think that a higher excitation probability of the atom, which corresponds in some sense to a ``better'' interaction between the atom and the pulse, would correspond to a higher QFI of the outgoing pulse of light. 
Our results show otherwise.
Firstly, the optimal pulse duration for a given pulse shape for the two quantities are different.
Secondly, while a rising and a decaying exponential pulse of the same duration yield the same asymptotic QFI, i.e., overlapping curves in Fig.~\ref{fig:QFIshapes}, the rising exponential is optimal to excite the atom~\cite{Stobinska2009a,Rag2017} (reaching one in the inset plot), while the decaying exponential performs much worse.

In Fig.~\ref{fig:QFIshapes}(c), we show how much information can be extracted by detecting the photon in the original temporal mode compared to the information available in the asymptotic state by plotting the ratio between the CFI of this detection strategy $\mathcal{C}(p_\mathrm{orig})$ (for $t\to \infty$) and the asymptotic QFI (plotted on its own in Fig.~\ref{fig:QFIshapes}(a)).
In the limit of short pulses $T_\sigma \to 0$ this ratio tends to the value $1/2$; curiously it is always $1/2$ for the rising and decaying exponentials.
This has been proven exactly for all pulse shapes except for Gaussian pulses, for which all the quantities must be evaluated by solving the integrals numerically (this is also the reason for the numerical noise for large $\Gamma T_\sigma$ in Fig.~\ref{fig:QFIshapes}(c)).
We conjecture this to be a general feature of this metrological problem in the $T_\sigma \to 0$ limit, since the details of the pulse shape should be less relevant in this regime.

It is remarkable that for quantum light spectroscopy with single-photon pulses, this simple detection strategy yields a substantial fraction of the maximal information available, quantified by the asymptotic QFI, about the parameter.
As we show in the next section, measuring the photon in the original temporal mode has the advantage that the information can be obtained rather rapidly after the interaction (about femto- or picoseconds for ultrafast pulses), without the need to wait for the atom to decay (timescale of nanoseconds in standard atomic and molecular systems).
While this may be of limited appeal in spectroscopy, it may be exploited in quantum information processing.

Finally, one could in principle, optimize the pulse shape to maximise the estimation precision.
Given the recent advances in the experimental shaping of single-photon wavepackets~\cite{Farrera2016,Morin2019,Karpinski2021,Knall2022}, this could be of practical use.


\subsubsection{Finite time}

\begin{figure}
    \includegraphics{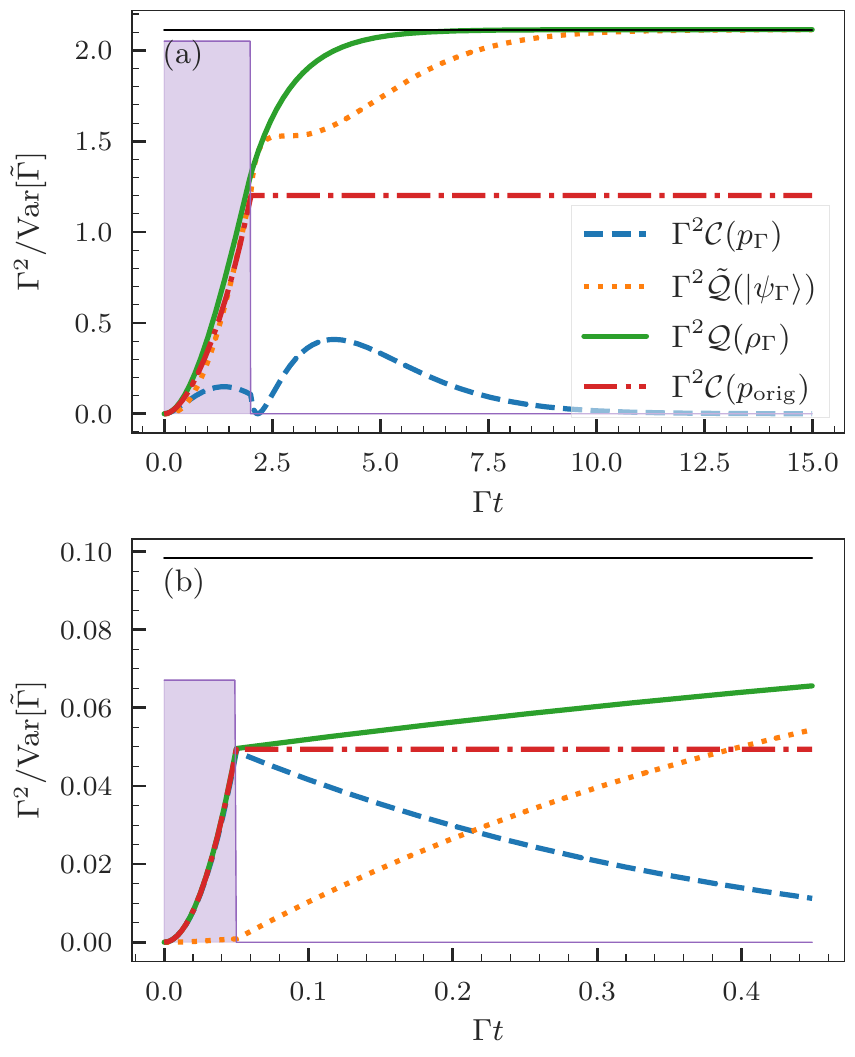}
    \caption{Upper bounds on the relative estimation precision $\Gamma^2 / \mathrm{Var}[\tilde{\Gamma}]$ as a function of time (in units of $1/\Gamma$) for a rectangular single-photon pulse with perfect atom coupling $\Gperp = 0$, panel (a): $\Gamma T =2$; panel (b): $\Gamma T = 1/20$.
    The curves represent: The classical contribution $\Gamma^2 \mathcal{C}(p_\Gamma)$ (dashed blue), the quantum contribution $\Gamma^2 \tilde{\mathcal{Q}} (\ket{\psi_\Gamma}) $ (dotted orange), the overall dimensionless QFI $\Gamma^2 \mathcal{Q}(\rho_\Gamma)$ (solid green, sum of the previous two), the dimensionless CFI $\Gamma^2 \mathcal{C}(p_{\mathrm{orig}})$ for photodetection in the original temporal mode (dot-dashed red) and the asymptotic value $\lim_{t \to \infty} \Gamma^2 \mathcal{Q}(\rho_\Gamma)$ (thin black line).
    The shaded purple region shows the pulse temporal distribution $|\xi(t)|^2$ as a guide for the eye, it is not to scale on the vertical axis.  
    }
    \label{fig:Rectangular_QFIcontrib_vs_time}
\end{figure}

For finite $t,$ the atom remains partially excited and the overall atom-pulse state entangled, so the classical contribution $\mathcal{C}(p_\Gamma)$ in Eq.~\eqref{eq:QFI_1ph_normalized} now plays a role.
While atom-pulse entanglement could in principle mean that not all the information is accessible by measuring the pulse subsystem only, at least for real-valued $\xi(t)$ and $\Gperp = 0$ this is not the case.
In fact, there is no information about $\Gamma$ in the relative phase of the atom-pulse state $\psi_e \ket{e} \ket{0^\mathrm{P}} + \ket{g} \ket*{\widetilde{\psi}_{g}^\mathrm{P}}$ and it is easy to verify that the QFI of this pure state is equal to the QFI of the reduced pulse state $\psi_e^2  \ket{0^\mathrm{P}}\bra{0^\mathrm{P}} + \ket*{\widetilde{\psi}_{g}^\mathrm{P}}\bra*{\widetilde{\psi}_{g}^\mathrm{P}}$, i.e., in Eq.~\eqref{eq:single_photon_PE_mixed} for $\Gperp = 0$.

To highlight the qualitative features in this regime, we focus on a rectangular pulse $\xi(t) = \sqrt{1/T } \Theta\left( t \right) \Theta\left(T  - t\right)$ supported on an interval of duration $T$\footnote{The duration parameter $T$ is a multiple of the one used in Fig.~\eqref{fig:QFIshapes}, $T_\sigma = T / \sqrt{12}$.}, starting at $t_0=0$, where $\Theta(x)$ is the Heaviside step function.
This choice makes both analytical calculations possible and the identification of the beginning and the end of the pulse unambiguous.

In Fig.~\ref{fig:Rectangular_QFIcontrib_vs_time} we plot the two contributions to the QFI in Eq.~\eqref{eq:QFI_1ph_normalized} separately as a function of time, as well as the CFI $\mathcal{C}(p_{\mathrm{orig}})$ obtained by detecting the photon in the original temporal mode.
For a pulse of duration comparable to the fluorescence lifetime, $\Gamma T = 2,$ Fig.~\ref{fig:Rectangular_QFIcontrib_vs_time}(a) shows 
that the total QFI approaches its asymptotic value by an interplay of the two contributions and for large times only the quantum contribution (ii) is relevant, as expected from the preceding long-time analysis.
The dot-dashed line represents the information obtained by detecting the photon wavepacket in its initial temporal mode, and it settles to a value around half of the asymptotic QFI, as previously shown in Fig.~\ref{fig:QFIshapes}(c).
Since the pulse has a finite duration $T$, the spontaneous emission
that happens after $T$ does not affect the dynamics in this temporal mode.
This effect is due to the abrupt cutoff of the rectangular pulse, but the same principle applies to other localized time envelopes $\xi(t)$ and the qualitative behaviour in this figure will be exhibited by other pulse shapes.

Fig.~\ref{fig:Rectangular_QFIcontrib_vs_time}(b) presents the results for a much shorter pulse, $\Gamma T=1/20$.
In this case the photon interacts with the atom for a short time and there is only a small distortion to the photon wavepacket.
This is witnessed by the fact that the dotted line representing the quantum contribution $\tilde{\mathcal{Q}}(\ket{\psi_\Gamma})$ increases only slightly while the pulse is interacting with the atom (shaded region).
On the other hand, during the interaction most of the information is obtained by measuring the absorption probability $p_\Gamma$, which in this regime practically coincides with $p_\mathrm{orig}$ and we have $ \mathcal{C}(p_\Gamma) \approx \mathcal{C}(p_\mathrm{orig}) $ in this region of the plot.
However, after the interaction is over $\mathcal{C}(p_\mathrm{orig})$ remains unchanged, exactly as in the upper panel, while $\mathcal{C}(p_\Gamma)$ decreases and $\tilde{\mathcal{Q}}(\ket{\psi_\Gamma})$ increases slowly to roughly $2 \mathcal{C}(p_\mathrm{orig})$ (thin black line at the top of the bottom panel) as the atom decays back to the ground state.

\subsection{Atom in free space}
\begin{figure}
    \includegraphics{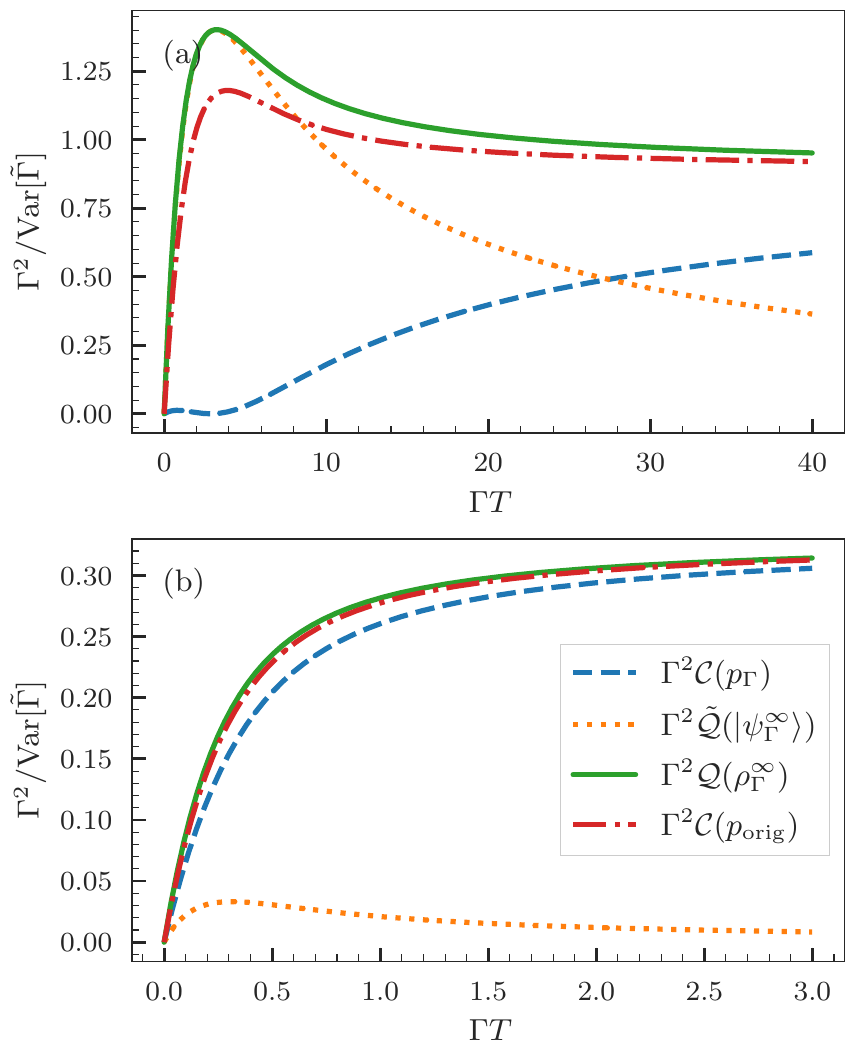}
    \caption{
        Upper bounds on the relative estimation precision $\Gamma^2 / \mathrm{Var}[\tilde{\Gamma}]$ for measurements on the asymptotic state of the pulse, as a function of the dimensionless pulse duration $\Gamma T$, for $\Gperp = \Gamma /2$ (top) and  $\Gperp = 10 \Gamma$ (bottom).
        The curves represent: The classical contribution $\Gamma^2 \mathcal{C}(p_\Gamma)$ (dashed blue), the quantum contribution $\Gamma^2 \tilde{\mathcal{Q}} (\ket{\psi_\Gamma^\infty}) $ (dotted orange), the total dimensionless QFI $\Gamma^2 \mathcal{Q}(\rho_\Gamma^\infty)$ (solid green, sum of the previous two) and the dimensionless CFI $\Gamma^2 \mathcal{C}(p_{\mathrm{orig}})$ for photodetection in the original temporal mode (dot-dashed red).
        }
    \label{fig:GammaPerpContribs}
\end{figure}

We now deal with a nonzero coupling to the additional environmental field modes, i.e., $\Gperp > 0$.
For simplicity, we focus on asymptotic results and present those for a rectangular pulse.
We expect qualitatively similar results for other shapes and we have confirmed this explicitly for decaying exponential pulses.
In Fig.~\ref{fig:GammaPerpContribs} we show two exemplary cases, one (top panel) where $\Gperp$ is smaller than but comparable to $\Gamma$, $\Gperp = \Gamma/2$ and another (bottom panel) where the coupling to the environment is significantly more relevant than to the pulse mode, $\Gperp = 10 \Gamma$.
For larger $\Gperp$ the quantum contribution due the perturbation of the photon wavepacket effected by spontaneous emission of the atom is less important and almost all the information can be retrieved by restricting measurements to the incoming temporal mode.

The choice $\Gperp / \Gamma = 10$ is intended to capture a pulse interacting with an atom in free space, without particular geometries to enhance the coupling.
It is comparable to that for the Na $D_2$ transition we consider in Sec.~\ref{sec:sodium}.
In general, for $\Gperp \gg \Gamma$ the atom is coupled far more strongly with the vacuum environment than with the pulse and thus after the excitation it will spontaneously emit predominantly into the environmental modes.

\begin{figure}
    \includegraphics{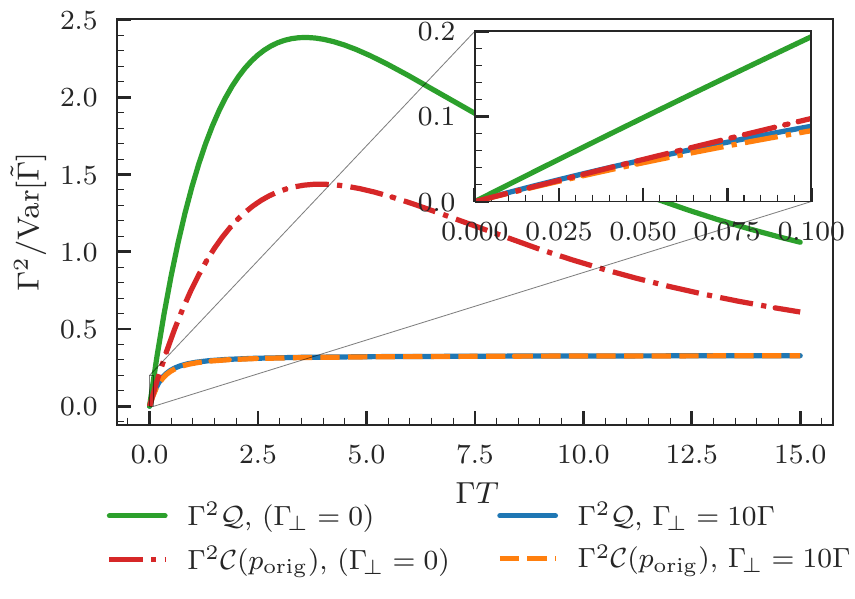}
    \caption{Upper bounds on the relative estimation precision $\Gamma^2 / \mathrm{Var}[\tilde{\Gamma}]$ for measurements on the asymptotic state of the pulse.
    The curves represent the dimensionless QFI $\Gamma^2 \mathcal{Q}(\rho_\Gamma^\infty)$ (two solid lines) and CFI $\Gamma^2 \mathcal{C}(p_\mathrm{orig})$ (two dot-dashed lines) for $\Gperp=10 \Gamma$ (two lower curves, blue and orange) and for perfect coupling $\Gperp=0$ (two top curves, green and red).
    Inset: zoom on the region close to $\Gamma T=0$.
    }
    \label{fig:QFIGammaPerpComparison}
\end{figure}
In Fig.~\ref{fig:QFIGammaPerpComparison} we compare the total QFI (including both contributions in Eq.~\eqref{eq:QFI_1ph_normalized}) and the CFI of the original temporal mode for $\Gperp / \Gamma = 10$ with that for the perfect-coupling case $\Gperp = 0$ considered in the previous section.
As expected, a larger $\Gperp$ decreases both the QFI and the CFI.
However, in the region of short pulses (shown in the inset), the CFI $\mathcal{C}(p_\mathrm{orig})$ of the perfect coupling case follows closely the CFI and the QFI of the curves for $\Gperp = 10 \Gamma$.

This observation is confirmed more generally. 
Indeed using the analytical expressions for the rectangular pulse in Table~\ref{tab:SinglePhotonShapes} of Appendix~\ref{app:1phShapes} and the expressions in Appendix~\ref{app:1photonQFI} we can show that 
 \begin{equation}
  \lim_{\Gperp \to \infty } \lim_{T \to 0} \frac{\mathcal{Q}(\rho_{\Gamma}(\Gperp),T)}{\mathcal{C}(p_\mathrm{orig}(\Gperp = 0,T ))} = 1 ,
 \end{equation}
where we have highlighted the dependence on the parameters $\Gperp$ and $T.$ 
These limits cannot be interchanged, since $\lim_{\Gperp \to \infty} \mathcal{Q}(\rho_{\Gamma}(\Gperp)) = 0 $.
Analogous results are observed numerically for the pulse shapes considered in the previous section.
These formal limits should be understood physically as the following separation of timescales $1/\omega_0 \ll T \ll 1/\Gperp \ll 1/\Gamma$ for short pulses in the free-space scenario of stronger coupling to vacuum modes than to the incoming pulse mode.
Indeed, if $\Gperp$ or $1/T$ are comparable with the optical carrier frequency $\omega_0$, the rotating-wave and slowly varying envelope approximations are typically invalid.

Physically, the above observation means the following:
The information obtained during the interaction between the atom and the pulse is the same regardless of the presence of additional environment modes because the pulse is so short (i.e., $ T \Gamma_{\mathrm{tot}} \ll 1$) that the spontaneous emission terms can be neglected during this part of the dynamics.
Moreover, in this limit all the information on $\Gamma$ is entirely retrieved by considering only the original temporal mode; this also motivates the next section where we introduce an effective single-mode model for the pulse-atom interaction.
However, by waiting until the atom decays by spontaneous emission, additional information can be obtained if the emitted photon can be measured ($\Gperp = 0$ case) but nothing more if it decays almost completely into inaccessible modes ($\Gperp \gg \Gamma$).

\section{Short time and short pulse regime}
\label{sec:short_pulses}

In this section we consider pulses with a real-valued temporal amplitude, which we rewrite as $\xi(t)= f \bigl(\frac{t-\bar{t}}{T} \bigr) / \sqrt{T}$, where $f(x)$ is a scale-invariant shape function~\cite{Ko2022}, dimensionless and squared-normalized $\int dx f(x)^2 = 1$.
We have factored out the parameter $T$ that mathematically represents a dilation of $f(x)$ and 
it captures the pulse duration when $f(x)$ is well-localized around $x = 0$, as we assume in this paper; $\bar{t}$ is a location parameter that essentially conveys the ``time of arrival'' of the pulse, often this is the peak of a unimodal temporal amplitude such as a Gaussian wavepacket.

The approximate approach presented in this section will be applied to estimating the Na $D_2$ dipole moment in Sec.~\ref{sec:sodium}.

\subsection{Approximate interaction Hamiltonian with a single temporal mode }

For short times $ t \ll 1/\Gamma_\mathrm{tot}$ spontaneous emission, either back into the pulse or into the environment, can be neglected.
Considering also short pulses $T \ll 1/\Gamma_\mathrm{tot}$, the evolution of the atom-field state according to the master equation in Eq.~\eqref{eq:LindbladPulseAtom} can be approximated by a unitary evolution obtained from a time-dependent Jaynes-Cummings (JC) interaction between the pulse temporal mode and the atom:
\begin{equation} \label{eq:tdepJC}
H_{\mathrm{JC}}(t) = \I \hbar \sqrt{\Gamma} \xi(t) \left(  A_\xi  \sigplus - A_\xi^\dag \sigmin \right).
\end{equation}
This is the same approximate model introduced in Ref.~\cite{vanEnk2002} and investigated in Refs.~\cite{Silberfarb2003,Silberfarb2004} for rectangular pulses, where its validity for paraxial beams propagating in free space was corroborated.
The idea is that in this limit the temporal mode is not distorted and the dynamics can be approximated as a coherent interaction between the atom and a single temporal mode of the field.

Considering a complete basis of orthonormal temporal modes that satisfy $\sum_k \xi_k(t) \xi^*_k(t') = \delta(t-t')$ we can formally express the white noise operators in the light-matter Hamiltonian in Eq.~\eqref{eq:interaction_hamiltonian_Gamma} as $a(t)= \int dt' \sum_k \xi_k(t) \xi_k^*(t') a(t') = \sum_k \xi_k(t) A_{\xi_k}$.
Fixing the zeroth temporal mode to be pulse temporal amplitude $\xi_0(t)=\xi(t)$, all the other temporal modes are initially empty and our approximation is tantamount to saying that for short evolution times and short pulses they continue to remain practically empty.
The dynamics is equivalent to that obtained by neglecting the terms $k>0$ in the summation.

We first checked the validity of this approximation by truncating the number of orthonormal modes and solving Scrödinger's equation numerically for Gaussian pulses and single-photon states~\cite{Bisketzi2022}.
However, extending this approach to multiphoton states is rather challenging numerically.
We have taken an alternative route and employed the methods of Refs.~\cite{Kiilerich2019,Kiilerich2020} that allow to obtain the quantum state of a specific temporal mode of the light after the interaction with the atom.
Interestingly, this method has recently been reformulated as an effective time-dependent JC interaction (as in Eq.~\eqref{eq:tdepJC}) between the atom and a fixed temporal mode, plus the interaction with an auxiliary orthogonal mode~\cite{Christiansen2022}.
We provide details on the method of Ref.~\cite{Christiansen2022} in Appendix~\ref{app:Molmer}, as well as a few plots suggesting the convergence to the time-dependent JC model as the pulse duration (and consequently the final time of the experiment) decreases, showing that not only the reduced states of the pulse coincide, but also that the auxiliary orthogonal mode remains practically empty. 
All the numerical calculations have been performed for real temporal amplitudes $\xi(t)$ and we restrict ourselves to this case for the rest of this section.

Since we are considering zero detuning, the time-dependence of the Hamiltonian in Eq.~\eqref{eq:tdepJC} is trivial and no-time ordering is needed; the solution is the same as for the standard JC model (describing a discrete cavity mode) with a redefined ``time'' variable $\int_{t_0}^t  \xi(t')  dt' \equiv G_t$~\cite{vanEnk2002}.
We assume that the pulse amplitude is localized at a time much later than $t_0$ so that formally we can set $t_0 = -\infty$ when needed, as in Sec.~\ref{sec:1photon}; we also assume that the integral $\int_{-\infty}^\infty  \xi(t')  dt'$, i.e., the total pulse-atom ``interaction time'' is finite.
As in the previous sections, we also assume the atom to be initially in the ground state.
For an arbitrary initial state of the pulse $\ket{\psi_0^\mathrm{P}}= \sum_{n=0}^\infty \psi_n \ket{n_\xi}$, where $\ket{n_\xi}$ are the Fock states in the pulse temporal mode in Eq.~\eqref{eq:fock_state_definition_MT}, the atom-pulse initial state is thus $\ket{\psi_0^{\mathrm{AP}}} = \ket{g} \ket{\psi_0^\mathrm{P}}$.
The evolved state is~\cite{Larson2021b}
\begin{equation}
    \label{eq:JCsolution}
	\begin{split}
	\ket{\psi^\mathrm{AP}(t)} =& - i \ket{e} \sum_{n=0}^\infty \sin \left( \sqrt{\Gamma} G_t \sqrt{n+1} \right) \psi_{n+1} \ket {n_\xi} \\
	 &+ \ket{g} \sum_{n=0}^\infty \cos \left( \sqrt{\Gamma} G_t \sqrt{n} \right) \psi_n \ket{n_\xi} \\
	\equiv & \ket{e} \ket*{\widetilde{\psi}_e(t)} + \ket{g} \ket*{\widetilde{\psi}_g(t)},
	\end{split}
\end{equation}
where we have introduced two unnormalized field states that also appear in the rank-2 reduced state of the field 
\begin{equation}
    \label{eq:JCreducedFieldState}
	\begin{split}
 	\rho^\mathrm{P}(t) &= \Tr_\mathrm{A} \left[ \ket*{\Psi^\mathrm{AP}(t)}\bra*{\Psi^\mathrm{AP}(t)} \right] \\
    &=  \ket*{\widetilde{\psi}_e(t)}\bra*{\widetilde{\psi}_e(t)} + \ket*{\widetilde{\psi}_g(t)}\bra*{\widetilde{\psi}_g(t)}. 
	\end{split}
\end{equation}
This state is mixed since the atom-field state in Eq.~\eqref{eq:JCsolution} is entangled, as predicted also by perturbative calculations, see e.g.~Ref.~\cite[Sec.~V]{Raymer2021}.

If $\sqrt{\Gamma} G_t \sqrt{n}$ is large enough, this model predicts coherent Rabi oscillations between the two-level atom and a \emph{single} temporal mode, a non-trivial result, given the intrinsic multimode nature of the problem.
This behaviour was suggested by the atom's reduced dynamics~\cite{Baragiola2012,Baragiola2014a} and it has been confirmed rigorously using quantum stochastic calculus in Ref.~\cite{Fischer2018a}, reproducing the approximated model of Refs.~\cite{vanEnk2002,Silberfarb2004} that we also employ here.

\subsection{QFI expressions}
\label{subsec:QFIapproxJC}

The main advantage of this approximate model is that we can easily evaluate the QFI and apply existing results regarding the estimation of the coupling constant of the JC Hamiltonian~\cite{Genoni2012b}.
Firstly, the overall atom-field QFI is proportional to the average number of photons $\bar{n}_\xi$ in the pulse (since the atom is initially in the ground state);
the time-dependent details of the problem enter only as a multiplicative factor:
\begin{equation}
    \label{eq:QFI_tdepJC}
    \mathcal{Q}(\ket{\Psi^\mathrm{AP}(t)}) = \frac{ G_t ^2}{\Gamma} \bra{\Psi_0^\mathrm{P}} A_\xi^\dag A_\xi \ket{\Psi_0^\mathrm{P}} \equiv \frac{ G_t ^2}{\Gamma} \bar{n}_\xi.
\end{equation}
For an arbitrary pulse shape $\xi(t) = f (\frac{t}{T} ) / {\sqrt{T}}$, centered around $\bar{t}=0$ without loss of generality, a change of variable gives $G_t = \sqrt{T} \int_{t_0}^{t/T} dx f(x) \equiv \sqrt{T} F_t$.
This means that within this approximation the global atom-field QFI $\mathcal{Q}(\ket{\Psi^\mathrm{AP}})$ is linear in the pulse duration $T$, and different shapes only induce different proportionality constants.

Secondly, for the atom initially in the ground state and a Fock state wavepacket $\ket{n_\xi},$ the QFI $\mathcal{Q}(\rho^\mathrm{P})$ of the reduced field state in Eq.~\eqref{eq:JCreducedFieldState} is equal to the pure-state QFI $\mathcal{Q}(\ket{\Psi^\mathrm{AP}})$ of the composite field-atom system~\cite{Genoni2012b}.
The same also holds for the reduced atomic state, but this is practically irrelevant as the atom cannot be measured directly.
Moreover, the reduced state of the field is always diagonal in the Fock basis and photon counting is thus the optimal measurement that attains the QFI.
Specifically, for an $n$-photon Fock pulse the QFI is
\begin{equation} \label{eq:QFI_tdepJCFock}
    \mathcal{Q}_{\mathrm{Fock}} = \frac{n G_t ^2}{\Gamma} = \frac{n T F_t^2}{\Gamma}.
\end{equation}
We show in Appendix~\ref{app:appShortSinglePh} that Eq.~\eqref{eq:QFI_tdepJCFock} for $n=1$ is consistent with the short time and short pulse limit of the single-photon QFI of Sec.~\ref{sec:1photon}; in particular the classical contribution is the only relevant one and coincides with the CFI of the probability $p_\mathrm{orig}$ of finding the photon in the original temporal mode.

For other initial states of the pulse we need to evaluate the QFI $\mathcal{Q}(\rho^\mathrm{P})$ for the rank-2 density matrix in Eq.~\eqref{eq:JCreducedFieldState} employing Eq.~\eqref{eq:QFIrank2} in Appendix~\ref{app:QFI_rank2}, derived using the methods of Ref.~\cite{Fiderer2021a}.
Unlike Fock states, arbitrary states do not always saturate the inequality $\mathcal{Q}(\rho^\mathrm{P}) \leq  \bar{n}_\xi G_t^2/\Gamma $.

\subsection{Linear absorption regime and connection to bosonic loss estimation}
\label{subsec:linearabsorption}

When the argument of the trigonometric functions in Eq.~\eqref{eq:JCsolution}, i.e., the effective pulse-atom interaction, is small $\sqrt{\Gamma T} F_t \sqrt{n} \ll 1$ we can ignore saturation effects and the atom excitation probability is approximately $p_e(t) \approx n \Gamma G_t^2 = n \Gamma T F_t^2.$
This is linear in the number of photons, and we call this the linear absorption regime. 
Note that an absorption probability approximately linear in $\Gamma T$ is a general feature in the short time and short pulse regime that holds also for more complex matter systems~\cite[Sec.~IV]{Ko2022}.
Similarly, for states with an indefinite number of photons in this regime we obtain $p_e(t) \approx  \bar{n}_\xi \Gamma T F_t^2$, showing that the details of the quantum state of the light are not important, as far as the excitation probability is concerned.

On the contrary, the QFI, i.e., the bound of the precision of the estimating the parameter of interest, is greatly influenced by the choice of the photonic probe state.
While in this linear absorption regime the dimensionless QFI is equal to the excitation probability for Fock states, this is not true for arbitrary photonic states with coherences, since their QFI does not saturate the upper bound $\bar{n}_\xi G_t^2/{\Gamma}$ in general.
We will see in Sec.~\ref{sec:sodium} that coherent states with the same average number of photons perform much worse, while squeezed states also saturate the upper bound.

In this linear absorption regime the estimation problem is very closely connected to absorption spectroscopy~\cite{Whittaker2017,Allen2020,Belsley2022}, and formally equivalent to loss estimation~\cite{Monras2007,Adesso2009}.
The correspondence between the two problems is evident when considering the estimation of a small bosonic loss rate $0 < \gamma \ll 1 $ appearing in the Lindblad master equation describing the loss of excitations of a bosonic mode $d \rho /dt = \gamma \left( a \rho a^\dag - 1/2 \left\{ a^\dag a , \rho\right\} \right)$.
For $\gamma \to 0$, the leading-order term of the QFI for an initial Fock probe state $\ket{n}$, evolved for a time $t$, is $n t/\gamma$~\cite{Monras2007,Adesso2009}.
This is very similar to the QCRB obtained from the QFI in Eq.~\eqref{eq:QFI_tdepJCFock} for the atom-pulse coupling $\Gamma$, the difference being that the pulse duration $T$ acts as an effective interaction time instead of $t$ and there is an additional proportionality constant encoding the details of the pulse shape.
Furthermore, for loss estimation, also measuring the environment into which the photons are lost gives no additional information about the loss parameter if Fock states (or other optimal probe states) are used~\cite{Monras2007,Nair2018}, just like for the estimation of the JC coupling parameter~\cite{Genoni2012b}.


\section{Entangled biphoton probes}\label{sec:biphoton}
\begin{figure}
    \includegraphics[width=.95\columnwidth]{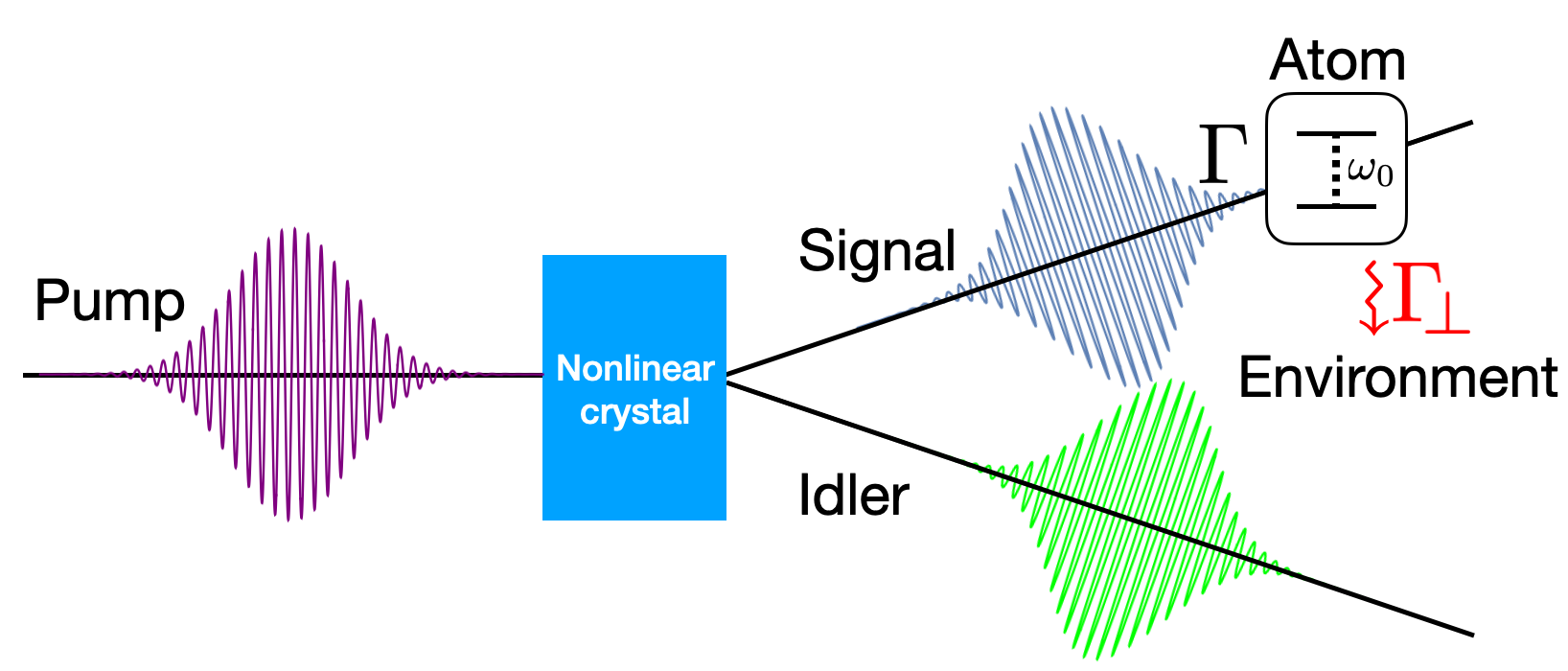}
    \caption{Schematic (not to scale) illustration of a linear biphoton setup.
    An entangled biphoton state is generated by a pulse interacting with a nonlinear crystal and the signal photon interacts with a two-level atom with a dipole coupling of strength $\Gamma$.
    The atom can also decay by spontaneous emission to other environment modes, with a rate $\Gamma_\perp$.}
    \label{fig:scheme_biphoton}
\end{figure}

In this section, we consider the so-called linear biphoton setup, illustrated in Fig.~\ref{fig:scheme_biphoton}, in which only one, labelled the signal (S), of the two mascroscopically distinct modes, i.e., two beams, of an entangled biphoton state interacts with the atom, while the other mode, labelled the idler (I), evolves freely.
This is the simplest instance of spectroscopy using entangled light--the archetypal instance of quantum light spectroscopy~\cite{Dorfman2016,Schlawin2016,Schlawin2017,li2017probing,Mukamel2020}.
The biphoton setup~(with coincidence detection at the end such that the idler photon serves as timing gate for signal photon) has been employed in absorption spectroscopy experiments~\cite{yabushita2004spectroscopy,kalachev2007biphoton,Kalashnikov2014}, where improved SNR vis-\'{a}-vis spectroscopy using single-mode detection was demonstrated.
Our objective is to quantify the performance of entangled states in the simple spectroscopic setup of Fig.~\ref{fig:scheme_biphoton}.

Theoretically, the statistics generated by a setup relying only on uncorrelated coincidence measurements can be reproduced exactly without the need of entanglement, as pointed out by Stefanov~\cite{Stefanov2017a}. 
Moreover, since only one of the entangled photons interacts with the sample in Fig.~\ref{fig:scheme_biphoton}, the setup is formally equivalent to the use of noiseless ancilla in quantum metrology.
Therein, it is well-known that entanglement with ancilla is not advantageous in the case of noiseless unitary dynamics, but may be useful in presence of noise~\cite{Demkowicz-Dobrzanski2014}.
The exact conditions when noiseless ancilla improve the optimal attainable precision in quantum metrology, however, remain unknown~\cite{Layden2019}.
For our problem, since the biphoton pulse becomes entangled with the atom, the dynamics of the field is not unitary and the initial entanglement between the signal and idler modes may be useful.

The most general biphoton state, entangled over the macroscopically distinct signal and idler modes, is written as
\begin{equation}\label{eq:two_photon_entangled}
	\ket{\Phi_{\mathrm{biph}}}=\int d\omega_{\mathrm{S}} d\omega_{\mathrm{I}} \tilde{\Phi}(\omega_{\mathrm{S}},\omega_{\mathrm{I}}) a_{\mathrm{S}}^{\dagger}(\omega_{\mathrm{S}}) a^{\dagger}_{\mathrm{I}}(\omega_{\mathrm{I}}) \ket{0^\mathrm{S}}\ket{0^\mathrm{I}},
\end{equation}
where $\tilde{\Phi}(\omega_{\mathrm{S}},\omega_{\mathrm{I}})$ is the joint spectral amplitude (JSA) that captures the spectral~(or equivalently temporal) correlations of the biphoton state, and  $\omega_{\mathrm{S}}$ and $\omega_{\mathrm{I}}$ denote the signal and idler frequencies respectively.

We first present an expression for the QFI without specifying a particular form for the  JSA so as to preserve generality.
Subsequently, in Sec.~\ref{subsec:advantage_biphoton} we make an additional assumption on the form of JSA that applies, for instance, to the specific example of entangled biphoton states produced as a result of type-II spontaneous parametric downconversion~(PDC) in birefringent $\chi^{(2)}$-nonlinear crystals in the weak downconversion limit~\cite{Grice2001a,URen2003,Christ2013}.
In the PDC setting, the JSA is a product of the envelope of the pump field and a phase-matching function, which can be approximated using a Gaussian\footnote{The validity of approximating the sinc phase-matching function as a Gaussian was studied by experimentally measuring joint temporal intensities using time-resolved femtosecond upsconversion~\cite{kuzucu2008joint}.
Both the sinc and Gaussian phase-matching function were in rough agreement with experimental values.}~\cite{Grice2001a,URen2003}.
Numerical results under this assumption will presented in Sec.~\ref{sec:sodium}. 

Recent years have seen considerable experimental efforts devoted to developing methods to shape the JSA of biphoton states~\cite{Peer2005,Lukens2015}.
PDC states with novel JSAs have been proposed or reported in experiments by domain-engineering the nonlinear crystal~\cite{Graffitti2020,Morrison2022}, as well as fabricating multipole nonlinear crystals to generate $n$-mode frequency bin entanglement~\cite{kaneda2019direct}. Time-frequency entangled states can also be produced using $\chi^{(3)}$(Kerr)-nonlinear interaction of spontaneous four-wave mixing~(FWM)~\cite{cohenSFWM2016} using pulsed~\cite{li2004all,fulconis2005high,chen2005two,sharping2006generation,harada2008generation,kues2017chip} or continuous-wave~\cite{chen2011frequency,clemmen2009continuous} pumping in conventional optical fibres, photonic crystal fibres, and silicon-on-insulator~(SOI) waveguides.
Another curious source of correlated pairs is the biexciton-exciton cascade~\cite{simon2005creating} that was used to produced time-bin entangled states using quantum dot emitters~\cite{jayakumar2014time}.
It thus makes sense to optimize the JSA directly~\cite{Carnio2021}, or more practically the pump profile~\cite{Arzani2018}, for quantum information processing tasks. For our purposes, optimizing the JSA in order to obtain maximal QFI represents a quantum metrological recipe for source engineering the time-energy entangled states employed to estimate the $\Gamma$ parameter.

The JSA admits a Schmidt decomposition in terms of discrete Schmidt modes $\tilde{\Phi}(\omega_{\mathrm{S}},\omega_{\mathrm{I}}) = \sum_k r_k \tilde{\xi}_k^\mathrm{S}(\omega_\mathrm{S})\tilde{\xi}_k^\mathrm{I}(\omega_\mathrm{I}),$ whereby
\begin{equation}\label{eq:two_photon _entangled_schmidt_expression}
	\ket{\Phi_{\mathrm{biph}}}=\sum_{k} r_{k} a^{\dagger}_{k,\,\mathrm{S}} a^{\dagger}_{k,\mathrm{I}} \ket{0^\mathrm{S}} \ket{0^\mathrm{I}} = \sum_{k} r_{k} \ket{\xi^\mathrm{S}_k} \ket{\xi^\mathrm{I}_k} ,
\end{equation}
where $a^{\dagger}_{k,\mathrm{S}} = \int d\omega_\mathrm{S}\, \tilde{\xi}_k^\mathrm{S}(\omega_\mathrm{S}) a_{\mathrm{S}}^\dag(\omega_\mathrm{S}) $ and $a^{\dagger}_{k,\mathrm{I}} = \int d\omega_\mathrm{I}\, \tilde{\xi}_k^\mathrm{I}(\omega_\mathrm{I}) a_{\mathrm{I}}^\dag(\omega_\mathrm{I}) $ are photon-wavepacket creation operators for each Schmidt mode of the signal and idler photons respectively and $\ket{\xi_k^\mathrm{S}}$, $\ket{\xi_k^\mathrm{I}}$ are the respective single-photon wavepackets, and $r_k$ are positive Schmidt weights~\cite{Parker2000,Lamata2005}.
For instance, a Gaussian JSA has Hermite-Gauss functions as Schmidt modes~\cite{law2000}. 

As we have done previously, we assume that the JSA is peaked around two carrier frequencies $\bar{\omega}_\mathrm{S}$ and $\bar{\omega}_\mathrm{I}$, in addition to $\bar{\omega}_\mathrm{S} = \omega_0.$
In analogy with the single-photon case, we define the time-domain envelope, $\Phi(t_\mathrm{S},t_\mathrm{I})$ as the Fourier transform of $\tilde{\Phi}(\omega_\mathrm{I}+\bar{\omega}_\mathrm{I},\omega_\mathrm{S}+\bar{\omega}_\mathrm{S})$.
This in turn defines temporal amplitudes of the Schmidt modes $\xi_k^\mathrm{S}(t_\mathrm{S})=\int_{-\infty}^\infty \frac{d \omega_\mathrm{S}}{\sqrt{2\pi}} \tilde{\xi}^\mathrm{S}_k(\omega_\mathrm{S}) e^{-\I (\omega_\mathrm{S}-\bar{\omega}_\mathrm{S}) t_\mathrm{S}}$ so that $a^{\dagger}_{k,\mathrm{S}} = \int d t_\mathrm{S}\, \xi_k^\mathrm{S}(t_\mathrm{S}) a_{\mathrm{S}}^\dag(t_\mathrm{S}).$
Analogous definitions hold for the idler.

We can thus make the same approximations explained in Sec.~\ref{sec:theory} and consider an interaction-picture Hamiltonian identical to Eq.~\eqref{eq:interaction_hamiltonian_Gamma} with the substitution $a(t) \mapsto a_{\mathrm{S}}(t) \otimes \id^\mathrm{I} $, i.e., only the signal beam interacts with the atom, as in Fig.~\ref{fig:scheme_biphoton}.
By linearity we can employ the previous single-photon solution given by Eqs.~\eqref{eq:1ph_pulse_full_state}, \eqref{eq:1ph_pulse_psie} and \eqref{eq:1ph_pulse_unnorm_state} denoting the atom-signal-environment unitary operator corresponding to integrating the Schrödinger equation as $U^\mathrm{ASE} (t)$.
Applying it to the wavepackets $\xi_k^{\mathrm{S}}(t)$ of the signal mode we have the overall atom-signal-idler-environment global state
\begin{align}
    & U^\mathrm{ASIE}(t) \ket{g} \ket{\Phi_{\mathrm{biph}}} \ket{0^\mathrm{E}} = \sum_k r_k  \left( U^\mathrm{ASE} (t) \ket{g} \ket{\xi^\mathrm{S}_k} \ket{0^{\mathrm{E}}} \right) \ket{\xi^\mathrm{I}_k}  \nonumber \\
    & = \sum_k r_k  \biggl( \psi_{e,k}(t)\ket{e} \ket{0^\mathrm{S}} \ket{0^\mathrm{E}}  + \ket{g} \ket*{\widetilde{\psi}^\mathrm{S}_{g,k}(t)}\ket{0^\mathrm{E}} \nonumber \\ 
    & \qquad \qquad + \ket{g}\ket{0^\mathrm{S}} \ket*{\widetilde{\psi}^\mathrm{E}_{g,k}(t)} \biggr) \ket{\xi^\mathrm{I}_k}. 
    \label{eq:global_entangled}
\end{align} 
We can trace out the atomic and environmental degrees of freedom to obtain
\begin{align}
\rho^\mathrm{SI} =& \sum_{jk} r_j r^{*}_k \left( \psi_{e,j} \psi_{e,k}^* + \braket*{ \widetilde{\psi}^\mathrm{E}_{g,k} }{ \widetilde{\psi}^\mathrm{E}_{g,j} } \right) | 0^\mathrm{S} \rangle \langle 0^\mathrm{S} | \otimes | \xi^\mathrm{I}_j \rangle \langle \xi^\mathrm{I}_k |  \nonumber  \\
+ &  \biggl( \sum_{k} r_k \ket*{\widetilde{\psi}_{g,k}^\mathrm{S}} \ket*{\xi^\mathrm{I}_k} \biggr) \biggl(  \sum_{j} r_j^* \bra*{\widetilde{\psi}_{g,j}^\mathrm{S}} \bra*{\xi^\mathrm{I}_j} \biggr),
\label{eq:rhoSIbiphoton}
\end{align}
which has the form $ p_\Gamma \rho_\Gamma^{(0)} + (1-p_\Gamma) \ket*{\psi_\Gamma^{(1)}}\bra*{\psi_\Gamma^{(1)}}$, where we have introduced the normalized density matrix in the vacuum subspace of the signal photon 
\begin{equation}
\rho_\Gamma^{(0)}=\frac{1}{p_\Gamma}\sum_{jk} r_j r^{*}_k \left( \psi_{e,j} \psi_{e,k}^* + \braket*{ \widetilde{\psi}^\mathrm{E}_{g,k} }{ \widetilde{\psi}^\mathrm{E}_{g,j} } \right) | 0^\mathrm{S} \rangle \langle 0^\mathrm{S} | \otimes | \xi^\mathrm{I}_j \rangle \langle \xi^\mathrm{I}_k |
\end{equation}
and the normalized pure state in the signal single-photon subspace 
\begin{equation}
\ket*{\psi_\Gamma^{(1)}}=\frac{1}{\sqrt{1-p_\Gamma}} \sum_{k} r_k \ket*{\widetilde{\psi}_{g,k}^\mathrm{S}} \ket*{\xi^\mathrm{I}_k},
\end{equation}
while $p_\Gamma = \sum_k |r_k|^2 \left( |\psi_{e,k}|^2  + \braket*{ \widetilde{\psi}^\mathrm{E}_{g,k} }{ \widetilde{\psi}^\mathrm{E}_{g,k} }  \right)$ 
is the probability of losing a photon from the signal beam.
Since these are normalized states living in orthogonal subspaces, the QFI saturates the upper bound in Eq.~\eqref{eq:ExtendedConvexityQFI} and is composed of a classical and a quantum contribution
\begin{equation}
    \label{eq:QFI_biph_normalized}
    \mathcal{Q}(\rho^\mathrm{SI}) = \mathcal{C}(p_\Gamma) +  p_\Gamma \mathcal{Q}\left(\rho_\Gamma^{(0)}\right) + (1-p_\Gamma) \mathcal{Q}\left(\ket*{\psi^{(1)}_\Gamma}\right),
\end{equation}
with the classical being $ \mathcal{C}(p_\Gamma) = (d p_\Gamma / d\Gamma )^2 / [ p_\Gamma (1-p_\Gamma) ]$ 
and the quantum $\tilde{\mathcal{Q}}= p_\Gamma \mathcal{Q}(\rho_0) + (1-p_\Gamma) \mathcal{Q}(\ket{\psi_1})$.
Eq.~ \eqref{eq:QFI_biph_normalized} is similar to Eq.~\eqref{eq:QFI_1ph_normalized} for single-photon pulses, the main difference being that $\rho_\Gamma^{(0)}$ can now carry information on the parameter due to the entanglement with idler modes, while in the single-photon case one would have just the vacuum, which carries no information.

If we neglect the emission into environment modes (by setting $\Gperp = 0$), the mixed state in Eq.~\eqref{eq:rhoSIbiphoton} becomes rank 2 and we can take evaluate the QFI using Eq.~\eqref{eq:rank2ortho} in Appendix~\ref{app:QFI_rank2}, obtaining
$\mathcal{Q}(\rho^\mathrm{SI}) = 4  \sum_k |r_k|^2 \biggl( |\partial_\Gamma \psi_{e,k} |^2 +  \braket*{\partial_\Gamma \widetilde{\psi}_{g,k}^\mathrm{S}}{\partial_\Gamma \widetilde{\psi}_{g,k}^\mathrm{S}} + \Im[ \psi^*_{e,k} \partial_\Gamma \psi_{e,k} +  \braket*{ \widetilde{\psi}_{g,k}^\mathrm{S}}{\partial_\Gamma \widetilde{\psi}_{g,k}^\mathrm{S}} ] 
\biggr).$

\subsection{No advantage from entanglement for real-valued joint temporal amplitudes}
\label{subsec:advantage_biphoton}

In this section, we limit ourselves to the case of perfect coupling ($\Gperp = 0$), but the argument also applies to $\Gperp > 0$ for short times when spontaneous emission can be neglected.

We also assume that the temporal amplitudes of the Schmidt modes are of the form $e^{\I \varphi_k} \xi_k^\mathrm{S} (t_\mathrm{S})$ with real $\xi_k^\mathrm{S} (t_\mathrm{S}) $ and $\varphi_k$ (i.e., they have no temporal phases). 
The Hermite-Gauss modes obtained as Schmidt basis functions of a two-dimensional Gaussian JSA have, for instance, this form.
More generally, this is also true when the time-domain envelope $\Phi(t_\mathrm{S},t_\mathrm{I})$ is real-valued:
\begin{align}
    &\sum_n\,r_n\,\xi_n^\mathrm{S}(t_\mathrm{S})^*\,\xi_n^\mathrm{I}(t_\mathrm{I})^* =  \sum_n\,r_n\,\xi_n^\mathrm{S}(t_\mathrm{S})\,\xi_n^\mathrm{I}(t_\mathrm{I}) \nonumber\noindent\\
    &\implies\xi_n^\mathrm{S}(t_\mathrm{S})^* = p\xi_n^\mathrm{S}(t_\mathrm{S}),\, \xi_n^\mathrm{I}(t_\mathrm{I})^* = p\xi_n^\mathrm{I}(t_\mathrm{I}),~p=\pm1,
\end{align}
i.e., the Schmidt signal and idler functions are either both real, or completely imaginary.
In either case, this implies the lack of a relative temporal phase for the signal Schmidt modes, which are of interest here\footnote{A real $\Phi(t_\mathrm{S},t_\mathrm{I})$ only constitutes a sufficient condition, and it is possible to construct more general JSAs whose Schmidt bases do not have a temporal phase.}.
Since the overall phases $\phi_k$ do not depend on time, they will also factor out of $\psi_{e,k}(t)$ and $\ket*{\widetilde{\psi}_{g,k}^\mathrm{P}}$ and since it does not depend on the parameter $\Gamma$ it will also factor out when taking the derivatives.

Under these assumptions,
\begin{align}
    \label{eq:QFIbiphotonNoUseEntang}
    & \mathcal{Q} \left( U^\mathrm{ASI} \ket{g} \ket{\Phi_{\mathrm{biph}}} \right) \\
    & = \sum_k |r_k|^2 \mathcal{Q}\left( \psi_{e,k}\ket{e} \ket{0^\mathrm{S}}  + \ket{g} \ket*{\widetilde{\psi}_{g,k}^\mathrm{S}} \right) \\
    & = \sum_k |r_k|^2 \mathcal{Q}\left( |\psi_{e,k}|^2  |0^\mathrm{S} \rangle\langle 0^\mathrm{S}|  + \ket*{\widetilde{\psi}_{g,k}^\mathrm{S}} \bra*{ \widetilde{\psi}_{g,k}^\mathrm{S}}  \right) \label{eq:QFIbiphotonNoUseEntang3} \\ 
    & =  \mathcal{Q} \biggl( \sum_k |r_k|^2 \left[ |\psi_{e,k}|^2  |0^\mathrm{S} \rangle\langle 0^\mathrm{S}|  + \ket*{\widetilde{\psi}^\mathrm{S}_{g,k}} \bra*{\widetilde{\psi}^\mathrm{S}_{g,k}}  \right] \otimes \ket{\xi^\mathrm{I}_k} \bra{\xi^\mathrm{I}_k}  \biggr). \nonumber
\end{align}
The first equality 
holds because all the components in the superposition live in mutually orthogonal subspaces, thanks to the idler modes, and because  the normalized pure states $\psi_{e,k}\ket{e} \ket{0^\mathrm{P}}  + \ket{g} \ket*{\widetilde{\psi}_{g,k}^\mathrm{P}}$ are orthogonal to their $\Gamma$-derivatives thanks to the assumption on the absence of temporal phases for the Schmidt modes.
This assumption justifies also the second equality, together with the fact that we are considering single-photon wavepackets of the signal beam. 
Physically, it means that for each single-photon wavepacket the information on $\Gamma$ is fully available in the reduced state of the field subsystem.
In the final equality, we have stressed that the QFI obtained in the previous line corresponds to the QFI of an initial classically correlated state $\sum_k r_k^2 \ket{\xi_k^\mathrm{S}}\bra{\xi_k^\mathrm{S}} \otimes \ket{\xi_k^\mathrm{I}}\bra{\xi_k^\mathrm{I}}$ instead of an entangled state.

Eqns. \eqref{eq:QFIbiphotonNoUseEntang}--\eqref{eq:QFIbiphotonNoUseEntang3} mean that using a biphoton probe state whose Schmidt temporal modes have no temporal phases is equivalent to probing the atom with randomly chosen single photon states $\ket{\xi_k^\mathrm{S}}$ with probability $|r_k|^2$, but retaining the knowledge on each value $k$, e.g., by detecting the idler photons in the Schmidt modes to perform heralded state preparation of single-photon wavepackets in the signal mode.
If the knowledge on the values $k$ is not available we are left with the mixed single-photon state $\sum_k r_k^2 \ket{\xi_k^\mathrm{S}}\bra{\xi_k^\mathrm{S}}$ obtained by tracing over the idler mode.
Such a mixed single-photon state yields in general less information on $\Gamma$ as shown by the convexity property of the QFI in Eq.~\eqref{eq:ExtendedConvexityQFI}.

Since the QFI in Eq.~\eqref{eq:QFIbiphotonNoUseEntang} is a convex sum of the QFI of the different Schmidt-modes, it is,  in principle, always better to deterministically prepare the single-photon wavepacket in the mode $\xi_k^\mathrm{S}$ with the largest QFI $\max_k \mathcal{Q}( \left| \psi_{e,k} \right|^2  |0 \rangle\langle 0|  + | \widetilde{\psi}_{g,k}^\mathrm{S} \rangle \langle  \widetilde{\psi}_{g,k}^\mathrm{S}|  )$.
This clearly shows that entanglement is not a fundamental resource, since there is always a single-photon wavepacket that gives at least as high a precision.
However, we note that it could be more practical to implement entangled-state strategies rather than some theoretically superior non-entangled one.
More specifically, in the next section we show that for a realistic Gaussian joint spectral density coming from PDC, the additional entanglement actually decreases the short-time QFI and it is better to employ a Gaussian single-photon wavepacket.

\subsection{Short-time and short signal photons regime}
\label{subsec:shortbiphoton}

The idler and signal photons being entangled in time, the temporal properties of one of the two subsystems cannot be defined unambiguously.
However, we can get a sense of the relevant time scales from the arrival-time distribution of the signal photon $p(t_\mathrm{S})=\sum_k |r_k|^2 | \xi_k^\mathrm{S}(t_\mathrm{S})|^2 $ (where the idler beam is traced out), since this distribution will have a well-defined temporal width.
We can write each Schmidt temporal mode in term of scale-invariant orthonormal functions as $\xi_k^\mathrm{S}(t_\mathrm{S})= f^\mathrm{S}_k(t_\mathrm{S}/T) /\sqrt{T}$, introducing an overall scale parameter $T$ for the whole basis of functions.
Even if being a complete basis implies that the functions $\xi_k^\mathrm{S}(t_\mathrm{S})$ will eventually spread over the whole real axis, we can still think of the parameter $T$ as a duration when there is moderate entanglement, so that a limited number of Schmidt modes are sufficient to describe the state and all of them have a temporal duration still captured by $T$.

Making these assumptions, when $ \Gamma t \ll 1$ and $\Gamma T \ll 1 $ we see that, just like in the single-photon case, the excitation probability is linear in $\Gamma T $: $p_e(t) = \sum_k | r_k|^2 | \psi_{e,k}(t) |^2 = \Gamma T \sum_k |r_k|^2 | F_{t,k} |^2$ where $F_t = \int_{-\infty}^{t/T} dx f_k^\mathrm{S}(x) $ and thus also very small.
There is approximately no perturbation to the shape of each Schmidt temporal mode, and in this limit all the information is contained in the classical term so that the QFI reads
\begin{equation}
    \label{eq:QFIbiphotonShort}
    \mathcal{Q}(\rho^\mathrm{SI}) \approx \mathcal{C}(p_\Gamma) \approx \frac{p_e(t)}{\Gamma^2} = \frac{T \sum_k |r_k|^2 | F_{t,k} |^2}{\Gamma}.
\end{equation}
This expression will be used in the next section, as its predictions match the results obtained from solving Schrödinger equation numerically~\cite{Bisketzi2022} for the relevant time-scale.

\section{Dipole moment estimation of a sodium atom in free space}
\label{sec:sodium}

In this section we rephrase estimation of $\Gamma$ as the more physical problem of estimating the EDM $\mu = \bm{\mu}_{eg}\cdot \bm{\epsilon}$.
For simplicity, we further assume that $ \bm{\mu}_{eg}$ and $ \bm{\epsilon}$ are parallel.
Then the EDM is related to the parameter we have considered in previous sections as $\Gamma= \mu^2 \mathcal{A}(\bar{\omega})^2$, where we assume that the constant $\mathcal{A}(\bar{\omega}) = \sqrt{\bar{\omega}/(4 \pi \hbar \epsilon_{0}c A)}$ of the propagating field is known perfectly, so that estimating $\Gamma$ or $\mu$ are formally equivalent problems.
Such a reparameterization entails the relation $ \mathcal{Q}_\mu =  ( d \Gamma / d \mu )^2 \mathcal{Q}_\Gamma =  4 \mu^2 \mathcal{A}(\bar{\omega})^4 \mathcal{Q}_\Gamma $ between the QFI for the two different parameters (and analogously for any CFI of particular measurements).

To obtain concrete numbers, we use the experimental data reported in Ref.~\cite{Steck2019} for the $D_{2}$ transition of a sodium atom.
Specifically, we set the dipole moment $ \mu = 2.988 \times 10^{-29} \, \mathrm{C} \cdot \mathrm{m} = 1.868 \times  10^{-8} \, \mathrm{e}\cdot \mathrm{cm} $, the transition frequency $\omega_{0} = 2 \pi \times  508.333  \, \mathrm{THz}$ (also equal to the carrier frequency of the pulse) and the decay constant $\Gamma_{\mathrm{tot}} = 61.542 \times  10^{6} \,\mathrm{s}^{-1}$ corresponding to a lifetime $1/\Gamma_{\mathrm{tot}} = 16.249 \, \mathrm{ns}$.
We compute the value of $\mathcal{A}(\bar{\omega})$ by considering the transverse quantisation area $A$ to be equal to the effective scattering of the light $\mathit{\sigma} = \lambda_{0}^{2}/2\pi$, with $ \lambda_{0} = 2 \pi c / \omega_0$ the central wavelength of the light.
The aim of this example is to capture realistic parameter regimes, without performing a full modelization of an experiment; thus we do not employ the more accurate expression for the constant $A$ stemming from the transverse spatial mode function mentioned in Sec.~\ref{sec:theory}.

With these parameter values, we obtain the ratio $\Gperp / \Gamma = 11.56$, similar to the value previously considered in Sec.~\ref{sec:1photon}; the decay rate into the perpendicular modes is obtained by subtracting the decay rate into the propagating pulse modes from the total free-space decay rate $\Gperp = \Gamma_{\mathrm{tot}} - \Gamma $.

\begin{figure}[t]
    \includegraphics{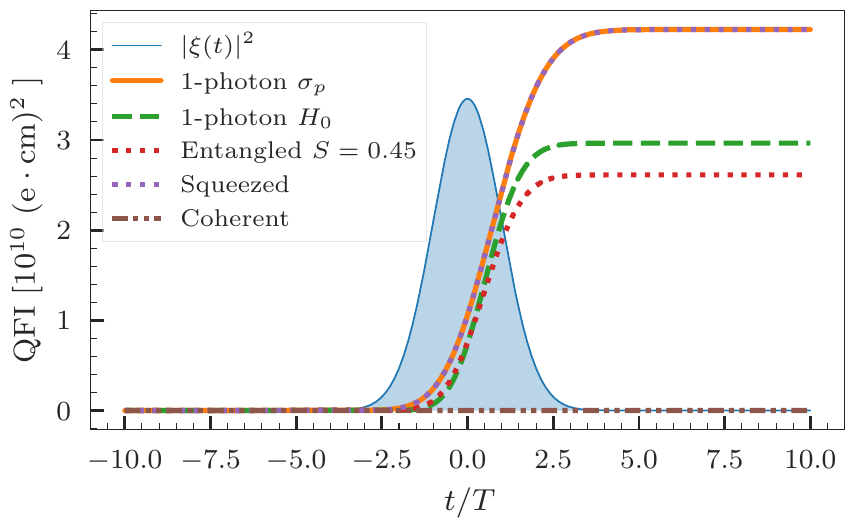}
    \caption{QFI for different states: Single-photon Gaussian pulse (orange) of duration $T=1/\sigma_{\mathrm{p}}=0.15\mathrm{ps}$ (defined as the variance of $|\xi(t)|^2$), squeezed vacuum (brown) and coherent (purple) state with one mean photon ($\bar{n}_\xi = 1$) and the same Gaussian temporal mode, entangled biphoton with a Gaussian phase-matching function and a Gaussian pump profile of duration $T$ (red) and a single-photon pulse corresponding to the $0$-th Schmidt mode of the entangled biphoton (green).
    The shaded Gaussian represents the pulse profile (not to scale on the vertical axis).} 
    \label{fig:EvaPlot}
\end{figure}

\begin{figure}[t]
    \includegraphics{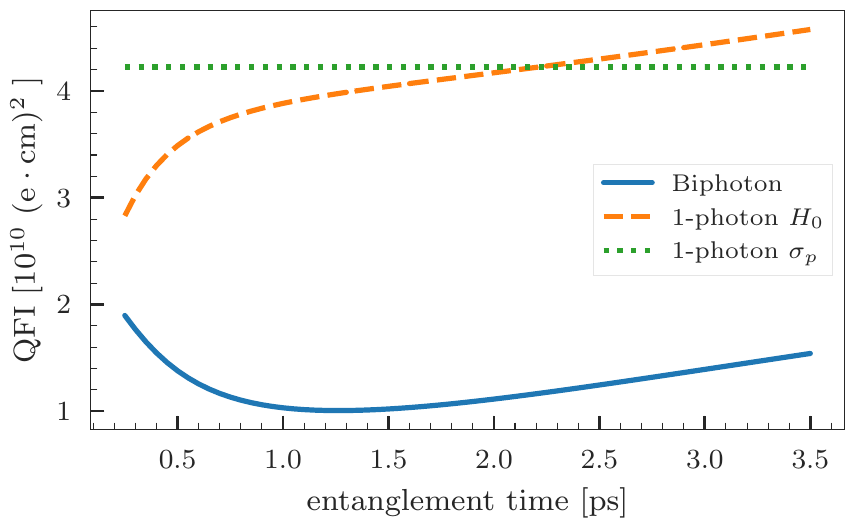}
    \caption{QFI for a fixed pump width $1/\sigma_{\mathrm{p}} = 0.15 \mathrm{ps}$ and varying entanglement time. 
        The solid line represents the biphoton state, the green dotted line represents a single-photon Gaussian pulse with duration $\sigma_{\mathrm{p}}$, while the orange dashed represents a singele-photon pulse in the $0$-th Schmidt temporal mode, this is also Gaussian but the duration depends on the entanglement time (see Appendix~\ref{app:PDC}).
        In all cases the QFI is evaluated at $t=10 T$ for a pulse peaked at $t=0$.
    }
    \label{fig:Tent}
\end{figure}

We fix the pulse shape to be an ultrashort Gaussian of duration $T=1/\sigma_{\mathrm{p}} = 0.15 \, \mathrm{ps}$ and we consider different single temporal-mode states: Single-photon Fock (denoted as ``1-photon $\sigma_{\mathrm{p}}$'' in Figs.~\ref{fig:EvaPlot} and~\ref{fig:Tent}), coherent and squeezed vacuum, as defined in Sec.~\ref{subsec:model} (the complex phases of the coherent state parameter $\alpha$ and of the squeezing parameter $r$ have no effect on the results).
These parameters put us well into the short-pulse regime defined previously: $ \Gamma_{\mathrm{tot}} T= 9.2313 \times  10^{-6}$ and $ \Gamma T= 7.34995 \times  10^{-7}$ and we can thus neglect all spontaneous emission effects by considering the dynamics of the system up to shortly after the interaction.
If we considered a regime where spontaneous emission is not negligible the fact that $\Gperp$ is also proportional to $\mu^2$ would also need to be accounted for, making the problem different from the one studied in previous sections.

We also consider entangled biphoton states obtained with a Gaussian pump pulse with spectral width $\sigma_{\mathrm{p}} = 1/T$ and with a Gaussian phase-matching function, so that the overall JSA is a bivariate Gaussian and the Schmidt modes are Hermite-Gauss polynomials.
We fix the entanglement time $T_\mathrm{qent}=2.09 \, \mathrm{ps}.$
See Appendix~\ref{app:PDC} for the definition of $T_\mathrm{qent}$ and other details of the PDC process.
This corresponds to an entanglement entropy $S=0.62$.
Notice that we are fixing the pump to have the same temporal profile as the temporal mode considered for the unentangled probe states.
However, the scale parameter of the family of signal Schmidt modes, as introduced in Sec.~\ref{subsec:shortbiphoton}, is not the pump pulse duration $T$, but the parameter $1/k_S$ introduced in Eq.~\eqref{eq:kappa_Schmidt} in Appendix~\ref{app:PDC}, which depends on the details of the PDC process\footnote{A fuller discussion of the tradeoffs in the precision of estimation between the pump pulse and crystal parameters in the PDC process will be provided in following publications~\cite{Khan2022,Khan2022a}.}.
For this reason we also consider a single-photon state having a Gaussian shape corresponding to the $0$-th Schmidt mode of the entangled state (denoted as ``1-photon $H_0$'' in Figs.~\ref{fig:EvaPlot} and~\ref{fig:Tent}).

With this choice of parameters we are safely in the regime of validity of the approximation presented in Sec.~\ref{sec:short_pulses}.
Since we will not consider intense pulses with high photon numbers, we are also working in the linear absorption regime.
We can thus evaluate the QFI using Eq.~\eqref{eq:QFIbiphotonShort}.

The comparison of these states is shown in Fig.~\ref{fig:EvaPlot}, where the coherent and squeezed vacuum pulses have a mean photon number $\bar{n}_\xi=1$.
In this short-pulse and linear absorption regime the excitation probability obtained for single-photon, coherent and squeezed states are essentially identical, since such probability depends only on the average photon number and on the shape of the temporal mode, as argued in Sec.~\ref{subsec:linearabsorption}.
On the contrary, the behaviour of the QFI of the light subsystem may be very different.
Indeed, Fig.~\ref{fig:EvaPlot} shows that while single-photon and squeezed states perform almost identically, the information extracted by coherent states is very small and indistinguishable from zero at this scale.\footnote{
This corrects a few erroneous results reported in Ref.~\cite{Bisketzi2022}, in particular Figs.~5.4, 5.5 and 5.13 therein.}
This is confirmed more generally for other parameter configurations for which the approximate model is valid.
These results are consistent with the conclusion of Sec.~\ref{subsec:linearabsorption}, which draws connection to estimation of an optical loss rate, for which fixed photon number states are known to be optimal and for which it is known that squeezed vacuum states perform optimally in the small loss regime~\cite{Monras2007}.

In Fig.~\ref{fig:EvaPlot} we see that the $0$-th HG mode alone carries more information than the entire biphoton state.
This behaviour is confirmed more generally in Fig.~\ref{fig:Tent}, where we show the QFI for a fixed pump pulse duration $T=0.15$ps, identical to the previous figure, but now varying the entanglement time. 
Notice that in this regime, the amount of entanglement in the biphoton state has the same qualitative behaviour as the entanglement time $T_\mathrm{qent}$.
In this figure we also show the QFI of a single photon state with the same Gaussian shape as the pump and the QFI of $0$-th Schmidt mode single-photon states for the different values of $T_\mathrm{qent}$.
In the plotted region we see that considering just a single-photon state prepared in the $0$-th Schmidt modes always outperforms the correspoding entangled state.
This is consistent with the argument in Sec.~\ref{subsec:advantage_biphoton} on the suboptimality of entangled biphoton probes, for real-valued joint temporal amplitudes (such as the Gaussian considered in this section).

\section{Discussion and conclusions}
\label{sec:discussion}

We have introduced a quantum information theoretic methodology for analysing and understanding spectroscopy with pulses of quantum light.
Focussing on the simplest quantum matter system and employing a fully quantum model of light-matter interaction,
we have elucidated the origins of the classical and quantum information that lead to precision spectroscopy. 
Along the way, we have recognized connections to existing spectroscopic techniques.

Our first step towards the understanding of spectroscopy with pulsed quantum light can serve as the foundation for numerous explorations. 
Evident theoretical questions on the utility of non-resonant pulses~\cite{Khan2022,Khan2022a} in the spectroscopy of simple and more complex matter systems such as those affected by a phononic bath~\cite{Ko2022} remain open.
Going forward, to understand the fundamental limits to spectrosopic precision, it may also prove useful to properly take into account that more than one parameter may be unknown, e.g. the position of the atom in the beam,
by applying the, admittedly more involved, theories of quantum multiparameter~\cite{Szczykulska2016,Albarelli2019c} or semiparametric~\cite{Tsang2019} estimation.
In conclusion, we hope that our work will clear a path towards tangible quantum advantages in spectroscopy experiments with pulsed quantum light.

\emph{Acknowledgements}:
We thank Elnaz Darsheshdar for fruitful discussions and feedback on the manuscript.
This work was supported, in part, by an EPSRC New Horizons grant (EP/V04818X/1).
EB was supported by the UK Quantum Technology Hub in Quantum Enhanced Imaging - QuantIC (EP/M01326X/1) 
and AK by a Chancellor's International Scholarship from the University of Warwick.
Computing facilities were provided by the Scientific Computing Research Technology Platform of the University of Warwick.

\bibliography{2022QuantumSpectroscopyAnimesh}

\begin{thebibliography}{125}%
\makeatletter
\providecommand \@ifxundefined [1]{%
 \@ifx{#1\undefined}
}%
\providecommand \@ifnum [1]{%
 \ifnum #1\expandafter \@firstoftwo
 \else \expandafter \@secondoftwo
 \fi
}%
\providecommand \@ifx [1]{%
 \ifx #1\expandafter \@firstoftwo
 \else \expandafter \@secondoftwo
 \fi
}%
\providecommand \natexlab [1]{#1}%
\providecommand \enquote  [1]{``#1''}%
\providecommand \bibnamefont  [1]{#1}%
\providecommand \bibfnamefont [1]{#1}%
\providecommand \citenamefont [1]{#1}%
\providecommand \href@noop [0]{\@secondoftwo}%
\providecommand \href [0]{\begingroup \@sanitize@url \@href}%
\providecommand \@href[1]{\@@startlink{#1}\@@href}%
\providecommand \@@href[1]{\endgroup#1\@@endlink}%
\providecommand \@sanitize@url [0]{\catcode `\\12\catcode `\$12\catcode `\&12\catcode `\#12\catcode `\^12\catcode `\_12\catcode `\%12\relax}%
\providecommand \@@startlink[1]{}%
\providecommand \@@endlink[0]{}%
\providecommand \url  [0]{\begingroup\@sanitize@url \@url }%
\providecommand \@url [1]{\endgroup\@href {#1}{\urlprefix }}%
\providecommand \urlprefix  [0]{URL }%
\providecommand \Eprint [0]{\href }%
\providecommand \doibase [0]{https://doi.org/}%
\providecommand \selectlanguage [0]{\@gobble}%
\providecommand \bibinfo  [0]{\@secondoftwo}%
\providecommand \bibfield  [0]{\@secondoftwo}%
\providecommand \translation [1]{[#1]}%
\providecommand \BibitemOpen [0]{}%
\providecommand \bibitemStop [0]{}%
\providecommand \bibitemNoStop [0]{.\EOS\space}%
\providecommand \EOS [0]{\spacefactor3000\relax}%
\providecommand \BibitemShut  [1]{\csname bibitem#1\endcsname}%
\let\auto@bib@innerbib\@empty
\bibitem [{\citenamefont {Mukamel}\ \emph {et~al.}(2020)\citenamefont {Mukamel}, \citenamefont {Freyberger}, \citenamefont {Schleich}, \citenamefont {Bellini}, \citenamefont {Zavatta}, \citenamefont {Leuchs}, \citenamefont {Silberhorn}, \citenamefont {Boyd}, \citenamefont {{S{\'a}nchez-Soto}}, \citenamefont {Stefanov}, \citenamefont {Barbieri}, \citenamefont {Paterova}, \citenamefont {Krivitsky}, \citenamefont {Shwartz}, \citenamefont {Tamasaku} \emph {et~al.}}]{Mukamel2020}%
  \BibitemOpen
  \bibfield  {author} {\bibinfo {author} {\bibfnamefont {S.}~\bibnamefont {Mukamel}}, \bibinfo {author} {\bibfnamefont {M.}~\bibnamefont {Freyberger}}, \bibinfo {author} {\bibfnamefont {W.}~\bibnamefont {Schleich}}, \bibinfo {author} {\bibfnamefont {M.}~\bibnamefont {Bellini}}, \bibinfo {author} {\bibfnamefont {A.}~\bibnamefont {Zavatta}}, \bibinfo {author} {\bibfnamefont {G.}~\bibnamefont {Leuchs}}, \bibinfo {author} {\bibfnamefont {C.}~\bibnamefont {Silberhorn}}, \bibinfo {author} {\bibfnamefont {R.~W.}\ \bibnamefont {Boyd}}, \bibinfo {author} {\bibfnamefont {L.~L.}\ \bibnamefont {{S{\'a}nchez-Soto}}}, \bibinfo {author} {\bibfnamefont {A.}~\bibnamefont {Stefanov}}, \bibinfo {author} {\bibfnamefont {M.}~\bibnamefont {Barbieri}}, \bibinfo {author} {\bibfnamefont {A.}~\bibnamefont {Paterova}}, \bibinfo {author} {\bibfnamefont {L.}~\bibnamefont {Krivitsky}}, \bibinfo {author} {\bibfnamefont {S.}~\bibnamefont {Shwartz}}, \bibinfo {author} {\bibfnamefont {K.}~\bibnamefont {Tamasaku}}, \emph {et~al.},\ }\bibfield  {title} {\bibinfo {title} {Roadmap on quantum light spectroscopy},\ }\href {https://doi.org/10.1088/1361-6455/ab69a8} {\bibfield  {journal} {\bibinfo  {journal} {J. Phys. B}\ }\textbf {\bibinfo {volume} {53}},\ \bibinfo {pages} {072002} (\bibinfo {year} {2020})}\BibitemShut {NoStop}%
\bibitem [{\citenamefont {Walmsley}(2015)}]{Walmsley2015}%
  \BibitemOpen
  \bibfield  {author} {\bibinfo {author} {\bibfnamefont {I.~A.}\ \bibnamefont {Walmsley}},\ }\bibfield  {title} {\bibinfo {title} {Quantum optics: {{Science}} and technology in a new light},\ }\href {https://doi.org/10.1126/science.aab0097} {\bibfield  {journal} {\bibinfo  {journal} {Science}\ }\textbf {\bibinfo {volume} {348}},\ \bibinfo {pages} {525} (\bibinfo {year} {2015})}\BibitemShut {NoStop}%
\bibitem [{\citenamefont {Polzik}\ \emph {et~al.}(1992)\citenamefont {Polzik}, \citenamefont {Carri},\ and\ \citenamefont {Kimble}}]{polzik1992spectroscopy}%
  \BibitemOpen
  \bibfield  {author} {\bibinfo {author} {\bibfnamefont {E.~S.}\ \bibnamefont {Polzik}}, \bibinfo {author} {\bibfnamefont {J.}~\bibnamefont {Carri}},\ and\ \bibinfo {author} {\bibfnamefont {H.~J.}\ \bibnamefont {Kimble}},\ }\bibfield  {title} {\bibinfo {title} {Spectroscopy with squeezed light},\ }\href {https://doi.org/10.1103/PhysRevLett.68.3020} {\bibfield  {journal} {\bibinfo  {journal} {Phys. Rev. Lett.}\ }\textbf {\bibinfo {volume} {68}},\ \bibinfo {pages} {3020} (\bibinfo {year} {1992})}\BibitemShut {NoStop}%
\bibitem [{\citenamefont {Kalachev}\ \emph {et~al.}(2007)\citenamefont {Kalachev}, \citenamefont {Kalashnikov}, \citenamefont {Kalinkin}, \citenamefont {Mitrofanova}, \citenamefont {Shkalikov},\ and\ \citenamefont {Samartsev}}]{kalachev2007biphoton}%
  \BibitemOpen
  \bibfield  {author} {\bibinfo {author} {\bibfnamefont {A.~A.}\ \bibnamefont {Kalachev}}, \bibinfo {author} {\bibfnamefont {D.~A.}\ \bibnamefont {Kalashnikov}}, \bibinfo {author} {\bibfnamefont {A.~A.}\ \bibnamefont {Kalinkin}}, \bibinfo {author} {\bibfnamefont {T.~G.}\ \bibnamefont {Mitrofanova}}, \bibinfo {author} {\bibfnamefont {A.~V.}\ \bibnamefont {Shkalikov}},\ and\ \bibinfo {author} {\bibfnamefont {V.~V.}\ \bibnamefont {Samartsev}},\ }\bibfield  {title} {\bibinfo {title} {Biphoton spectroscopy of {{YAG}}:{{Er3}}+ crystal},\ }\href {https://doi.org/10.1002/lapl.200710061} {\bibfield  {journal} {\bibinfo  {journal} {Laser Phys. Lett.}\ }\textbf {\bibinfo {volume} {4}},\ \bibinfo {pages} {722} (\bibinfo {year} {2007})}\BibitemShut {NoStop}%
\bibitem [{\citenamefont {Kalashnikov}\ \emph {et~al.}(2014)\citenamefont {Kalashnikov}, \citenamefont {Pan}, \citenamefont {Kuznetsov},\ and\ \citenamefont {Krivitsky}}]{Kalashnikov2014}%
  \BibitemOpen
  \bibfield  {author} {\bibinfo {author} {\bibfnamefont {D.~A.}\ \bibnamefont {Kalashnikov}}, \bibinfo {author} {\bibfnamefont {Z.}~\bibnamefont {Pan}}, \bibinfo {author} {\bibfnamefont {A.~I.}\ \bibnamefont {Kuznetsov}},\ and\ \bibinfo {author} {\bibfnamefont {L.~A.}\ \bibnamefont {Krivitsky}},\ }\bibfield  {title} {\bibinfo {title} {Quantum {{Spectroscopy}} of {{Plasmonic Nanostructures}}},\ }\href {https://doi.org/10.1103/PhysRevX.4.011049} {\bibfield  {journal} {\bibinfo  {journal} {Phys. Rev. X}\ }\textbf {\bibinfo {volume} {4}},\ \bibinfo {pages} {011049} (\bibinfo {year} {2014})}\BibitemShut {NoStop}%
\bibitem [{\citenamefont {Dorfman}\ \emph {et~al.}(2021)\citenamefont {Dorfman}, \citenamefont {Liu}, \citenamefont {Lou}, \citenamefont {Wei}, \citenamefont {Jing}, \citenamefont {Schlawin},\ and\ \citenamefont {Mukamel}}]{Dorfman2021b}%
  \BibitemOpen
  \bibfield  {author} {\bibinfo {author} {\bibfnamefont {K.}~\bibnamefont {Dorfman}}, \bibinfo {author} {\bibfnamefont {S.}~\bibnamefont {Liu}}, \bibinfo {author} {\bibfnamefont {Y.}~\bibnamefont {Lou}}, \bibinfo {author} {\bibfnamefont {T.}~\bibnamefont {Wei}}, \bibinfo {author} {\bibfnamefont {J.}~\bibnamefont {Jing}}, \bibinfo {author} {\bibfnamefont {F.}~\bibnamefont {Schlawin}},\ and\ \bibinfo {author} {\bibfnamefont {S.}~\bibnamefont {Mukamel}},\ }\bibfield  {title} {\bibinfo {title} {Multidimensional four-wave mixing signals detected by quantum squeezed light},\ }\href {https://doi.org/10.1073/pnas.2105601118} {\bibfield  {journal} {\bibinfo  {journal} {Proc. Natl. Acad. Sci.}\ }\textbf {\bibinfo {volume} {118}},\ \bibinfo {pages} {e2105601118} (\bibinfo {year} {2021})}\BibitemShut {NoStop}%
\bibitem [{\citenamefont {Dorfman}\ \emph {et~al.}(2016)\citenamefont {Dorfman}, \citenamefont {Schlawin},\ and\ \citenamefont {Mukamel}}]{Dorfman2016}%
  \BibitemOpen
  \bibfield  {author} {\bibinfo {author} {\bibfnamefont {K.~E.}\ \bibnamefont {Dorfman}}, \bibinfo {author} {\bibfnamefont {F.}~\bibnamefont {Schlawin}},\ and\ \bibinfo {author} {\bibfnamefont {S.}~\bibnamefont {Mukamel}},\ }\bibfield  {title} {\bibinfo {title} {Nonlinear optical signals and spectroscopy with quantum light},\ }\href {https://doi.org/10.1103/RevModPhys.88.045008} {\bibfield  {journal} {\bibinfo  {journal} {Rev. Mod. Phys.}\ }\textbf {\bibinfo {volume} {88}},\ \bibinfo {pages} {045008} (\bibinfo {year} {2016})}\BibitemShut {NoStop}%
\bibitem [{\citenamefont {Yabushita}\ and\ \citenamefont {Kobayashi}(2004)}]{yabushita2004spectroscopy}%
  \BibitemOpen
  \bibfield  {author} {\bibinfo {author} {\bibfnamefont {A.}~\bibnamefont {Yabushita}}\ and\ \bibinfo {author} {\bibfnamefont {T.}~\bibnamefont {Kobayashi}},\ }\bibfield  {title} {\bibinfo {title} {Spectroscopy by frequency-entangled photon pairs},\ }\href {https://doi.org/10.1103/PhysRevA.69.013806} {\bibfield  {journal} {\bibinfo  {journal} {Phys. Rev. A}\ }\textbf {\bibinfo {volume} {69}},\ \bibinfo {pages} {013806} (\bibinfo {year} {2004})}\BibitemShut {NoStop}%
\bibitem [{\citenamefont {Saleh}\ \emph {et~al.}(1998)\citenamefont {Saleh}, \citenamefont {Jost}, \citenamefont {Fei},\ and\ \citenamefont {Teich}}]{saleh1998entangled}%
  \BibitemOpen
  \bibfield  {author} {\bibinfo {author} {\bibfnamefont {B.~E.~A.}\ \bibnamefont {Saleh}}, \bibinfo {author} {\bibfnamefont {B.~M.}\ \bibnamefont {Jost}}, \bibinfo {author} {\bibfnamefont {H.-B.}\ \bibnamefont {Fei}},\ and\ \bibinfo {author} {\bibfnamefont {M.~C.}\ \bibnamefont {Teich}},\ }\bibfield  {title} {\bibinfo {title} {Entangled-{{Photon Virtual-State Spectroscopy}}},\ }\href {https://doi.org/10.1103/PhysRevLett.80.3483} {\bibfield  {journal} {\bibinfo  {journal} {Phys. Rev. Lett.}\ }\textbf {\bibinfo {volume} {80}},\ \bibinfo {pages} {3483} (\bibinfo {year} {1998})}\BibitemShut {NoStop}%
\bibitem [{\citenamefont {Raymer}\ \emph {et~al.}(2013)\citenamefont {Raymer}, \citenamefont {Marcus}, \citenamefont {Widom},\ and\ \citenamefont {Vitullo}}]{Raymer2013}%
  \BibitemOpen
  \bibfield  {author} {\bibinfo {author} {\bibfnamefont {M.~G.}\ \bibnamefont {Raymer}}, \bibinfo {author} {\bibfnamefont {A.~H.}\ \bibnamefont {Marcus}}, \bibinfo {author} {\bibfnamefont {J.~R.}\ \bibnamefont {Widom}},\ and\ \bibinfo {author} {\bibfnamefont {D.~L.~P.}\ \bibnamefont {Vitullo}},\ }\bibfield  {title} {\bibinfo {title} {Entangled {{Photon-Pair Two-Dimensional Fluorescence Spectroscopy}} ({{EPP-2DFS}})},\ }\href {https://doi.org/10.1021/jp405829n} {\bibfield  {journal} {\bibinfo  {journal} {J. Phys. Chem. B}\ }\textbf {\bibinfo {volume} {117}},\ \bibinfo {pages} {15559} (\bibinfo {year} {2013})}\BibitemShut {NoStop}%
\bibitem [{\citenamefont {Ishizaki}(2020)}]{Ishizaki2020}%
  \BibitemOpen
  \bibfield  {author} {\bibinfo {author} {\bibfnamefont {A.}~\bibnamefont {Ishizaki}},\ }\bibfield  {title} {\bibinfo {title} {Probing excited-state dynamics with quantum entangled photons: {{Correspondence}} to coherent multidimensional spectroscopy},\ }\href {https://doi.org/10.1063/5.0015432} {\bibfield  {journal} {\bibinfo  {journal} {J. Chem. Phys.}\ }\textbf {\bibinfo {volume} {153}},\ \bibinfo {pages} {051102} (\bibinfo {year} {2020})}\BibitemShut {NoStop}%
\bibitem [{\citenamefont {Fujihashi}\ and\ \citenamefont {Ishizaki}(2021)}]{Fujihashi2021}%
  \BibitemOpen
  \bibfield  {author} {\bibinfo {author} {\bibfnamefont {Y.}~\bibnamefont {Fujihashi}}\ and\ \bibinfo {author} {\bibfnamefont {A.}~\bibnamefont {Ishizaki}},\ }\bibfield  {title} {\bibinfo {title} {Achieving two-dimensional optical spectroscopy with temporal and spectral resolution using quantum entangled three photons},\ }\href {https://doi.org/10.1063/5.0056808} {\bibfield  {journal} {\bibinfo  {journal} {J. Chem. Phys.}\ }\textbf {\bibinfo {volume} {155}},\ \bibinfo {pages} {044101} (\bibinfo {year} {2021})}\BibitemShut {NoStop}%
\bibitem [{\citenamefont {Giovannetti}\ \emph {et~al.}(2011)\citenamefont {Giovannetti}, \citenamefont {Lloyd},\ and\ \citenamefont {Maccone}}]{Giovannetti2011}%
  \BibitemOpen
  \bibfield  {author} {\bibinfo {author} {\bibfnamefont {V.}~\bibnamefont {Giovannetti}}, \bibinfo {author} {\bibfnamefont {S.}~\bibnamefont {Lloyd}},\ and\ \bibinfo {author} {\bibfnamefont {L.}~\bibnamefont {Maccone}},\ }\bibfield  {title} {\bibinfo {title} {Advances in quantum metrology},\ }\href {https://doi.org/10.1038/nphoton.2011.35} {\bibfield  {journal} {\bibinfo  {journal} {Nat. Photonics}\ }\textbf {\bibinfo {volume} {5}},\ \bibinfo {pages} {222} (\bibinfo {year} {2011})}\BibitemShut {NoStop}%
\bibitem [{\citenamefont {Polino}\ \emph {et~al.}(2020)\citenamefont {Polino}, \citenamefont {Valeri}, \citenamefont {Spagnolo},\ and\ \citenamefont {Sciarrino}}]{Polino2020}%
  \BibitemOpen
  \bibfield  {author} {\bibinfo {author} {\bibfnamefont {E.}~\bibnamefont {Polino}}, \bibinfo {author} {\bibfnamefont {M.}~\bibnamefont {Valeri}}, \bibinfo {author} {\bibfnamefont {N.}~\bibnamefont {Spagnolo}},\ and\ \bibinfo {author} {\bibfnamefont {F.}~\bibnamefont {Sciarrino}},\ }\bibfield  {title} {\bibinfo {title} {Photonic quantum metrology},\ }\href {https://doi.org/10.1116/5.0007577} {\bibfield  {journal} {\bibinfo  {journal} {AVS Quantum Sci.}\ }\textbf {\bibinfo {volume} {2}},\ \bibinfo {pages} {024703} (\bibinfo {year} {2020})}\BibitemShut {NoStop}%
\bibitem [{\citenamefont {Srimath~Kandada}\ and\ \citenamefont {Cerullo}(2021)}]{SrimathKandada2021}%
  \BibitemOpen
  \bibfield  {author} {\bibinfo {author} {\bibfnamefont {A.~R.}\ \bibnamefont {Srimath~Kandada}}\ and\ \bibinfo {author} {\bibfnamefont {G.}~\bibnamefont {Cerullo}},\ }\bibfield  {title} {\bibinfo {title} {The path toward quantum advantage in optical spectroscopy of materials},\ }\href {https://doi.org/10.1073/pnas.2112897118} {\bibfield  {journal} {\bibinfo  {journal} {Proc. Natl. Acad. Sci.}\ }\textbf {\bibinfo {volume} {118}},\ \bibinfo {pages} {e2112897118} (\bibinfo {year} {2021})}\BibitemShut {NoStop}%
\bibitem [{\citenamefont {Whittaker}\ \emph {et~al.}(2017)\citenamefont {Whittaker}, \citenamefont {Erven}, \citenamefont {Neville}, \citenamefont {Berry}, \citenamefont {O'Brien}, \citenamefont {Cable},\ and\ \citenamefont {Matthews}}]{Whittaker2017}%
  \BibitemOpen
  \bibfield  {author} {\bibinfo {author} {\bibfnamefont {R.}~\bibnamefont {Whittaker}}, \bibinfo {author} {\bibfnamefont {C.}~\bibnamefont {Erven}}, \bibinfo {author} {\bibfnamefont {A.}~\bibnamefont {Neville}}, \bibinfo {author} {\bibfnamefont {M.}~\bibnamefont {Berry}}, \bibinfo {author} {\bibfnamefont {J.~L.}\ \bibnamefont {O'Brien}}, \bibinfo {author} {\bibfnamefont {H.}~\bibnamefont {Cable}},\ and\ \bibinfo {author} {\bibfnamefont {J.~C.~F.}\ \bibnamefont {Matthews}},\ }\bibfield  {title} {\bibinfo {title} {Absorption spectroscopy at the ultimate quantum limit from single-photon states},\ }\href {https://doi.org/10.1088/1367-2630/aa5512} {\bibfield  {journal} {\bibinfo  {journal} {New J. Phys.}\ }\textbf {\bibinfo {volume} {19}},\ \bibinfo {pages} {023013} (\bibinfo {year} {2017})}\BibitemShut {NoStop}%
\bibitem [{\citenamefont {Mitchell}\ and\ \citenamefont {Backlund}(2022)}]{Mitchell2022}%
  \BibitemOpen
  \bibfield  {author} {\bibinfo {author} {\bibfnamefont {C.~S.}\ \bibnamefont {Mitchell}}\ and\ \bibinfo {author} {\bibfnamefont {M.~P.}\ \bibnamefont {Backlund}},\ }\bibfield  {title} {\bibinfo {title} {Quantum limits to resolution and discrimination of spontaneous emission lifetimes},\ }\href {https://doi.org/10.1103/PhysRevA.105.062603} {\bibfield  {journal} {\bibinfo  {journal} {Phys. Rev. A}\ }\textbf {\bibinfo {volume} {105}},\ \bibinfo {pages} {062603} (\bibinfo {year} {2022})}\BibitemShut {NoStop}%
\bibitem [{\citenamefont {Schlawin}(2017{\natexlab{a}})}]{Schlawin2017a}%
  \BibitemOpen
  \bibfield  {author} {\bibinfo {author} {\bibfnamefont {F.}~\bibnamefont {Schlawin}},\ }\href {https://doi.org/10.1007/978-3-319-44397-3} {\emph {\bibinfo {title} {Quantum-{{Enhanced Nonlinear Spectroscopy}}}}}\ (\bibinfo  {publisher} {{Springer International Publishing}},\ \bibinfo {address} {{Cham}},\ \bibinfo {year} {2017})\BibitemShut {NoStop}%
\bibitem [{\citenamefont {Genoni}\ and\ \citenamefont {Invernizzi}(2012)}]{Genoni2012b}%
  \BibitemOpen
  \bibfield  {author} {\bibinfo {author} {\bibfnamefont {M.~G.}\ \bibnamefont {Genoni}}\ and\ \bibinfo {author} {\bibfnamefont {C.}~\bibnamefont {Invernizzi}},\ }\bibfield  {title} {\bibinfo {title} {Optimal quantum estimation of the coupling constant of {{Jaynes-Cummings}} interaction},\ }\href {https://doi.org/10.1140/epjst/e2012-01534-2} {\bibfield  {journal} {\bibinfo  {journal} {Eur. Phys. J. Spec. Top.}\ }\textbf {\bibinfo {volume} {203}},\ \bibinfo {pages} {49} (\bibinfo {year} {2012})}\BibitemShut {NoStop}%
\bibitem [{\citenamefont {Bern{\'a}d}\ \emph {et~al.}(2019)\citenamefont {Bern{\'a}d}, \citenamefont {Sanavio},\ and\ \citenamefont {Xuereb}}]{Bernad2019}%
  \BibitemOpen
  \bibfield  {author} {\bibinfo {author} {\bibfnamefont {J.~Z.}\ \bibnamefont {Bern{\'a}d}}, \bibinfo {author} {\bibfnamefont {C.}~\bibnamefont {Sanavio}},\ and\ \bibinfo {author} {\bibfnamefont {A.}~\bibnamefont {Xuereb}},\ }\bibfield  {title} {\bibinfo {title} {Optimal estimation of matter-field coupling strength in the dipole approximation},\ }\href {https://doi.org/10.1103/PhysRevA.99.062106} {\bibfield  {journal} {\bibinfo  {journal} {Phys. Rev. A}\ }\textbf {\bibinfo {volume} {99}},\ \bibinfo {pages} {062106} (\bibinfo {year} {2019})}\BibitemShut {NoStop}%
\bibitem [{\citenamefont {Schlawin}(2017{\natexlab{b}})}]{Schlawin2017}%
  \BibitemOpen
  \bibfield  {author} {\bibinfo {author} {\bibfnamefont {F.}~\bibnamefont {Schlawin}},\ }\bibfield  {title} {\bibinfo {title} {Entangled photon spectroscopy},\ }\href {https://doi.org/10.1088/1361-6455/aa8a7a} {\bibfield  {journal} {\bibinfo  {journal} {J. Phys. B}\ }\textbf {\bibinfo {volume} {50}},\ \bibinfo {pages} {203001} (\bibinfo {year} {2017}{\natexlab{b}})}\BibitemShut {NoStop}%
\bibitem [{\citenamefont {Dinani}\ \emph {et~al.}(2016)\citenamefont {Dinani}, \citenamefont {Gupta}, \citenamefont {Dowling},\ and\ \citenamefont {Berry}}]{Dinani2016}%
  \BibitemOpen
  \bibfield  {author} {\bibinfo {author} {\bibfnamefont {H.~T.}\ \bibnamefont {Dinani}}, \bibinfo {author} {\bibfnamefont {M.~K.}\ \bibnamefont {Gupta}}, \bibinfo {author} {\bibfnamefont {J.~P.}\ \bibnamefont {Dowling}},\ and\ \bibinfo {author} {\bibfnamefont {D.~W.}\ \bibnamefont {Berry}},\ }\bibfield  {title} {\bibinfo {title} {Quantum-enhanced spectroscopy with entangled multiphoton states},\ }\href {https://doi.org/10.1103/PhysRevA.93.063804} {\bibfield  {journal} {\bibinfo  {journal} {Phys. Rev. A}\ }\textbf {\bibinfo {volume} {93}},\ \bibinfo {pages} {063804} (\bibinfo {year} {2016})}\BibitemShut {NoStop}%
\bibitem [{\citenamefont {Birchall}\ \emph {et~al.}(2020)\citenamefont {Birchall}, \citenamefont {Allen}, \citenamefont {Stace}, \citenamefont {O'Brien}, \citenamefont {Matthews},\ and\ \citenamefont {Cable}}]{Birchall2019}%
  \BibitemOpen
  \bibfield  {author} {\bibinfo {author} {\bibfnamefont {P.~M.}\ \bibnamefont {Birchall}}, \bibinfo {author} {\bibfnamefont {E.~J.}\ \bibnamefont {Allen}}, \bibinfo {author} {\bibfnamefont {T.~M.}\ \bibnamefont {Stace}}, \bibinfo {author} {\bibfnamefont {J.~L.}\ \bibnamefont {O'Brien}}, \bibinfo {author} {\bibfnamefont {J.~C.~F.}\ \bibnamefont {Matthews}},\ and\ \bibinfo {author} {\bibfnamefont {H.}~\bibnamefont {Cable}},\ }\bibfield  {title} {\bibinfo {title} {Quantum {{Optical Metrology}} of {{Correlated Phase}} and {{Loss}}},\ }\href {https://doi.org/10.1103/PhysRevLett.124.140501} {\bibfield  {journal} {\bibinfo  {journal} {Phys. Rev. Lett.}\ }\textbf {\bibinfo {volume} {124}},\ \bibinfo {pages} {140501} (\bibinfo {year} {2020})}\BibitemShut {NoStop}%
\bibitem [{\citenamefont {Biele}\ \emph {et~al.}(2021)\citenamefont {Biele}, \citenamefont {Wollmann}, \citenamefont {Silverstone}, \citenamefont {Matthews},\ and\ \citenamefont {Allen}}]{Biele2021}%
  \BibitemOpen
  \bibfield  {author} {\bibinfo {author} {\bibfnamefont {J.}~\bibnamefont {Biele}}, \bibinfo {author} {\bibfnamefont {S.}~\bibnamefont {Wollmann}}, \bibinfo {author} {\bibfnamefont {J.~W.}\ \bibnamefont {Silverstone}}, \bibinfo {author} {\bibfnamefont {J.~C.~F.}\ \bibnamefont {Matthews}},\ and\ \bibinfo {author} {\bibfnamefont {E.~J.}\ \bibnamefont {Allen}},\ }\bibfield  {title} {\bibinfo {title} {Maximizing precision in saturation-limited absorption measurements},\ }\href {https://doi.org/10.1103/PhysRevA.104.053717} {\bibfield  {journal} {\bibinfo  {journal} {Phys. Rev. A}\ }\textbf {\bibinfo {volume} {104}},\ \bibinfo {pages} {053717} (\bibinfo {year} {2021})}\BibitemShut {NoStop}%
\bibitem [{\citenamefont {Blow}\ \emph {et~al.}(1990)\citenamefont {Blow}, \citenamefont {Loudon}, \citenamefont {Phoenix},\ and\ \citenamefont {Shepherd}}]{Blow1990a}%
  \BibitemOpen
  \bibfield  {author} {\bibinfo {author} {\bibfnamefont {K.~J.}\ \bibnamefont {Blow}}, \bibinfo {author} {\bibfnamefont {R.}~\bibnamefont {Loudon}}, \bibinfo {author} {\bibfnamefont {S.~J.~D.}\ \bibnamefont {Phoenix}},\ and\ \bibinfo {author} {\bibfnamefont {T.~J.}\ \bibnamefont {Shepherd}},\ }\bibfield  {title} {\bibinfo {title} {Continuum fields in quantum optics},\ }\href {https://doi.org/10.1103/PhysRevA.42.4102} {\bibfield  {journal} {\bibinfo  {journal} {Phys. Rev. A}\ }\textbf {\bibinfo {volume} {42}},\ \bibinfo {pages} {4102} (\bibinfo {year} {1990})}\BibitemShut {NoStop}%
\bibitem [{\citenamefont {Ko}\ \emph {et~al.}(2022)\citenamefont {Ko}, \citenamefont {Cook},\ and\ \citenamefont {Whaley}}]{Ko2022}%
  \BibitemOpen
  \bibfield  {author} {\bibinfo {author} {\bibfnamefont {L.}~\bibnamefont {Ko}}, \bibinfo {author} {\bibfnamefont {R.~L.}\ \bibnamefont {Cook}},\ and\ \bibinfo {author} {\bibfnamefont {K.~B.}\ \bibnamefont {Whaley}},\ }\bibfield  {title} {\bibinfo {title} {Dynamics of photosynthetic light harvesting systems interacting with {{N-photon Fock}} states},\ }\href {https://doi.org/10.1063/5.0082822} {\bibfield  {journal} {\bibinfo  {journal} {J. Chem. Phys.}\ }\textbf {\bibinfo {volume} {156}},\ \bibinfo {pages} {244108} (\bibinfo {year} {2022})}\BibitemShut {NoStop}%
\bibitem [{\citenamefont {Scully}\ and\ \citenamefont {Zubairy}(1997)}]{Scully1997}%
  \BibitemOpen
  \bibfield  {author} {\bibinfo {author} {\bibfnamefont {M.}~\bibnamefont {Scully}}\ and\ \bibinfo {author} {\bibfnamefont {M.~S.}\ \bibnamefont {Zubairy}},\ }\href@noop {} {\emph {\bibinfo {title} {Quantum {{Optics}}}}}\ (\bibinfo  {publisher} {{Cambridge University Press}},\ \bibinfo {year} {1997})\BibitemShut {NoStop}%
\bibitem [{\citenamefont {Gardiner}\ \emph {et~al.}(2004)\citenamefont {Gardiner}, \citenamefont {Zoller},\ and\ \citenamefont {Zoller}}]{gardiner2004quantum}%
  \BibitemOpen
  \bibfield  {author} {\bibinfo {author} {\bibfnamefont {C.}~\bibnamefont {Gardiner}}, \bibinfo {author} {\bibfnamefont {P.}~\bibnamefont {Zoller}},\ and\ \bibinfo {author} {\bibfnamefont {P.}~\bibnamefont {Zoller}},\ }\href@noop {} {\emph {\bibinfo {title} {Quantum noise: a handbook of Markovian and non-Markovian quantum stochastic methods with applications to quantum optics}}}\ (\bibinfo  {publisher} {Springer Science \& Business Media},\ \bibinfo {year} {2004})\BibitemShut {NoStop}%
\bibitem [{\citenamefont {Baragiola}(2014)}]{Baragiola2014a}%
  \BibitemOpen
  \bibfield  {author} {\bibinfo {author} {\bibfnamefont {B.~Q.}\ \bibnamefont {Baragiola}},\ }\emph {\bibinfo {title} {Open {{Systems Dynamics}} for {{Propagating Quantum Fields}}}},\ \href {https://digitalrepository.unm.edu/phyc_etds/7} {Ph.D. thesis},\ \bibinfo  {school} {University of New Mexico} (\bibinfo {year} {2014}),\ \Eprint {https://arxiv.org/abs/1408.4447} {arXiv:1408.4447} \BibitemShut {NoStop}%
\bibitem [{\citenamefont {Fischer}(2018)}]{Fischer2018a}%
  \BibitemOpen
  \bibfield  {author} {\bibinfo {author} {\bibfnamefont {K.~A.}\ \bibnamefont {Fischer}},\ }\bibfield  {title} {\bibinfo {title} {Exact calculation of stimulated emission driven by pulsed light},\ }\href {https://doi.org/10.1364/OSAC.1.000772} {\bibfield  {journal} {\bibinfo  {journal} {OSA Continuum}\ }\textbf {\bibinfo {volume} {1}},\ \bibinfo {pages} {772} (\bibinfo {year} {2018})}\BibitemShut {NoStop}%
\bibitem [{\citenamefont {D{\k{a}}browska}\ \emph {et~al.}(2021)\citenamefont {D{\k{a}}browska}, \citenamefont {Chru{\'s}ci{\'n}ski}, \citenamefont {Chakraborty},\ and\ \citenamefont {Sarbicki}}]{Dabrowska2021a}%
  \BibitemOpen
  \bibfield  {author} {\bibinfo {author} {\bibfnamefont {A.}~\bibnamefont {D{\k{a}}browska}}, \bibinfo {author} {\bibfnamefont {D.}~\bibnamefont {Chru{\'s}ci{\'n}ski}}, \bibinfo {author} {\bibfnamefont {S.}~\bibnamefont {Chakraborty}},\ and\ \bibinfo {author} {\bibfnamefont {G.}~\bibnamefont {Sarbicki}},\ }\bibfield  {title} {\bibinfo {title} {Eternally non-{{Markovian}} dynamics of a qubit interacting with a single-photon wavepacket},\ }\href {https://doi.org/10.1088/1367-2630/ac3c60} {\bibfield  {journal} {\bibinfo  {journal} {New J. Phys.}\ }\textbf {\bibinfo {volume} {23}},\ \bibinfo {pages} {123019} (\bibinfo {year} {2021})}\BibitemShut {NoStop}%
\bibitem [{\citenamefont {Baragiola}\ \emph {et~al.}(2012)\citenamefont {Baragiola}, \citenamefont {Cook}, \citenamefont {Bra{\'n}czyk},\ and\ \citenamefont {Combes}}]{Baragiola2012}%
  \BibitemOpen
  \bibfield  {author} {\bibinfo {author} {\bibfnamefont {B.~Q.}\ \bibnamefont {Baragiola}}, \bibinfo {author} {\bibfnamefont {R.~L.}\ \bibnamefont {Cook}}, \bibinfo {author} {\bibfnamefont {A.~M.}\ \bibnamefont {Bra{\'n}czyk}},\ and\ \bibinfo {author} {\bibfnamefont {J.}~\bibnamefont {Combes}},\ }\bibfield  {title} {\bibinfo {title} {N-photon wave packets interacting with an arbitrary quantum system},\ }\href {https://doi.org/https://doi.org/10.1103/PhysRevA.86.013811} {\bibfield  {journal} {\bibinfo  {journal} {Phys. Rev. A}\ }\textbf {\bibinfo {volume} {86}},\ \bibinfo {pages} {013811} (\bibinfo {year} {2012})}\BibitemShut {NoStop}%
\bibitem [{\citenamefont {Gross}\ \emph {et~al.}(2022)\citenamefont {Gross}, \citenamefont {Baragiola}, \citenamefont {Stace},\ and\ \citenamefont {Combes}}]{Gross2022}%
  \BibitemOpen
  \bibfield  {author} {\bibinfo {author} {\bibfnamefont {J.~A.}\ \bibnamefont {Gross}}, \bibinfo {author} {\bibfnamefont {B.~Q.}\ \bibnamefont {Baragiola}}, \bibinfo {author} {\bibfnamefont {T.~M.}\ \bibnamefont {Stace}},\ and\ \bibinfo {author} {\bibfnamefont {J.}~\bibnamefont {Combes}},\ }\bibfield  {title} {\bibinfo {title} {Master equations and quantum trajectories for squeezed wave packets},\ }\href {https://doi.org/10.1103/PhysRevA.105.023721} {\bibfield  {journal} {\bibinfo  {journal} {Phys. Rev. A}\ }\textbf {\bibinfo {volume} {105}},\ \bibinfo {pages} {023721} (\bibinfo {year} {2022})}\BibitemShut {NoStop}%
\bibitem [{\citenamefont {Kiilerich}\ and\ \citenamefont {M{\o}lmer}(2019)}]{Kiilerich2019}%
  \BibitemOpen
  \bibfield  {author} {\bibinfo {author} {\bibfnamefont {A.~H.}\ \bibnamefont {Kiilerich}}\ and\ \bibinfo {author} {\bibfnamefont {K.}~\bibnamefont {M{\o}lmer}},\ }\bibfield  {title} {\bibinfo {title} {Input-{{Output Theory}} with {{Quantum Pulses}}},\ }\href {https://doi.org/10.1103/PhysRevLett.123.123604} {\bibfield  {journal} {\bibinfo  {journal} {Phys. Rev. Lett.}\ }\textbf {\bibinfo {volume} {123}},\ \bibinfo {pages} {123604} (\bibinfo {year} {2019})}\BibitemShut {NoStop}%
\bibitem [{\citenamefont {Kiilerich}\ and\ \citenamefont {M{\o}lmer}(2020)}]{Kiilerich2020}%
  \BibitemOpen
  \bibfield  {author} {\bibinfo {author} {\bibfnamefont {A.~H.}\ \bibnamefont {Kiilerich}}\ and\ \bibinfo {author} {\bibfnamefont {K.}~\bibnamefont {M{\o}lmer}},\ }\bibfield  {title} {\bibinfo {title} {Quantum interactions with pulses of radiation},\ }\href {https://doi.org/10.1103/PhysRevA.102.023717} {\bibfield  {journal} {\bibinfo  {journal} {Phys. Rev. A}\ }\textbf {\bibinfo {volume} {102}},\ \bibinfo {pages} {023717} (\bibinfo {year} {2020})}\BibitemShut {NoStop}%
\bibitem [{\citenamefont {Silberfarb}\ and\ \citenamefont {Deutsch}(2003)}]{Silberfarb2003}%
  \BibitemOpen
  \bibfield  {author} {\bibinfo {author} {\bibfnamefont {A.}~\bibnamefont {Silberfarb}}\ and\ \bibinfo {author} {\bibfnamefont {I.~H.}\ \bibnamefont {Deutsch}},\ }\bibfield  {title} {\bibinfo {title} {Continuous measurement with traveling-wave probes},\ }\href {https://doi.org/10.1103/PhysRevA.68.013817} {\bibfield  {journal} {\bibinfo  {journal} {Phys. Rev. A}\ }\textbf {\bibinfo {volume} {68}},\ \bibinfo {pages} {013817} (\bibinfo {year} {2003})}\BibitemShut {NoStop}%
\bibitem [{\citenamefont {Wang}\ \emph {et~al.}(2011)\citenamefont {Wang}, \citenamefont {Min{\'a}{\v r}}, \citenamefont {Sheridan},\ and\ \citenamefont {Scarani}}]{Wang2011b}%
  \BibitemOpen
  \bibfield  {author} {\bibinfo {author} {\bibfnamefont {Y.}~\bibnamefont {Wang}}, \bibinfo {author} {\bibfnamefont {J.}~\bibnamefont {Min{\'a}{\v r}}}, \bibinfo {author} {\bibfnamefont {L.}~\bibnamefont {Sheridan}},\ and\ \bibinfo {author} {\bibfnamefont {V.}~\bibnamefont {Scarani}},\ }\bibfield  {title} {\bibinfo {title} {Efficient excitation of a two-level atom by a single photon in a propagating mode},\ }\href {https://doi.org/10.1103/PhysRevA.83.063842} {\bibfield  {journal} {\bibinfo  {journal} {Phys. Rev. A}\ }\textbf {\bibinfo {volume} {83}},\ \bibinfo {pages} {063842} (\bibinfo {year} {2011})}\BibitemShut {NoStop}%
\bibitem [{\citenamefont {Konyk}\ and\ \citenamefont {{Gea-Banacloche}}(2016)}]{Konyk2016}%
  \BibitemOpen
  \bibfield  {author} {\bibinfo {author} {\bibfnamefont {W.}~\bibnamefont {Konyk}}\ and\ \bibinfo {author} {\bibfnamefont {J.}~\bibnamefont {{Gea-Banacloche}}},\ }\bibfield  {title} {\bibinfo {title} {Quantum multimode treatment of light scattering by an atom in a waveguide},\ }\href {https://doi.org/10.1103/PhysRevA.93.063807} {\bibfield  {journal} {\bibinfo  {journal} {Phys. Rev. A}\ }\textbf {\bibinfo {volume} {93}},\ \bibinfo {pages} {063807} (\bibinfo {year} {2016})}\BibitemShut {NoStop}%
\bibitem [{\citenamefont {Roulet}\ and\ \citenamefont {Scarani}(2016)}]{Roulet2016a}%
  \BibitemOpen
  \bibfield  {author} {\bibinfo {author} {\bibfnamefont {A.}~\bibnamefont {Roulet}}\ and\ \bibinfo {author} {\bibfnamefont {V.}~\bibnamefont {Scarani}},\ }\bibfield  {title} {\bibinfo {title} {Solving the scattering of {{N}} photons on a two-level atom without computation},\ }\href {https://doi.org/10.1088/1367-2630/18/9/093035} {\bibfield  {journal} {\bibinfo  {journal} {New J. Phys.}\ }\textbf {\bibinfo {volume} {18}},\ \bibinfo {pages} {093035} (\bibinfo {year} {2016})}\BibitemShut {NoStop}%
\bibitem [{\citenamefont {Silberfarb}\ and\ \citenamefont {Deutsch}(2004)}]{Silberfarb2004}%
  \BibitemOpen
  \bibfield  {author} {\bibinfo {author} {\bibfnamefont {A.}~\bibnamefont {Silberfarb}}\ and\ \bibinfo {author} {\bibfnamefont {I.~H.}\ \bibnamefont {Deutsch}},\ }\bibfield  {title} {\bibinfo {title} {Entanglement generated between a single atom and a laser pulse},\ }\href {https://doi.org/10.1103/PhysRevA.69.042308} {\bibfield  {journal} {\bibinfo  {journal} {Phys. Rev. A}\ }\textbf {\bibinfo {volume} {69}},\ \bibinfo {pages} {042308} (\bibinfo {year} {2004})}\BibitemShut {NoStop}%
\bibitem [{\citenamefont {Brecht}\ \emph {et~al.}(2015)\citenamefont {Brecht}, \citenamefont {Reddy}, \citenamefont {Silberhorn},\ and\ \citenamefont {Raymer}}]{Brecht2015}%
  \BibitemOpen
  \bibfield  {author} {\bibinfo {author} {\bibfnamefont {B.}~\bibnamefont {Brecht}}, \bibinfo {author} {\bibfnamefont {D.~V.}\ \bibnamefont {Reddy}}, \bibinfo {author} {\bibfnamefont {C.}~\bibnamefont {Silberhorn}},\ and\ \bibinfo {author} {\bibfnamefont {M.~G.}\ \bibnamefont {Raymer}},\ }\bibfield  {title} {\bibinfo {title} {Photon {{Temporal Modes}}: {{A Complete Framework}} for {{Quantum Information Science}}},\ }\href {https://doi.org/10.1103/PhysRevX.5.041017} {\bibfield  {journal} {\bibinfo  {journal} {Phys. Rev. X}\ }\textbf {\bibinfo {volume} {5}},\ \bibinfo {pages} {041017} (\bibinfo {year} {2015})}\BibitemShut {NoStop}%
\bibitem [{\citenamefont {Raymer}\ and\ \citenamefont {Walmsley}(2020)}]{Raymer2020b}%
  \BibitemOpen
  \bibfield  {author} {\bibinfo {author} {\bibfnamefont {M.~G.}\ \bibnamefont {Raymer}}\ and\ \bibinfo {author} {\bibfnamefont {I.~A.}\ \bibnamefont {Walmsley}},\ }\bibfield  {title} {\bibinfo {title} {Temporal modes in quantum optics: Then and now},\ }\href {https://doi.org/10.1088/1402-4896/ab6153} {\bibfield  {journal} {\bibinfo  {journal} {Phys. Scr.}\ }\textbf {\bibinfo {volume} {95}},\ \bibinfo {pages} {064002} (\bibinfo {year} {2020})}\BibitemShut {NoStop}%
\bibitem [{\citenamefont {Fabre}\ and\ \citenamefont {Treps}(2020)}]{Fabre2020}%
  \BibitemOpen
  \bibfield  {author} {\bibinfo {author} {\bibfnamefont {C.}~\bibnamefont {Fabre}}\ and\ \bibinfo {author} {\bibfnamefont {N.}~\bibnamefont {Treps}},\ }\bibfield  {title} {\bibinfo {title} {Modes and states in quantum optics},\ }\href {https://doi.org/10.1103/RevModPhys.92.035005} {\bibfield  {journal} {\bibinfo  {journal} {Rev. Mod. Phys.}\ }\textbf {\bibinfo {volume} {92}},\ \bibinfo {pages} {035005} (\bibinfo {year} {2020})}\BibitemShut {NoStop}%
\bibitem [{\citenamefont {Rohde}\ \emph {et~al.}(2007)\citenamefont {Rohde}, \citenamefont {Mauerer},\ and\ \citenamefont {Silberhorn}}]{Rohde2007}%
  \BibitemOpen
  \bibfield  {author} {\bibinfo {author} {\bibfnamefont {P.~P.}\ \bibnamefont {Rohde}}, \bibinfo {author} {\bibfnamefont {W.}~\bibnamefont {Mauerer}},\ and\ \bibinfo {author} {\bibfnamefont {C.}~\bibnamefont {Silberhorn}},\ }\bibfield  {title} {\bibinfo {title} {Spectral structure and decompositions of optical states, and their applications},\ }\href {https://doi.org/10.1088/1367-2630/9/4/091} {\bibfield  {journal} {\bibinfo  {journal} {New J. Phys.}\ }\textbf {\bibinfo {volume} {9}},\ \bibinfo {pages} {91} (\bibinfo {year} {2007})}\BibitemShut {NoStop}%
\bibitem [{\citenamefont {Christ}\ \emph {et~al.}(2011)\citenamefont {Christ}, \citenamefont {Laiho}, \citenamefont {Eckstein}, \citenamefont {Cassemiro},\ and\ \citenamefont {Silberhorn}}]{Christ2011}%
  \BibitemOpen
  \bibfield  {author} {\bibinfo {author} {\bibfnamefont {A.}~\bibnamefont {Christ}}, \bibinfo {author} {\bibfnamefont {K.}~\bibnamefont {Laiho}}, \bibinfo {author} {\bibfnamefont {A.}~\bibnamefont {Eckstein}}, \bibinfo {author} {\bibfnamefont {K.~N.}\ \bibnamefont {Cassemiro}},\ and\ \bibinfo {author} {\bibfnamefont {C.}~\bibnamefont {Silberhorn}},\ }\bibfield  {title} {\bibinfo {title} {Probing multimode squeezing with correlation functions},\ }\href {https://doi.org/10.1088/1367-2630/13/3/033027} {\bibfield  {journal} {\bibinfo  {journal} {New J. Phys.}\ }\textbf {\bibinfo {volume} {13}},\ \bibinfo {pages} {033027} (\bibinfo {year} {2011})}\BibitemShut {NoStop}%
\bibitem [{\citenamefont {Karpi{\'n}ski}\ \emph {et~al.}(2021)\citenamefont {Karpi{\'n}ski}, \citenamefont {Davis}, \citenamefont {So{\'s}nicki}, \citenamefont {Thiel},\ and\ \citenamefont {Smith}}]{Karpinski2021}%
  \BibitemOpen
  \bibfield  {author} {\bibinfo {author} {\bibfnamefont {M.}~\bibnamefont {Karpi{\'n}ski}}, \bibinfo {author} {\bibfnamefont {A.~O.~C.}\ \bibnamefont {Davis}}, \bibinfo {author} {\bibfnamefont {F.}~\bibnamefont {So{\'s}nicki}}, \bibinfo {author} {\bibfnamefont {V.}~\bibnamefont {Thiel}},\ and\ \bibinfo {author} {\bibfnamefont {B.~J.}\ \bibnamefont {Smith}},\ }\bibfield  {title} {\bibinfo {title} {Control and {{Measurement}} of {{Quantum Light Pulses}} for {{Quantum Information Science}} and {{Technology}}},\ }\href {https://doi.org/10.1002/qute.202000150} {\bibfield  {journal} {\bibinfo  {journal} {Adv. Quantum Technol.}\ }\textbf {\bibinfo {volume} {4}},\ \bibinfo {pages} {2000150} (\bibinfo {year} {2021})}\BibitemShut {NoStop}%
\bibitem [{\citenamefont {Rag}\ and\ \citenamefont {{Gea-Banacloche}}(2017)}]{Rag2017}%
  \BibitemOpen
  \bibfield  {author} {\bibinfo {author} {\bibfnamefont {H.~S.}\ \bibnamefont {Rag}}\ and\ \bibinfo {author} {\bibfnamefont {J.}~\bibnamefont {{Gea-Banacloche}}},\ }\bibfield  {title} {\bibinfo {title} {Two-level-atom excitation probability for single- and {{N}} -photon wave packets},\ }\href {https://doi.org/10.1103/PhysRevA.96.033817} {\bibfield  {journal} {\bibinfo  {journal} {Phys. Rev. A}\ }\textbf {\bibinfo {volume} {96}},\ \bibinfo {pages} {033817} (\bibinfo {year} {2017})}\BibitemShut {NoStop}%
\bibitem [{\citenamefont {Khan}\ \emph {et~al.}(2022)\citenamefont {Khan}, \citenamefont {Bisketzi}, \citenamefont {Albarelli},\ and\ \citenamefont {Datta}}]{Khan2022}%
  \BibitemOpen
  \bibfield  {author} {\bibinfo {author} {\bibfnamefont {A.}~\bibnamefont {Khan}}, \bibinfo {author} {\bibfnamefont {E.}~\bibnamefont {Bisketzi}}, \bibinfo {author} {\bibfnamefont {F.}~\bibnamefont {Albarelli}},\ and\ \bibinfo {author} {\bibfnamefont {A.}~\bibnamefont {Datta}},\ }\href@noop {} {\bibinfo {title} {In preparation}} (\bibinfo {year} {2022})\BibitemShut {NoStop}%
\bibitem [{\citenamefont {Khan}(2022)}]{Khan2022a}%
  \BibitemOpen
  \bibfield  {author} {\bibinfo {author} {\bibfnamefont {A.}~\bibnamefont {Khan}},\ }\emph {\bibinfo {title} {{Characterisation of Complex Systems Using Quantum Information and Sensing Techniques}}},\ \href@noop {} {Ph.D. thesis},\ \bibinfo  {school} {University of Warwick} (\bibinfo {year} {2022})\BibitemShut {NoStop}%
\bibitem [{\citenamefont {Heinosaari}\ and\ \citenamefont {Ziman}(2011)}]{Heinosaari2011a}%
  \BibitemOpen
  \bibfield  {author} {\bibinfo {author} {\bibfnamefont {T.}~\bibnamefont {Heinosaari}}\ and\ \bibinfo {author} {\bibfnamefont {M.}~\bibnamefont {Ziman}},\ }\href {https://doi.org/10.1017/CBO9781139031103} {\emph {\bibinfo {title} {The {{Mathematical}} Language of {{Quantum Theory}}}}}\ (\bibinfo  {publisher} {{Cambridge University Press}},\ \bibinfo {address} {{Cambridge}},\ \bibinfo {year} {2011})\BibitemShut {NoStop}%
\bibitem [{\citenamefont {Van~Trees}\ \emph {et~al.}(2013)\citenamefont {Van~Trees}, \citenamefont {Bell},\ and\ \citenamefont {Tian}}]{VanTrees2013}%
  \BibitemOpen
  \bibfield  {author} {\bibinfo {author} {\bibfnamefont {H.~L.}\ \bibnamefont {Van~Trees}}, \bibinfo {author} {\bibfnamefont {K.~L.}\ \bibnamefont {Bell}},\ and\ \bibinfo {author} {\bibfnamefont {Z.}~\bibnamefont {Tian}},\ }\href@noop {} {\emph {\bibinfo {title} {Detection Estimation and Modulation Theory}}},\ \bibinfo {edition} {second edition}\ ed.,\ Vol.\ \bibinfo {volume} {Volume 1. Detection, estimation, and filtering theory}\ (\bibinfo  {publisher} {{John Wiley {$\&$} Sons, Inc}},\ \bibinfo {address} {{Hoboken, N.J}},\ \bibinfo {year} {2013})\BibitemShut {NoStop}%
\bibitem [{\citenamefont {Helstrom}(1976)}]{helstrom1976quantum}%
  \BibitemOpen
  \bibfield  {author} {\bibinfo {author} {\bibfnamefont {C.~W.}\ \bibnamefont {Helstrom}},\ }\href@noop {} {\emph {\bibinfo {title} {Quantum Detection and Estimation Theory}}}\ (\bibinfo  {publisher} {{Academic Press}},\ \bibinfo {address} {{New York}},\ \bibinfo {year} {1976})\BibitemShut {NoStop}%
\bibitem [{\citenamefont {Holevo}(2011)}]{Holevo2011b}%
  \BibitemOpen
  \bibfield  {author} {\bibinfo {author} {\bibfnamefont {A.~S.}\ \bibnamefont {Holevo}},\ }\href {http://link.springer.com/10.1007/978-88-7642-378-9} {\emph {\bibinfo {title} {Probabilistic and {{Statistical Aspects}} of {{Quantum Theory}}}}},\ \bibinfo {edition} {2nd}\ ed.\ (\bibinfo  {publisher} {{Edizioni della Normale}},\ \bibinfo {address} {{Pisa}},\ \bibinfo {year} {2011})\BibitemShut {NoStop}%
\bibitem [{\citenamefont {Braunstein}\ and\ \citenamefont {Caves}(1994)}]{Braunstein1994}%
  \BibitemOpen
  \bibfield  {author} {\bibinfo {author} {\bibfnamefont {S.~L.}\ \bibnamefont {Braunstein}}\ and\ \bibinfo {author} {\bibfnamefont {C.~M.}\ \bibnamefont {Caves}},\ }\bibfield  {title} {\bibinfo {title} {Statistical distance and the geometry of quantum states},\ }\href {https://doi.org/10.1103/PhysRevLett.72.3439} {\bibfield  {journal} {\bibinfo  {journal} {Phys. Rev. Lett.}\ }\textbf {\bibinfo {volume} {72}},\ \bibinfo {pages} {3439} (\bibinfo {year} {1994})}\BibitemShut {NoStop}%
\bibitem [{\citenamefont {Paris}(2009)}]{Paris2009}%
  \BibitemOpen
  \bibfield  {author} {\bibinfo {author} {\bibfnamefont {M.~G.~A.}\ \bibnamefont {Paris}},\ }\bibfield  {title} {\bibinfo {title} {Quantum estimation for quantum technology},\ }\href {https://doi.org/10.1142/S0219749909004839} {\bibfield  {journal} {\bibinfo  {journal} {Int. J. Quantum Inf.}\ }\textbf {\bibinfo {volume} {07}},\ \bibinfo {pages} {125} (\bibinfo {year} {2009})}\BibitemShut {NoStop}%
\bibitem [{\citenamefont {{Barndorff-Nielsen}}\ and\ \citenamefont {Gill}(2000)}]{Barndorff-Nielsen2000}%
  \BibitemOpen
  \bibfield  {author} {\bibinfo {author} {\bibfnamefont {O.~E.}\ \bibnamefont {{Barndorff-Nielsen}}}\ and\ \bibinfo {author} {\bibfnamefont {R.~D.}\ \bibnamefont {Gill}},\ }\bibfield  {title} {\bibinfo {title} {Fisher information in quantum statistics},\ }\href {https://doi.org/10.1088/0305-4470/33/24/306} {\bibfield  {journal} {\bibinfo  {journal} {J. Phys. A}\ }\textbf {\bibinfo {volume} {33}},\ \bibinfo {pages} {4481} (\bibinfo {year} {2000})}\BibitemShut {NoStop}%
\bibitem [{\citenamefont {Liu}\ \emph {et~al.}(2014)\citenamefont {Liu}, \citenamefont {Jing}, \citenamefont {Zhong},\ and\ \citenamefont {Wang}}]{Liu2014a}%
  \BibitemOpen
  \bibfield  {author} {\bibinfo {author} {\bibfnamefont {J.}~\bibnamefont {Liu}}, \bibinfo {author} {\bibfnamefont {X.-X.}\ \bibnamefont {Jing}}, \bibinfo {author} {\bibfnamefont {W.}~\bibnamefont {Zhong}},\ and\ \bibinfo {author} {\bibfnamefont {X.}~\bibnamefont {Wang}},\ }\bibfield  {title} {\bibinfo {title} {Quantum {{Fisher Information}} for {{Density Matrices}} with {{Arbitrary Ranks}}},\ }\href {https://doi.org/10.1088/0253-6102/61/1/08} {\bibfield  {journal} {\bibinfo  {journal} {Commun. Theor. Phys.}\ }\textbf {\bibinfo {volume} {61}},\ \bibinfo {pages} {45} (\bibinfo {year} {2014})}\BibitemShut {NoStop}%
\bibitem [{\citenamefont {Genoni}\ and\ \citenamefont {Tufarelli}(2019)}]{Genoni2019}%
  \BibitemOpen
  \bibfield  {author} {\bibinfo {author} {\bibfnamefont {M.~G.}\ \bibnamefont {Genoni}}\ and\ \bibinfo {author} {\bibfnamefont {T.}~\bibnamefont {Tufarelli}},\ }\bibfield  {title} {\bibinfo {title} {Non-orthogonal bases for quantum metrology},\ }\href {https://doi.org/10.1088/1751-8121/ab3fe0} {\bibfield  {journal} {\bibinfo  {journal} {J. Phys. A}\ }\textbf {\bibinfo {volume} {52}},\ \bibinfo {pages} {434002} (\bibinfo {year} {2019})}\BibitemShut {NoStop}%
\bibitem [{\citenamefont {Bisketzi}\ \emph {et~al.}(2019)\citenamefont {Bisketzi}, \citenamefont {Branford},\ and\ \citenamefont {Datta}}]{Bisketzi2019}%
  \BibitemOpen
  \bibfield  {author} {\bibinfo {author} {\bibfnamefont {E.}~\bibnamefont {Bisketzi}}, \bibinfo {author} {\bibfnamefont {D.}~\bibnamefont {Branford}},\ and\ \bibinfo {author} {\bibfnamefont {A.}~\bibnamefont {Datta}},\ }\bibfield  {title} {\bibinfo {title} {Quantum limits of localisation microscopy},\ }\href {https://doi.org/10.1088/1367-2630/ab58a0} {\bibfield  {journal} {\bibinfo  {journal} {New J. Phys.}\ }\textbf {\bibinfo {volume} {21}},\ \bibinfo {pages} {123032} (\bibinfo {year} {2019})}\BibitemShut {NoStop}%
\bibitem [{\citenamefont {Fiderer}\ \emph {et~al.}(2021)\citenamefont {Fiderer}, \citenamefont {Tufarelli}, \citenamefont {Piano},\ and\ \citenamefont {Adesso}}]{Fiderer2021a}%
  \BibitemOpen
  \bibfield  {author} {\bibinfo {author} {\bibfnamefont {L.~J.}\ \bibnamefont {Fiderer}}, \bibinfo {author} {\bibfnamefont {T.}~\bibnamefont {Tufarelli}}, \bibinfo {author} {\bibfnamefont {S.}~\bibnamefont {Piano}},\ and\ \bibinfo {author} {\bibfnamefont {G.}~\bibnamefont {Adesso}},\ }\bibfield  {title} {\bibinfo {title} {General {{Expressions}} for the {{Quantum Fisher Information Matrix}} with {{Applications}} to {{Discrete Quantum Imaging}}},\ }\href {https://doi.org/10.1103/PRXQuantum.2.020308} {\bibfield  {journal} {\bibinfo  {journal} {PRX Quantum}\ }\textbf {\bibinfo {volume} {2}},\ \bibinfo {pages} {020308} (\bibinfo {year} {2021})}\BibitemShut {NoStop}%
\bibitem [{\citenamefont {Alipour}\ and\ \citenamefont {Rezakhani}(2015)}]{Alipour2015}%
  \BibitemOpen
  \bibfield  {author} {\bibinfo {author} {\bibfnamefont {S.}~\bibnamefont {Alipour}}\ and\ \bibinfo {author} {\bibfnamefont {A.~T.}\ \bibnamefont {Rezakhani}},\ }\bibfield  {title} {\bibinfo {title} {Extended convexity of quantum {{Fisher}} information in quantum metrology},\ }\href {https://doi.org/10.1103/PhysRevA.91.042104} {\bibfield  {journal} {\bibinfo  {journal} {Phys. Rev. A}\ }\textbf {\bibinfo {volume} {91}},\ \bibinfo {pages} {042104} (\bibinfo {year} {2015})}\BibitemShut {NoStop}%
\bibitem [{\citenamefont {Ng}\ \emph {et~al.}(2016)\citenamefont {Ng}, \citenamefont {Ang}, \citenamefont {Wheatley}, \citenamefont {Yonezawa}, \citenamefont {Furusawa}, \citenamefont {Huntington},\ and\ \citenamefont {Tsang}}]{Ng2016}%
  \BibitemOpen
  \bibfield  {author} {\bibinfo {author} {\bibfnamefont {S.}~\bibnamefont {Ng}}, \bibinfo {author} {\bibfnamefont {S.~Z.}\ \bibnamefont {Ang}}, \bibinfo {author} {\bibfnamefont {T.~A.}\ \bibnamefont {Wheatley}}, \bibinfo {author} {\bibfnamefont {H.}~\bibnamefont {Yonezawa}}, \bibinfo {author} {\bibfnamefont {A.}~\bibnamefont {Furusawa}}, \bibinfo {author} {\bibfnamefont {E.~H.}\ \bibnamefont {Huntington}},\ and\ \bibinfo {author} {\bibfnamefont {M.}~\bibnamefont {Tsang}},\ }\bibfield  {title} {\bibinfo {title} {Spectrum analysis with quantum dynamical systems},\ }\href {https://doi.org/10.1103/PhysRevA.93.042121} {\bibfield  {journal} {\bibinfo  {journal} {Phys. Rev. A}\ }\textbf {\bibinfo {volume} {93}},\ \bibinfo {pages} {042121} (\bibinfo {year} {2016})}\BibitemShut {NoStop}%
\bibitem [{\citenamefont {Shitara}\ \emph {et~al.}(2016)\citenamefont {Shitara}, \citenamefont {Kuramochi},\ and\ \citenamefont {Ueda}}]{Shitara2016}%
  \BibitemOpen
  \bibfield  {author} {\bibinfo {author} {\bibfnamefont {T.}~\bibnamefont {Shitara}}, \bibinfo {author} {\bibfnamefont {Y.}~\bibnamefont {Kuramochi}},\ and\ \bibinfo {author} {\bibfnamefont {M.}~\bibnamefont {Ueda}},\ }\bibfield  {title} {\bibinfo {title} {Trade-off relation between information and disturbance in quantum measurement},\ }\href {https://doi.org/10.1103/PhysRevA.93.032134} {\bibfield  {journal} {\bibinfo  {journal} {Phys. Rev. A}\ }\textbf {\bibinfo {volume} {93}},\ \bibinfo {pages} {032134} (\bibinfo {year} {2016})}\BibitemShut {NoStop}%
\bibitem [{\citenamefont {Combes}\ \emph {et~al.}(2014)\citenamefont {Combes}, \citenamefont {Ferrie}, \citenamefont {Jiang},\ and\ \citenamefont {Caves}}]{Combes2014}%
  \BibitemOpen
  \bibfield  {author} {\bibinfo {author} {\bibfnamefont {J.}~\bibnamefont {Combes}}, \bibinfo {author} {\bibfnamefont {C.}~\bibnamefont {Ferrie}}, \bibinfo {author} {\bibfnamefont {Z.}~\bibnamefont {Jiang}},\ and\ \bibinfo {author} {\bibfnamefont {C.~M.}\ \bibnamefont {Caves}},\ }\bibfield  {title} {\bibinfo {title} {Quantum limits on postselected, probabilistic quantum metrology},\ }\href {https://doi.org/10.1103/PhysRevA.89.052117} {\bibfield  {journal} {\bibinfo  {journal} {Phys. Rev. A}\ }\textbf {\bibinfo {volume} {89}},\ \bibinfo {pages} {052117} (\bibinfo {year} {2014})}\BibitemShut {NoStop}%
\bibitem [{\citenamefont {{Demkowicz-Dobrza{\'n}ski}}\ \emph {et~al.}(2009)\citenamefont {{Demkowicz-Dobrza{\'n}ski}}, \citenamefont {Dorner}, \citenamefont {Smith}, \citenamefont {Lundeen}, \citenamefont {Wasilewski}, \citenamefont {Banaszek},\ and\ \citenamefont {Walmsley}}]{Demkowicz-Dobrzanski2009}%
  \BibitemOpen
  \bibfield  {author} {\bibinfo {author} {\bibfnamefont {R.}~\bibnamefont {{Demkowicz-Dobrza{\'n}ski}}}, \bibinfo {author} {\bibfnamefont {U.}~\bibnamefont {Dorner}}, \bibinfo {author} {\bibfnamefont {B.~J.}\ \bibnamefont {Smith}}, \bibinfo {author} {\bibfnamefont {J.~S.}\ \bibnamefont {Lundeen}}, \bibinfo {author} {\bibfnamefont {W.}~\bibnamefont {Wasilewski}}, \bibinfo {author} {\bibfnamefont {K.}~\bibnamefont {Banaszek}},\ and\ \bibinfo {author} {\bibfnamefont {I.~A.}\ \bibnamefont {Walmsley}},\ }\bibfield  {title} {\bibinfo {title} {Quantum phase estimation with lossy interferometers},\ }\href {https://doi.org/10.1103/PhysRevA.80.013825} {\bibfield  {journal} {\bibinfo  {journal} {Phys. Rev. A}\ }\textbf {\bibinfo {volume} {80}},\ \bibinfo {pages} {013825} (\bibinfo {year} {2009})}\BibitemShut {NoStop}%
\bibitem [{\citenamefont {Farrera}\ \emph {et~al.}(2016)\citenamefont {Farrera}, \citenamefont {Heinze}, \citenamefont {Albrecht}, \citenamefont {Ho}, \citenamefont {Ch{\'a}vez}, \citenamefont {Teo}, \citenamefont {Sangouard},\ and\ \citenamefont {{de Riedmatten}}}]{Farrera2016}%
  \BibitemOpen
  \bibfield  {author} {\bibinfo {author} {\bibfnamefont {P.}~\bibnamefont {Farrera}}, \bibinfo {author} {\bibfnamefont {G.}~\bibnamefont {Heinze}}, \bibinfo {author} {\bibfnamefont {B.}~\bibnamefont {Albrecht}}, \bibinfo {author} {\bibfnamefont {M.}~\bibnamefont {Ho}}, \bibinfo {author} {\bibfnamefont {M.}~\bibnamefont {Ch{\'a}vez}}, \bibinfo {author} {\bibfnamefont {C.}~\bibnamefont {Teo}}, \bibinfo {author} {\bibfnamefont {N.}~\bibnamefont {Sangouard}},\ and\ \bibinfo {author} {\bibfnamefont {H.}~\bibnamefont {{de Riedmatten}}},\ }\bibfield  {title} {\bibinfo {title} {Generation of single photons with highly tunable wave shape from a cold atomic ensemble},\ }\href {https://doi.org/10.1038/ncomms13556} {\bibfield  {journal} {\bibinfo  {journal} {Nat. Commun.}\ }\textbf {\bibinfo {volume} {7}},\ \bibinfo {pages} {13556} (\bibinfo {year} {2016})}\BibitemShut {NoStop}%
\bibitem [{\citenamefont {Pursley}\ \emph {et~al.}(2018)\citenamefont {Pursley}, \citenamefont {Carter}, \citenamefont {Yakes}, \citenamefont {Bracker},\ and\ \citenamefont {Gammon}}]{Pursley2018}%
  \BibitemOpen
  \bibfield  {author} {\bibinfo {author} {\bibfnamefont {B.~C.}\ \bibnamefont {Pursley}}, \bibinfo {author} {\bibfnamefont {S.~G.}\ \bibnamefont {Carter}}, \bibinfo {author} {\bibfnamefont {M.~K.}\ \bibnamefont {Yakes}}, \bibinfo {author} {\bibfnamefont {A.~S.}\ \bibnamefont {Bracker}},\ and\ \bibinfo {author} {\bibfnamefont {D.}~\bibnamefont {Gammon}},\ }\bibfield  {title} {\bibinfo {title} {Picosecond pulse shaping of single photons using quantum dots},\ }\href {https://doi.org/10.1038/s41467-017-02552-7} {\bibfield  {journal} {\bibinfo  {journal} {Nat. Commun.}\ }\textbf {\bibinfo {volume} {9}},\ \bibinfo {pages} {115} (\bibinfo {year} {2018})}\BibitemShut {NoStop}%
\bibitem [{\citenamefont {Morin}\ \emph {et~al.}(2019)\citenamefont {Morin}, \citenamefont {K{\"o}rber}, \citenamefont {Langenfeld},\ and\ \citenamefont {Rempe}}]{Morin2019}%
  \BibitemOpen
  \bibfield  {author} {\bibinfo {author} {\bibfnamefont {O.}~\bibnamefont {Morin}}, \bibinfo {author} {\bibfnamefont {M.}~\bibnamefont {K{\"o}rber}}, \bibinfo {author} {\bibfnamefont {S.}~\bibnamefont {Langenfeld}},\ and\ \bibinfo {author} {\bibfnamefont {G.}~\bibnamefont {Rempe}},\ }\bibfield  {title} {\bibinfo {title} {Deterministic {{Shaping}} and {{Reshaping}} of {{Single-Photon Temporal Wave Functions}}},\ }\href {https://doi.org/10.1103/PhysRevLett.123.133602} {\bibfield  {journal} {\bibinfo  {journal} {Phys. Rev. Lett.}\ }\textbf {\bibinfo {volume} {123}},\ \bibinfo {pages} {133602} (\bibinfo {year} {2019})}\BibitemShut {NoStop}%
\bibitem [{\citenamefont {Lipka}\ and\ \citenamefont {Parniak}(2021)}]{Lipka2021b}%
  \BibitemOpen
  \bibfield  {author} {\bibinfo {author} {\bibfnamefont {M.}~\bibnamefont {Lipka}}\ and\ \bibinfo {author} {\bibfnamefont {M.}~\bibnamefont {Parniak}},\ }\bibfield  {title} {\bibinfo {title} {Single-{{Photon Hologram}} of a {{Zero-Area Pulse}}},\ }\href {https://doi.org/10.1103/PhysRevLett.127.163601} {\bibfield  {journal} {\bibinfo  {journal} {Phys. Rev. Lett.}\ }\textbf {\bibinfo {volume} {127}},\ \bibinfo {pages} {163601} (\bibinfo {year} {2021})}\BibitemShut {NoStop}%
\bibitem [{\citenamefont {Zamir}(1998)}]{Zamir1998}%
  \BibitemOpen
  \bibfield  {author} {\bibinfo {author} {\bibfnamefont {R.}~\bibnamefont {Zamir}},\ }\bibfield  {title} {\bibinfo {title} {A proof of the {{Fisher Information Inequality}} via a data processing argument},\ }\href {https://doi.org/10.1109/18.669301} {\bibfield  {journal} {\bibinfo  {journal} {IEEE Trans. Inf. Theory}\ }\textbf {\bibinfo {volume} {44}},\ \bibinfo {pages} {1246} (\bibinfo {year} {1998})}\BibitemShut {NoStop}%
\bibitem [{\citenamefont {Macr{\`i}}\ \emph {et~al.}(2016)\citenamefont {Macr{\`i}}, \citenamefont {Smerzi},\ and\ \citenamefont {Pezz{\`e}}}]{Macri2016}%
  \BibitemOpen
  \bibfield  {author} {\bibinfo {author} {\bibfnamefont {T.}~\bibnamefont {Macr{\`i}}}, \bibinfo {author} {\bibfnamefont {A.}~\bibnamefont {Smerzi}},\ and\ \bibinfo {author} {\bibfnamefont {L.}~\bibnamefont {Pezz{\`e}}},\ }\bibfield  {title} {\bibinfo {title} {Loschmidt echo for quantum metrology},\ }\href {https://doi.org/10.1103/PhysRevA.94.010102} {\bibfield  {journal} {\bibinfo  {journal} {Phys. Rev. A}\ }\textbf {\bibinfo {volume} {94}},\ \bibinfo {pages} {010102} (\bibinfo {year} {2016})}\BibitemShut {NoStop}%
\bibitem [{\citenamefont {Kurdzialek}\ and\ \citenamefont {{Demkowicz-Dobrzanski}}(2022)}]{Kurdzialek2022}%
  \BibitemOpen
  \bibfield  {author} {\bibinfo {author} {\bibfnamefont {S.}~\bibnamefont {Kurdzialek}}\ and\ \bibinfo {author} {\bibfnamefont {R.}~\bibnamefont {{Demkowicz-Dobrzanski}}},\ }\bibfield  {title} {\bibinfo {title} {Measurement noise susceptibility in quantum estimation},\ }\href {https://arxiv.org/abs/2206.12430} {\bibfield  {journal} {\bibinfo  {journal} {arXiv:2206.12430}\ } (\bibinfo {year} {2022})}\BibitemShut {NoStop}%
\bibitem [{\citenamefont {Becker}\ \emph {et~al.}(2004)\citenamefont {Becker}, \citenamefont {Bergmann}, \citenamefont {Hink}, \citenamefont {K{\"o}nig}, \citenamefont {Benndorf},\ and\ \citenamefont {Biskup}}]{becker2004fluorescence}%
  \BibitemOpen
  \bibfield  {author} {\bibinfo {author} {\bibfnamefont {W.}~\bibnamefont {Becker}}, \bibinfo {author} {\bibfnamefont {A.}~\bibnamefont {Bergmann}}, \bibinfo {author} {\bibfnamefont {M.}~\bibnamefont {Hink}}, \bibinfo {author} {\bibfnamefont {K.}~\bibnamefont {K{\"o}nig}}, \bibinfo {author} {\bibfnamefont {K.}~\bibnamefont {Benndorf}},\ and\ \bibinfo {author} {\bibfnamefont {C.}~\bibnamefont {Biskup}},\ }\bibfield  {title} {\bibinfo {title} {Fluorescence lifetime imaging by time-correlated single-photon counting},\ }\href {https://doi.org/https://doi.org/10.1002/jemt.10421} {\bibfield  {journal} {\bibinfo  {journal} {Microsc. Res. Tech.}\ }\textbf {\bibinfo {volume} {63}},\ \bibinfo {pages} {58} (\bibinfo {year} {2004})}\BibitemShut {NoStop}%
\bibitem [{\citenamefont {Eckstein}\ \emph {et~al.}(2011)\citenamefont {Eckstein}, \citenamefont {Brecht},\ and\ \citenamefont {Silberhorn}}]{eckstein2011quantum}%
  \BibitemOpen
  \bibfield  {author} {\bibinfo {author} {\bibfnamefont {A.}~\bibnamefont {Eckstein}}, \bibinfo {author} {\bibfnamefont {B.}~\bibnamefont {Brecht}},\ and\ \bibinfo {author} {\bibfnamefont {C.}~\bibnamefont {Silberhorn}},\ }\bibfield  {title} {\bibinfo {title} {A quantum pulse gate based on spectrally engineered sum frequency generation},\ }\href {https://doi.org/10.1364/OE.19.013770} {\bibfield  {journal} {\bibinfo  {journal} {Optics express}\ }\textbf {\bibinfo {volume} {19}},\ \bibinfo {pages} {13770} (\bibinfo {year} {2011})}\BibitemShut {NoStop}%
\bibitem [{\citenamefont {Donohue}\ \emph {et~al.}(2018)\citenamefont {Donohue}, \citenamefont {Ansari}, \citenamefont {{\v{R}}eh{\'a}{\v{c}}ek}, \citenamefont {Hradil}, \citenamefont {Stoklasa}, \citenamefont {Pa{\'u}r}, \citenamefont {S{\'a}nchez-Soto},\ and\ \citenamefont {Silberhorn}}]{donohue2018quantum}%
  \BibitemOpen
  \bibfield  {author} {\bibinfo {author} {\bibfnamefont {J.~M.}\ \bibnamefont {Donohue}}, \bibinfo {author} {\bibfnamefont {V.}~\bibnamefont {Ansari}}, \bibinfo {author} {\bibfnamefont {J.}~\bibnamefont {{\v{R}}eh{\'a}{\v{c}}ek}}, \bibinfo {author} {\bibfnamefont {Z.}~\bibnamefont {Hradil}}, \bibinfo {author} {\bibfnamefont {B.}~\bibnamefont {Stoklasa}}, \bibinfo {author} {\bibfnamefont {M.}~\bibnamefont {Pa{\'u}r}}, \bibinfo {author} {\bibfnamefont {L.~L.}\ \bibnamefont {S{\'a}nchez-Soto}},\ and\ \bibinfo {author} {\bibfnamefont {C.}~\bibnamefont {Silberhorn}},\ }\bibfield  {title} {\bibinfo {title} {Quantum-limited time-frequency estimation through mode-selective photon measurement},\ }\href {https://doi.org/10.1103/PhysRevLett.121.090501} {\bibfield  {journal} {\bibinfo  {journal} {Physical review letters}\ }\textbf {\bibinfo {volume} {121}},\ \bibinfo {pages} {090501} (\bibinfo {year} {2018})}\BibitemShut {NoStop}%
\bibitem [{\citenamefont {De}\ \emph {et~al.}(2021)\citenamefont {De}, \citenamefont {Gil-Lopez}, \citenamefont {Brecht}, \citenamefont {Silberhorn}, \citenamefont {S{\'a}nchez-Soto}, \citenamefont {Hradil},\ and\ \citenamefont {{\v{R}}eh{\'a}{\v{c}}ek}}]{de2021effects}%
  \BibitemOpen
  \bibfield  {author} {\bibinfo {author} {\bibfnamefont {S.}~\bibnamefont {De}}, \bibinfo {author} {\bibfnamefont {J.}~\bibnamefont {Gil-Lopez}}, \bibinfo {author} {\bibfnamefont {B.}~\bibnamefont {Brecht}}, \bibinfo {author} {\bibfnamefont {C.}~\bibnamefont {Silberhorn}}, \bibinfo {author} {\bibfnamefont {L.~L.}\ \bibnamefont {S{\'a}nchez-Soto}}, \bibinfo {author} {\bibfnamefont {Z.}~\bibnamefont {Hradil}},\ and\ \bibinfo {author} {\bibfnamefont {J.}~\bibnamefont {{\v{R}}eh{\'a}{\v{c}}ek}},\ }\bibfield  {title} {\bibinfo {title} {Effects of coherence on temporal resolution},\ }\href {https://doi.org/10.1103/PhysRevResearch.3.033082} {\bibfield  {journal} {\bibinfo  {journal} {Physical Review Research}\ }\textbf {\bibinfo {volume} {3}},\ \bibinfo {pages} {033082} (\bibinfo {year} {2021})}\BibitemShut {NoStop}%
\bibitem [{\citenamefont {Ansari}\ \emph {et~al.}(2021)\citenamefont {Ansari}, \citenamefont {Brecht}, \citenamefont {Gil-L{\'o}pez}, \citenamefont {Donohue}, \citenamefont {{\v{R}}eh{\'a}{\v{c}}ek}, \citenamefont {Hradil}, \citenamefont {S{\'a}nchez-Soto},\ and\ \citenamefont {Silberhorn}}]{ansari2021achieving}%
  \BibitemOpen
  \bibfield  {author} {\bibinfo {author} {\bibfnamefont {V.}~\bibnamefont {Ansari}}, \bibinfo {author} {\bibfnamefont {B.}~\bibnamefont {Brecht}}, \bibinfo {author} {\bibfnamefont {J.}~\bibnamefont {Gil-L{\'o}pez}}, \bibinfo {author} {\bibfnamefont {J.~M.}\ \bibnamefont {Donohue}}, \bibinfo {author} {\bibfnamefont {J.}~\bibnamefont {{\v{R}}eh{\'a}{\v{c}}ek}}, \bibinfo {author} {\bibfnamefont {Z.}~\bibnamefont {Hradil}}, \bibinfo {author} {\bibfnamefont {L.~L.}\ \bibnamefont {S{\'a}nchez-Soto}},\ and\ \bibinfo {author} {\bibfnamefont {C.}~\bibnamefont {Silberhorn}},\ }\bibfield  {title} {\bibinfo {title} {Achieving the ultimate quantum timing resolution},\ }\href {https://doi.org/10.1103/PRXQuantum.2.010301} {\bibfield  {journal} {\bibinfo  {journal} {PRX Quantum}\ }\textbf {\bibinfo {volume} {2}},\ \bibinfo {pages} {010301} (\bibinfo {year} {2021})}\BibitemShut {NoStop}%
\bibitem [{\citenamefont {Garikapati}\ \emph {et~al.}(2022)\citenamefont {Garikapati}, \citenamefont {Kumar}, \citenamefont {Zhang}, \citenamefont {Sua},\ and\ \citenamefont {Huang}}]{garikapati2022programmable}%
  \BibitemOpen
  \bibfield  {author} {\bibinfo {author} {\bibfnamefont {M.}~\bibnamefont {Garikapati}}, \bibinfo {author} {\bibfnamefont {S.}~\bibnamefont {Kumar}}, \bibinfo {author} {\bibfnamefont {H.}~\bibnamefont {Zhang}}, \bibinfo {author} {\bibfnamefont {Y.~M.}\ \bibnamefont {Sua}},\ and\ \bibinfo {author} {\bibfnamefont {Y.-P.}\ \bibnamefont {Huang}},\ }\bibfield  {title} {\bibinfo {title} {A programmable spatiotemporal quantum parametric mode sorter},\ }\href {https://arxiv.org/abs/2210.16517} {\bibfield  {journal} {\bibinfo  {journal} {arXiv:2210.16517}\ } (\bibinfo {year} {2022})}\BibitemShut {NoStop}%
\bibitem [{\citenamefont {Stobi{\'n}ska}\ \emph {et~al.}(2009)\citenamefont {Stobi{\'n}ska}, \citenamefont {Alber},\ and\ \citenamefont {Leuchs}}]{Stobinska2009a}%
  \BibitemOpen
  \bibfield  {author} {\bibinfo {author} {\bibfnamefont {M.}~\bibnamefont {Stobi{\'n}ska}}, \bibinfo {author} {\bibfnamefont {G.}~\bibnamefont {Alber}},\ and\ \bibinfo {author} {\bibfnamefont {G.}~\bibnamefont {Leuchs}},\ }\bibfield  {title} {\bibinfo {title} {Perfect excitation of a matter qubit by a single photon in free space},\ }\href {https://doi.org/10.1209/0295-5075/86/14007} {\bibfield  {journal} {\bibinfo  {journal} {EPL}\ }\textbf {\bibinfo {volume} {86}},\ \bibinfo {pages} {14007} (\bibinfo {year} {2009})}\BibitemShut {NoStop}%
\bibitem [{\citenamefont {Knall}\ \emph {et~al.}(2022)\citenamefont {Knall}, \citenamefont {Knaut}, \citenamefont {Bekenstein}, \citenamefont {Assumpcao}, \citenamefont {Stroganov}, \citenamefont {Gong}, \citenamefont {Huan}, \citenamefont {Stas}, \citenamefont {Machielse}, \citenamefont {Chalupnik}, \citenamefont {Levonian}, \citenamefont {Suleymanzade}, \citenamefont {Riedinger}, \citenamefont {Park}, \citenamefont {Lon{\v c}ar}, \citenamefont {Bhaskar},\ and\ \citenamefont {Lukin}}]{Knall2022}%
  \BibitemOpen
  \bibfield  {author} {\bibinfo {author} {\bibfnamefont {E.~N.}\ \bibnamefont {Knall}}, \bibinfo {author} {\bibfnamefont {C.~M.}\ \bibnamefont {Knaut}}, \bibinfo {author} {\bibfnamefont {R.}~\bibnamefont {Bekenstein}}, \bibinfo {author} {\bibfnamefont {D.~R.}\ \bibnamefont {Assumpcao}}, \bibinfo {author} {\bibfnamefont {P.~L.}\ \bibnamefont {Stroganov}}, \bibinfo {author} {\bibfnamefont {W.}~\bibnamefont {Gong}}, \bibinfo {author} {\bibfnamefont {Y.~Q.}\ \bibnamefont {Huan}}, \bibinfo {author} {\bibfnamefont {P.-J.}\ \bibnamefont {Stas}}, \bibinfo {author} {\bibfnamefont {B.}~\bibnamefont {Machielse}}, \bibinfo {author} {\bibfnamefont {M.}~\bibnamefont {Chalupnik}}, \bibinfo {author} {\bibfnamefont {D.}~\bibnamefont {Levonian}}, \bibinfo {author} {\bibfnamefont {A.}~\bibnamefont {Suleymanzade}}, \bibinfo {author} {\bibfnamefont {R.}~\bibnamefont {Riedinger}}, \bibinfo {author} {\bibfnamefont {H.}~\bibnamefont {Park}}, \bibinfo {author} {\bibfnamefont {M.}~\bibnamefont {Lon{\v c}ar}}, \bibinfo {author} {\bibfnamefont {M.~K.}\ \bibnamefont {Bhaskar}},\ and\ \bibinfo {author} {\bibfnamefont {M.~D.}\ \bibnamefont {Lukin}},\ }\bibfield  {title} {\bibinfo {title} {Efficient {{Source}} of {{Shaped Single Photons Based}} on an {{Integrated Diamond Nanophotonic System}}},\ }\href {https://doi.org/10.1103/PhysRevLett.129.053603} {\bibfield  {journal} {\bibinfo  {journal} {Phys. Rev. Lett.}\ }\textbf {\bibinfo {volume} {129}},\ \bibinfo {pages} {053603} (\bibinfo {year} {2022})}\BibitemShut {NoStop}%
\bibitem [{\citenamefont {{van Enk}}\ and\ \citenamefont {Kimble}(2002)}]{vanEnk2002}%
  \BibitemOpen
  \bibfield  {author} {\bibinfo {author} {\bibfnamefont {S.}~\bibnamefont {{van Enk}}}\ and\ \bibinfo {author} {\bibfnamefont {H.}~\bibnamefont {Kimble}},\ }\bibfield  {title} {\bibinfo {title} {On the classical character of control fields in quantum information processing},\ }\href {https://doi.org/10.26421/QIC2.1-1} {\bibfield  {journal} {\bibinfo  {journal} {Quantum Inf. Comput.}\ }\textbf {\bibinfo {volume} {2}},\ \bibinfo {pages} {1} (\bibinfo {year} {2002})}\BibitemShut {NoStop}%
\bibitem [{\citenamefont {Bisketzi}(2021)}]{Bisketzi2022}%
  \BibitemOpen
  \bibfield  {author} {\bibinfo {author} {\bibfnamefont {E.}~\bibnamefont {Bisketzi}},\ }\emph {\bibinfo {title} {Quantum Limits in Microscopy and Spectroscopy}},\ \href {http://wrap.warwick.ac.uk/170994} {Ph.D. thesis},\ \bibinfo  {school} {University of Warwick} (\bibinfo {year} {2021})\BibitemShut {NoStop}%
\bibitem [{\citenamefont {Christiansen}\ \emph {et~al.}(2023)\citenamefont {Christiansen}, \citenamefont {Kiilerich},\ and\ \citenamefont {M{\o}lmer}}]{Christiansen2022}%
  \BibitemOpen
  \bibfield  {author} {\bibinfo {author} {\bibfnamefont {V.~R.}\ \bibnamefont {Christiansen}}, \bibinfo {author} {\bibfnamefont {A.~H.}\ \bibnamefont {Kiilerich}},\ and\ \bibinfo {author} {\bibfnamefont {K.}~\bibnamefont {M{\o}lmer}},\ }\bibfield  {title} {\bibinfo {title} {Interactions of quantum systems with pulses of quantized radiation: {{From}} a cascaded master equation to a traveling mode perspective},\ }\href {https://doi.org/10.1103/PhysRevA.107.013706} {\bibfield  {journal} {\bibinfo  {journal} {Phys. Rev. A}\ }\textbf {\bibinfo {volume} {107}},\ \bibinfo {pages} {013706} (\bibinfo {year} {2023})}\BibitemShut {NoStop}%
\bibitem [{\citenamefont {Larson}\ and\ \citenamefont {Mavrogordatos}(2021)}]{Larson2021b}%
  \BibitemOpen
  \bibfield  {author} {\bibinfo {author} {\bibfnamefont {J.}~\bibnamefont {Larson}}\ and\ \bibinfo {author} {\bibfnamefont {T.}~\bibnamefont {Mavrogordatos}},\ }\href {https://doi.org/10.1088/978-0-7503-3447-1} {\emph {\bibinfo {title} {The {{Jaynes}}\textendash{{Cummings Model}} and {{Its Descendants}}}}}\ (\bibinfo  {publisher} {{IOP Publishing}},\ \bibinfo {address} {{Bristol, UK}},\ \bibinfo {year} {2021})\BibitemShut {NoStop}%
\bibitem [{\citenamefont {Raymer}\ \emph {et~al.}(2021)\citenamefont {Raymer}, \citenamefont {Landes},\ and\ \citenamefont {Marcus}}]{Raymer2021}%
  \BibitemOpen
  \bibfield  {author} {\bibinfo {author} {\bibfnamefont {M.~G.}\ \bibnamefont {Raymer}}, \bibinfo {author} {\bibfnamefont {T.}~\bibnamefont {Landes}},\ and\ \bibinfo {author} {\bibfnamefont {A.~H.}\ \bibnamefont {Marcus}},\ }\bibfield  {title} {\bibinfo {title} {Entangled two-photon absorption by atoms and molecules: {{A}} quantum optics tutorial},\ }\href {https://doi.org/10.1063/5.0049338} {\bibfield  {journal} {\bibinfo  {journal} {J. Chem. Phys.}\ }\textbf {\bibinfo {volume} {155}},\ \bibinfo {pages} {081501} (\bibinfo {year} {2021})}\BibitemShut {NoStop}%
\bibitem [{\citenamefont {Allen}\ \emph {et~al.}(2020)\citenamefont {Allen}, \citenamefont {{Sabines-Chesterking}}, \citenamefont {McMillan}, \citenamefont {Joshi}, \citenamefont {Turner},\ and\ \citenamefont {Matthews}}]{Allen2020}%
  \BibitemOpen
  \bibfield  {author} {\bibinfo {author} {\bibfnamefont {E.~J.}\ \bibnamefont {Allen}}, \bibinfo {author} {\bibfnamefont {J.}~\bibnamefont {{Sabines-Chesterking}}}, \bibinfo {author} {\bibfnamefont {A.~R.}\ \bibnamefont {McMillan}}, \bibinfo {author} {\bibfnamefont {S.~K.}\ \bibnamefont {Joshi}}, \bibinfo {author} {\bibfnamefont {P.~S.}\ \bibnamefont {Turner}},\ and\ \bibinfo {author} {\bibfnamefont {J.~C.~F.}\ \bibnamefont {Matthews}},\ }\bibfield  {title} {\bibinfo {title} {Approaching the quantum limit of precision in absorbance estimation using classical resources},\ }\href {https://doi.org/10.1103/PhysRevResearch.2.033243} {\bibfield  {journal} {\bibinfo  {journal} {Phys. Rev. Res.}\ }\textbf {\bibinfo {volume} {2}},\ \bibinfo {pages} {033243} (\bibinfo {year} {2020})}\BibitemShut {NoStop}%
\bibitem [{\citenamefont {Belsley}\ \emph {et~al.}(2022)\citenamefont {Belsley}, \citenamefont {Allen}, \citenamefont {Datta},\ and\ \citenamefont {Matthews}}]{Belsley2022}%
  \BibitemOpen
  \bibfield  {author} {\bibinfo {author} {\bibfnamefont {A.}~\bibnamefont {Belsley}}, \bibinfo {author} {\bibfnamefont {E.~J.}\ \bibnamefont {Allen}}, \bibinfo {author} {\bibfnamefont {A.}~\bibnamefont {Datta}},\ and\ \bibinfo {author} {\bibfnamefont {J.~C.~F.}\ \bibnamefont {Matthews}},\ }\bibfield  {title} {\bibinfo {title} {Advantage of {{Coherent States}} in {{Ring Resonators}} over {{Any Quantum Probe Single-Pass Absorption Estimation Strategy}}},\ }\href {https://doi.org/10.1103/PhysRevLett.128.230501} {\bibfield  {journal} {\bibinfo  {journal} {Phys. Rev. Lett.}\ }\textbf {\bibinfo {volume} {128}},\ \bibinfo {pages} {230501} (\bibinfo {year} {2022})}\BibitemShut {NoStop}%
\bibitem [{\citenamefont {Monras}\ and\ \citenamefont {Paris}(2007)}]{Monras2007}%
  \BibitemOpen
  \bibfield  {author} {\bibinfo {author} {\bibfnamefont {A.}~\bibnamefont {Monras}}\ and\ \bibinfo {author} {\bibfnamefont {M.~G.~A.}\ \bibnamefont {Paris}},\ }\bibfield  {title} {\bibinfo {title} {Optimal {{Quantum Estimation}} of {{Loss}} in {{Bosonic Channels}}},\ }\href {https://doi.org/10.1103/PhysRevLett.98.160401} {\bibfield  {journal} {\bibinfo  {journal} {Phys. Rev. Lett.}\ }\textbf {\bibinfo {volume} {98}},\ \bibinfo {pages} {160401} (\bibinfo {year} {2007})}\BibitemShut {NoStop}%
\bibitem [{\citenamefont {Adesso}\ \emph {et~al.}(2009)\citenamefont {Adesso}, \citenamefont {Dell'Anno}, \citenamefont {De~Siena}, \citenamefont {Illuminati},\ and\ \citenamefont {Souza}}]{Adesso2009}%
  \BibitemOpen
  \bibfield  {author} {\bibinfo {author} {\bibfnamefont {G.}~\bibnamefont {Adesso}}, \bibinfo {author} {\bibfnamefont {F.}~\bibnamefont {Dell'Anno}}, \bibinfo {author} {\bibfnamefont {S.}~\bibnamefont {De~Siena}}, \bibinfo {author} {\bibfnamefont {F.}~\bibnamefont {Illuminati}},\ and\ \bibinfo {author} {\bibfnamefont {L.~A.~M.}\ \bibnamefont {Souza}},\ }\bibfield  {title} {\bibinfo {title} {Optimal estimation of losses at the ultimate quantum limit with non-{{Gaussian}} states},\ }\href {https://doi.org/10.1103/PhysRevA.79.040305} {\bibfield  {journal} {\bibinfo  {journal} {Phys. Rev. A}\ }\textbf {\bibinfo {volume} {79}},\ \bibinfo {pages} {040305(R)} (\bibinfo {year} {2009})}\BibitemShut {NoStop}%
\bibitem [{\citenamefont {Nair}(2018)}]{Nair2018}%
  \BibitemOpen
  \bibfield  {author} {\bibinfo {author} {\bibfnamefont {R.}~\bibnamefont {Nair}},\ }\bibfield  {title} {\bibinfo {title} {Quantum-{{Limited Loss Sensing}}: {{Multiparameter Estimation}} and {{Bures Distance}} between {{Loss Channels}}},\ }\href {https://doi.org/10.1103/PhysRevLett.121.230801} {\bibfield  {journal} {\bibinfo  {journal} {Phys. Rev. Lett.}\ }\textbf {\bibinfo {volume} {121}},\ \bibinfo {pages} {230801} (\bibinfo {year} {2018})}\BibitemShut {NoStop}%
\bibitem [{\citenamefont {Schlawin}\ \emph {et~al.}(2016)\citenamefont {Schlawin}, \citenamefont {Dorfman},\ and\ \citenamefont {Mukamel}}]{Schlawin2016}%
  \BibitemOpen
  \bibfield  {author} {\bibinfo {author} {\bibfnamefont {F.}~\bibnamefont {Schlawin}}, \bibinfo {author} {\bibfnamefont {K.~E.}\ \bibnamefont {Dorfman}},\ and\ \bibinfo {author} {\bibfnamefont {S.}~\bibnamefont {Mukamel}},\ }\bibfield  {title} {\bibinfo {title} {Pump-probe spectroscopy using quantum light with two-photon coincidence detection},\ }\href {https://doi.org/10.1103/PhysRevA.93.023807} {\bibfield  {journal} {\bibinfo  {journal} {Phys. Rev. A}\ }\textbf {\bibinfo {volume} {93}},\ \bibinfo {pages} {023807} (\bibinfo {year} {2016})}\BibitemShut {NoStop}%
\bibitem [{\citenamefont {Li}\ \emph {et~al.}(2017)\citenamefont {Li}, \citenamefont {Piryatinski}, \citenamefont {Jerke}, \citenamefont {Kandada}, \citenamefont {Silva},\ and\ \citenamefont {Bittner}}]{li2017probing}%
  \BibitemOpen
  \bibfield  {author} {\bibinfo {author} {\bibfnamefont {H.}~\bibnamefont {Li}}, \bibinfo {author} {\bibfnamefont {A.}~\bibnamefont {Piryatinski}}, \bibinfo {author} {\bibfnamefont {J.}~\bibnamefont {Jerke}}, \bibinfo {author} {\bibfnamefont {A.~R.~S.}\ \bibnamefont {Kandada}}, \bibinfo {author} {\bibfnamefont {C.}~\bibnamefont {Silva}},\ and\ \bibinfo {author} {\bibfnamefont {E.~R.}\ \bibnamefont {Bittner}},\ }\bibfield  {title} {\bibinfo {title} {Probing dynamical symmetry breaking using quantum-entangled photons},\ }\href {https://doi.org/https://doi.org/10.1088/2058-9565/aa93b6} {\bibfield  {journal} {\bibinfo  {journal} {Quantum Sci. Technol.}\ }\textbf {\bibinfo {volume} {3}},\ \bibinfo {pages} {015003} (\bibinfo {year} {2017})}\BibitemShut {NoStop}%
\bibitem [{\citenamefont {Stefanov}(2017)}]{Stefanov2017a}%
  \BibitemOpen
  \bibfield  {author} {\bibinfo {author} {\bibfnamefont {A.}~\bibnamefont {Stefanov}},\ }\bibfield  {title} {\bibinfo {title} {On the role of entanglement in two-photon metrology},\ }\href {https://doi.org/10.1088/2058-9565/aa6ae1} {\bibfield  {journal} {\bibinfo  {journal} {Quantum Sci. Technol.}\ }\textbf {\bibinfo {volume} {2}},\ \bibinfo {pages} {025004} (\bibinfo {year} {2017})}\BibitemShut {NoStop}%
\bibitem [{\citenamefont {{Demkowicz-Dobrza{\'n}ski}}\ and\ \citenamefont {Maccone}(2014)}]{Demkowicz-Dobrzanski2014}%
  \BibitemOpen
  \bibfield  {author} {\bibinfo {author} {\bibfnamefont {R.}~\bibnamefont {{Demkowicz-Dobrza{\'n}ski}}}\ and\ \bibinfo {author} {\bibfnamefont {L.}~\bibnamefont {Maccone}},\ }\bibfield  {title} {\bibinfo {title} {Using {{Entanglement Against Noise}} in {{Quantum Metrology}}},\ }\href {https://doi.org/10.1103/PhysRevLett.113.250801} {\bibfield  {journal} {\bibinfo  {journal} {Phys. Rev. Lett.}\ }\textbf {\bibinfo {volume} {113}},\ \bibinfo {pages} {250801} (\bibinfo {year} {2014})}\BibitemShut {NoStop}%
\bibitem [{\citenamefont {Layden}\ \emph {et~al.}(2019)\citenamefont {Layden}, \citenamefont {Zhou}, \citenamefont {Cappellaro},\ and\ \citenamefont {Jiang}}]{Layden2019}%
  \BibitemOpen
  \bibfield  {author} {\bibinfo {author} {\bibfnamefont {D.}~\bibnamefont {Layden}}, \bibinfo {author} {\bibfnamefont {S.}~\bibnamefont {Zhou}}, \bibinfo {author} {\bibfnamefont {P.}~\bibnamefont {Cappellaro}},\ and\ \bibinfo {author} {\bibfnamefont {L.}~\bibnamefont {Jiang}},\ }\bibfield  {title} {\bibinfo {title} {Ancilla-{{Free Quantum Error Correction Codes}} for {{Quantum Metrology}}},\ }\href {https://doi.org/10.1103/PhysRevLett.122.040502} {\bibfield  {journal} {\bibinfo  {journal} {Phys. Rev. Lett.}\ }\textbf {\bibinfo {volume} {122}},\ \bibinfo {pages} {040502} (\bibinfo {year} {2019})}\BibitemShut {NoStop}%
\bibitem [{\citenamefont {Grice}\ \emph {et~al.}(2001)\citenamefont {Grice}, \citenamefont {U'Ren},\ and\ \citenamefont {Walmsley}}]{Grice2001a}%
  \BibitemOpen
  \bibfield  {author} {\bibinfo {author} {\bibfnamefont {W.~P.}\ \bibnamefont {Grice}}, \bibinfo {author} {\bibfnamefont {A.~B.}\ \bibnamefont {U'Ren}},\ and\ \bibinfo {author} {\bibfnamefont {I.~A.}\ \bibnamefont {Walmsley}},\ }\bibfield  {title} {\bibinfo {title} {Eliminating frequency and space-time correlations in multiphoton states},\ }\href {https://doi.org/10.1103/PhysRevA.64.063815} {\bibfield  {journal} {\bibinfo  {journal} {Phys. Rev. A}\ }\textbf {\bibinfo {volume} {64}},\ \bibinfo {pages} {063815} (\bibinfo {year} {2001})}\BibitemShut {NoStop}%
\bibitem [{\citenamefont {U'Ren}\ \emph {et~al.}(2003)\citenamefont {U'Ren}, \citenamefont {Banaszek},\ and\ \citenamefont {Walmsley}}]{URen2003}%
  \BibitemOpen
  \bibfield  {author} {\bibinfo {author} {\bibfnamefont {A.}~\bibnamefont {U'Ren}}, \bibinfo {author} {\bibfnamefont {K.}~\bibnamefont {Banaszek}},\ and\ \bibinfo {author} {\bibfnamefont {I.}~\bibnamefont {Walmsley}},\ }\bibfield  {title} {\bibinfo {title} {Photon engineering for quantum information processing},\ }\href {https://doi.org/10.26421/QIC3.s-3} {\bibfield  {journal} {\bibinfo  {journal} {QIC}\ }\textbf {\bibinfo {volume} {3}},\ \bibinfo {pages} {480} (\bibinfo {year} {2003})}\BibitemShut {NoStop}%
\bibitem [{\citenamefont {Christ}\ \emph {et~al.}(2013{\natexlab{a}})\citenamefont {Christ}, \citenamefont {Brecht}, \citenamefont {Mauerer},\ and\ \citenamefont {Silberhorn}}]{Christ2013}%
  \BibitemOpen
  \bibfield  {author} {\bibinfo {author} {\bibfnamefont {A.}~\bibnamefont {Christ}}, \bibinfo {author} {\bibfnamefont {B.}~\bibnamefont {Brecht}}, \bibinfo {author} {\bibfnamefont {W.}~\bibnamefont {Mauerer}},\ and\ \bibinfo {author} {\bibfnamefont {C.}~\bibnamefont {Silberhorn}},\ }\bibfield  {title} {\bibinfo {title} {Theory of quantum frequency conversion and type-{{II}} parametric down-conversion in the high-gain regime},\ }\href {https://doi.org/10.1088/1367-2630/15/5/053038} {\bibfield  {journal} {\bibinfo  {journal} {New J. Phys.}\ }\textbf {\bibinfo {volume} {15}},\ \bibinfo {pages} {053038} (\bibinfo {year} {2013}{\natexlab{a}})}\BibitemShut {NoStop}%
\bibitem [{\citenamefont {Kuzucu}\ \emph {et~al.}(2008)\citenamefont {Kuzucu}, \citenamefont {Wong}, \citenamefont {Kurimura},\ and\ \citenamefont {Tovstonog}}]{kuzucu2008joint}%
  \BibitemOpen
  \bibfield  {author} {\bibinfo {author} {\bibfnamefont {O.}~\bibnamefont {Kuzucu}}, \bibinfo {author} {\bibfnamefont {F.~N.}\ \bibnamefont {Wong}}, \bibinfo {author} {\bibfnamefont {S.}~\bibnamefont {Kurimura}},\ and\ \bibinfo {author} {\bibfnamefont {S.}~\bibnamefont {Tovstonog}},\ }\bibfield  {title} {\bibinfo {title} {Joint temporal density measurements for two-photon state characterization},\ }\href {https://doi.org/https://doi.org/10.1103/PhysRevLett.101.153602} {\bibfield  {journal} {\bibinfo  {journal} {Phys. Rev. Lett.}\ }\textbf {\bibinfo {volume} {101}},\ \bibinfo {pages} {153602} (\bibinfo {year} {2008})}\BibitemShut {NoStop}%
\bibitem [{\citenamefont {Pe'er}\ \emph {et~al.}(2005)\citenamefont {Pe'er}, \citenamefont {Dayan}, \citenamefont {Friesem},\ and\ \citenamefont {Silberberg}}]{Peer2005}%
  \BibitemOpen
  \bibfield  {author} {\bibinfo {author} {\bibfnamefont {A.}~\bibnamefont {Pe'er}}, \bibinfo {author} {\bibfnamefont {B.}~\bibnamefont {Dayan}}, \bibinfo {author} {\bibfnamefont {A.~A.}\ \bibnamefont {Friesem}},\ and\ \bibinfo {author} {\bibfnamefont {Y.}~\bibnamefont {Silberberg}},\ }\bibfield  {title} {\bibinfo {title} {Temporal {{Shaping}} of {{Entangled Photons}}},\ }\href {https://doi.org/10.1103/PhysRevLett.94.073601} {\bibfield  {journal} {\bibinfo  {journal} {Phys. Rev. Lett.}\ }\textbf {\bibinfo {volume} {94}},\ \bibinfo {pages} {073601} (\bibinfo {year} {2005})}\BibitemShut {NoStop}%
\bibitem [{\citenamefont {Lukens}\ and\ \citenamefont {Weiner}(2015)}]{Lukens2015}%
  \BibitemOpen
  \bibfield  {author} {\bibinfo {author} {\bibfnamefont {J.~M.}\ \bibnamefont {Lukens}}\ and\ \bibinfo {author} {\bibfnamefont {A.~M.}\ \bibnamefont {Weiner}},\ }\bibfield  {title} {\bibinfo {title} {Biphoton {{Pulse Shaping}}},\ }in\ \href {https://doi.org/10.1007/978-3-319-14992-9_13} {\emph {\bibinfo {booktitle} {All-{{Optical Signal Processing}}: {{Data Communication}} and {{Storage Applications}}}}},\ \bibinfo {series and number} {Springer {{Series}} in {{Optical Sciences}}},\ \bibinfo {editor} {edited by\ \bibinfo {editor} {\bibfnamefont {S.}~\bibnamefont {Wabnitz}}\ and\ \bibinfo {editor} {\bibfnamefont {B.~J.}\ \bibnamefont {Eggleton}}}\ (\bibinfo  {publisher} {{Springer International Publishing}},\ \bibinfo {address} {{Cham}},\ \bibinfo {year} {2015})\ pp.\ \bibinfo {pages} {423--448}\BibitemShut {NoStop}%
\bibitem [{\citenamefont {Graffitti}\ \emph {et~al.}(2020)\citenamefont {Graffitti}, \citenamefont {Barrow}, \citenamefont {Pickston}, \citenamefont {Bra{\'n}czyk},\ and\ \citenamefont {Fedrizzi}}]{Graffitti2020}%
  \BibitemOpen
  \bibfield  {author} {\bibinfo {author} {\bibfnamefont {F.}~\bibnamefont {Graffitti}}, \bibinfo {author} {\bibfnamefont {P.}~\bibnamefont {Barrow}}, \bibinfo {author} {\bibfnamefont {A.}~\bibnamefont {Pickston}}, \bibinfo {author} {\bibfnamefont {A.~M.}\ \bibnamefont {Bra{\'n}czyk}},\ and\ \bibinfo {author} {\bibfnamefont {A.}~\bibnamefont {Fedrizzi}},\ }\bibfield  {title} {\bibinfo {title} {Direct {{Generation}} of {{Tailored Pulse-Mode Entanglement}}},\ }\href {https://doi.org/10.1103/PhysRevLett.124.053603} {\bibfield  {journal} {\bibinfo  {journal} {Phys. Rev. Lett.}\ }\textbf {\bibinfo {volume} {124}},\ \bibinfo {pages} {053603} (\bibinfo {year} {2020})}\BibitemShut {NoStop}%
\bibitem [{\citenamefont {Morrison}\ \emph {et~al.}(2022)\citenamefont {Morrison}, \citenamefont {Graffitti}, \citenamefont {Barrow}, \citenamefont {Pickston}, \citenamefont {Ho},\ and\ \citenamefont {Fedrizzi}}]{Morrison2022}%
  \BibitemOpen
  \bibfield  {author} {\bibinfo {author} {\bibfnamefont {C.~L.}\ \bibnamefont {Morrison}}, \bibinfo {author} {\bibfnamefont {F.}~\bibnamefont {Graffitti}}, \bibinfo {author} {\bibfnamefont {P.}~\bibnamefont {Barrow}}, \bibinfo {author} {\bibfnamefont {A.}~\bibnamefont {Pickston}}, \bibinfo {author} {\bibfnamefont {J.}~\bibnamefont {Ho}},\ and\ \bibinfo {author} {\bibfnamefont {A.}~\bibnamefont {Fedrizzi}},\ }\bibfield  {title} {\bibinfo {title} {Frequency-bin entanglement from domain-engineered down-conversion},\ }\href {https://doi.org/10.1063/5.0089313} {\bibfield  {journal} {\bibinfo  {journal} {APL Photonics}\ }\textbf {\bibinfo {volume} {7}},\ \bibinfo {pages} {066102} (\bibinfo {year} {2022})}\BibitemShut {NoStop}%
\bibitem [{\citenamefont {Kaneda}\ \emph {et~al.}(2019)\citenamefont {Kaneda}, \citenamefont {Suzuki}, \citenamefont {Shimizu},\ and\ \citenamefont {Edamatsu}}]{kaneda2019direct}%
  \BibitemOpen
  \bibfield  {author} {\bibinfo {author} {\bibfnamefont {F.}~\bibnamefont {Kaneda}}, \bibinfo {author} {\bibfnamefont {H.}~\bibnamefont {Suzuki}}, \bibinfo {author} {\bibfnamefont {R.}~\bibnamefont {Shimizu}},\ and\ \bibinfo {author} {\bibfnamefont {K.}~\bibnamefont {Edamatsu}},\ }\bibfield  {title} {\bibinfo {title} {Direct generation of frequency-bin entangled photons via two-period quasi-phase-matched parametric downconversion},\ }\href {https://doi.org/https://doi.org/10.1364/OE.27.001416} {\bibfield  {journal} {\bibinfo  {journal} {Opt. Express}\ }\textbf {\bibinfo {volume} {27}},\ \bibinfo {pages} {1416} (\bibinfo {year} {2019})}\BibitemShut {NoStop}%
\bibitem [{\citenamefont {Christ}\ \emph {et~al.}(2013{\natexlab{b}})\citenamefont {Christ}, \citenamefont {Brecht}, \citenamefont {Mauerer},\ and\ \citenamefont {Silberhorn}}]{cohenSFWM2016}%
  \BibitemOpen
  \bibfield  {author} {\bibinfo {author} {\bibfnamefont {A.}~\bibnamefont {Christ}}, \bibinfo {author} {\bibfnamefont {B.}~\bibnamefont {Brecht}}, \bibinfo {author} {\bibfnamefont {W.}~\bibnamefont {Mauerer}},\ and\ \bibinfo {author} {\bibfnamefont {C.}~\bibnamefont {Silberhorn}},\ }\bibfield  {title} {\bibinfo {title} {Theory of quantum frequency conversion and type-ii parametric down-conversion in the high-gain regime},\ }\href {https://doi.org/https://doi.org/10.1088/1367-2630/15/5/053038} {\bibfield  {journal} {\bibinfo  {journal} {New J. Phys.}\ }\textbf {\bibinfo {volume} {15}},\ \bibinfo {pages} {053038} (\bibinfo {year} {2013}{\natexlab{b}})}\BibitemShut {NoStop}%
\bibitem [{\citenamefont {Li}\ \emph {et~al.}(2004)\citenamefont {Li}, \citenamefont {Chen}, \citenamefont {Voss}, \citenamefont {Sharping},\ and\ \citenamefont {Kumar}}]{li2004all}%
  \BibitemOpen
  \bibfield  {author} {\bibinfo {author} {\bibfnamefont {X.}~\bibnamefont {Li}}, \bibinfo {author} {\bibfnamefont {J.}~\bibnamefont {Chen}}, \bibinfo {author} {\bibfnamefont {P.}~\bibnamefont {Voss}}, \bibinfo {author} {\bibfnamefont {J.}~\bibnamefont {Sharping}},\ and\ \bibinfo {author} {\bibfnamefont {P.}~\bibnamefont {Kumar}},\ }\bibfield  {title} {\bibinfo {title} {All-fiber photon-pair source for quantum communications: Improved generation of correlated photons},\ }\href {https://doi.org/https://doi.org/10.1364/OPEX.12.003737} {\bibfield  {journal} {\bibinfo  {journal} {Opt. Express}\ }\textbf {\bibinfo {volume} {12}},\ \bibinfo {pages} {3737} (\bibinfo {year} {2004})}\BibitemShut {NoStop}%
\bibitem [{\citenamefont {Fulconis}\ \emph {et~al.}(2005)\citenamefont {Fulconis}, \citenamefont {Alibart}, \citenamefont {Wadsworth}, \citenamefont {Russell},\ and\ \citenamefont {Rarity}}]{fulconis2005high}%
  \BibitemOpen
  \bibfield  {author} {\bibinfo {author} {\bibfnamefont {J.}~\bibnamefont {Fulconis}}, \bibinfo {author} {\bibfnamefont {O.}~\bibnamefont {Alibart}}, \bibinfo {author} {\bibfnamefont {W.}~\bibnamefont {Wadsworth}}, \bibinfo {author} {\bibfnamefont {P.~S.~J.}\ \bibnamefont {Russell}},\ and\ \bibinfo {author} {\bibfnamefont {J.}~\bibnamefont {Rarity}},\ }\bibfield  {title} {\bibinfo {title} {High brightness single mode source of correlated photon pairs using a photonic crystal fiber},\ }\href {https://doi.org/https://doi.org/10.1364/OPEX.13.007572} {\bibfield  {journal} {\bibinfo  {journal} {Opt. Express}\ }\textbf {\bibinfo {volume} {13}},\ \bibinfo {pages} {7572} (\bibinfo {year} {2005})}\BibitemShut {NoStop}%
\bibitem [{\citenamefont {Chen}\ \emph {et~al.}(2005)\citenamefont {Chen}, \citenamefont {Li},\ and\ \citenamefont {Kumar}}]{chen2005two}%
  \BibitemOpen
  \bibfield  {author} {\bibinfo {author} {\bibfnamefont {J.}~\bibnamefont {Chen}}, \bibinfo {author} {\bibfnamefont {X.}~\bibnamefont {Li}},\ and\ \bibinfo {author} {\bibfnamefont {P.}~\bibnamefont {Kumar}},\ }\bibfield  {title} {\bibinfo {title} {Two-photon-state generation via four-wave mixing in optical fibers},\ }\href {https://doi.org/https://doi.org/10.1103/PhysRevA.72.033801} {\bibfield  {journal} {\bibinfo  {journal} {Phys. Rev. A}\ }\textbf {\bibinfo {volume} {72}},\ \bibinfo {pages} {033801} (\bibinfo {year} {2005})}\BibitemShut {NoStop}%
\bibitem [{\citenamefont {Sharping}\ \emph {et~al.}(2006)\citenamefont {Sharping}, \citenamefont {Lee}, \citenamefont {Foster}, \citenamefont {Turner}, \citenamefont {Schmidt}, \citenamefont {Lipson}, \citenamefont {Gaeta},\ and\ \citenamefont {Kumar}}]{sharping2006generation}%
  \BibitemOpen
  \bibfield  {author} {\bibinfo {author} {\bibfnamefont {J.~E.}\ \bibnamefont {Sharping}}, \bibinfo {author} {\bibfnamefont {K.~F.}\ \bibnamefont {Lee}}, \bibinfo {author} {\bibfnamefont {M.~A.}\ \bibnamefont {Foster}}, \bibinfo {author} {\bibfnamefont {A.~C.}\ \bibnamefont {Turner}}, \bibinfo {author} {\bibfnamefont {B.~S.}\ \bibnamefont {Schmidt}}, \bibinfo {author} {\bibfnamefont {M.}~\bibnamefont {Lipson}}, \bibinfo {author} {\bibfnamefont {A.~L.}\ \bibnamefont {Gaeta}},\ and\ \bibinfo {author} {\bibfnamefont {P.}~\bibnamefont {Kumar}},\ }\bibfield  {title} {\bibinfo {title} {Generation of correlated photons in nanoscale silicon waveguides},\ }\href {https://doi.org/https://doi.org/10.1364/OE.14.012388} {\bibfield  {journal} {\bibinfo  {journal} {Opt. Express}\ }\textbf {\bibinfo {volume} {14}},\ \bibinfo {pages} {12388} (\bibinfo {year} {2006})}\BibitemShut {NoStop}%
\bibitem [{\citenamefont {Harada}\ \emph {et~al.}(2008)\citenamefont {Harada}, \citenamefont {Takesue}, \citenamefont {Fukuda}, \citenamefont {Tsuchizawa}, \citenamefont {Watanabe}, \citenamefont {Yamada}, \citenamefont {Tokura},\ and\ \citenamefont {Itabashi}}]{harada2008generation}%
  \BibitemOpen
  \bibfield  {author} {\bibinfo {author} {\bibfnamefont {K.-i.}\ \bibnamefont {Harada}}, \bibinfo {author} {\bibfnamefont {H.}~\bibnamefont {Takesue}}, \bibinfo {author} {\bibfnamefont {H.}~\bibnamefont {Fukuda}}, \bibinfo {author} {\bibfnamefont {T.}~\bibnamefont {Tsuchizawa}}, \bibinfo {author} {\bibfnamefont {T.}~\bibnamefont {Watanabe}}, \bibinfo {author} {\bibfnamefont {K.}~\bibnamefont {Yamada}}, \bibinfo {author} {\bibfnamefont {Y.}~\bibnamefont {Tokura}},\ and\ \bibinfo {author} {\bibfnamefont {S.-i.}\ \bibnamefont {Itabashi}},\ }\bibfield  {title} {\bibinfo {title} {Generation of high-purity entangled photon pairs using silicon wire waveguide},\ }\href {https://doi.org/https://doi.org/10.1364/OE.16.020368} {\bibfield  {journal} {\bibinfo  {journal} {Opt. Express}\ }\textbf {\bibinfo {volume} {16}},\ \bibinfo {pages} {20368} (\bibinfo {year} {2008})}\BibitemShut {NoStop}%
\bibitem [{\citenamefont {Kues}\ \emph {et~al.}(2017)\citenamefont {Kues}, \citenamefont {Reimer}, \citenamefont {Roztocki}, \citenamefont {Cort{\'e}s}, \citenamefont {Sciara}, \citenamefont {Wetzel}, \citenamefont {Zhang}, \citenamefont {Cino}, \citenamefont {Chu}, \citenamefont {Little} \emph {et~al.}}]{kues2017chip}%
  \BibitemOpen
  \bibfield  {author} {\bibinfo {author} {\bibfnamefont {M.}~\bibnamefont {Kues}}, \bibinfo {author} {\bibfnamefont {C.}~\bibnamefont {Reimer}}, \bibinfo {author} {\bibfnamefont {P.}~\bibnamefont {Roztocki}}, \bibinfo {author} {\bibfnamefont {L.~R.}\ \bibnamefont {Cort{\'e}s}}, \bibinfo {author} {\bibfnamefont {S.}~\bibnamefont {Sciara}}, \bibinfo {author} {\bibfnamefont {B.}~\bibnamefont {Wetzel}}, \bibinfo {author} {\bibfnamefont {Y.}~\bibnamefont {Zhang}}, \bibinfo {author} {\bibfnamefont {A.}~\bibnamefont {Cino}}, \bibinfo {author} {\bibfnamefont {S.~T.}\ \bibnamefont {Chu}}, \bibinfo {author} {\bibfnamefont {B.~E.}\ \bibnamefont {Little}}, \emph {et~al.},\ }\bibfield  {title} {\bibinfo {title} {On-chip generation of high-dimensional entangled quantum states and their coherent control},\ }\href {https://doi.org/https://doi.org/10.1038/nature22986} {\bibfield  {journal} {\bibinfo  {journal} {Nature}\ }\textbf {\bibinfo {volume} {546}},\ \bibinfo {pages} {622} (\bibinfo {year} {2017})}\BibitemShut {NoStop}%
\bibitem [{\citenamefont {Chen}\ \emph {et~al.}(2011)\citenamefont {Chen}, \citenamefont {Levine}, \citenamefont {Fan},\ and\ \citenamefont {Migdall}}]{chen2011frequency}%
  \BibitemOpen
  \bibfield  {author} {\bibinfo {author} {\bibfnamefont {J.}~\bibnamefont {Chen}}, \bibinfo {author} {\bibfnamefont {Z.~H.}\ \bibnamefont {Levine}}, \bibinfo {author} {\bibfnamefont {J.}~\bibnamefont {Fan}},\ and\ \bibinfo {author} {\bibfnamefont {A.~L.}\ \bibnamefont {Migdall}},\ }\bibfield  {title} {\bibinfo {title} {Frequency-bin entangled comb of photon pairs from a silicon-on-insulator micro-resonator},\ }\href {https://doi.org/https://doi.org/10.1364/OE.19.001470} {\bibfield  {journal} {\bibinfo  {journal} {Opt. Express}\ }\textbf {\bibinfo {volume} {19}},\ \bibinfo {pages} {1470} (\bibinfo {year} {2011})}\BibitemShut {NoStop}%
\bibitem [{\citenamefont {Clemmen}\ \emph {et~al.}(2009)\citenamefont {Clemmen}, \citenamefont {Huy}, \citenamefont {Bogaerts}, \citenamefont {Baets}, \citenamefont {Emplit},\ and\ \citenamefont {Massar}}]{clemmen2009continuous}%
  \BibitemOpen
  \bibfield  {author} {\bibinfo {author} {\bibfnamefont {S.}~\bibnamefont {Clemmen}}, \bibinfo {author} {\bibfnamefont {K.~P.}\ \bibnamefont {Huy}}, \bibinfo {author} {\bibfnamefont {W.}~\bibnamefont {Bogaerts}}, \bibinfo {author} {\bibfnamefont {R.~G.}\ \bibnamefont {Baets}}, \bibinfo {author} {\bibfnamefont {P.}~\bibnamefont {Emplit}},\ and\ \bibinfo {author} {\bibfnamefont {S.}~\bibnamefont {Massar}},\ }\bibfield  {title} {\bibinfo {title} {Continuous wave photon pair generation in silicon-on-insulator waveguides and ring resonators},\ }\href {https://doi.org/https://doi.org/10.1364/OE.17.016558} {\bibfield  {journal} {\bibinfo  {journal} {Opt. Express}\ }\textbf {\bibinfo {volume} {17}},\ \bibinfo {pages} {16558} (\bibinfo {year} {2009})}\BibitemShut {NoStop}%
\bibitem [{\citenamefont {Simon}\ and\ \citenamefont {Poizat}(2005)}]{simon2005creating}%
  \BibitemOpen
  \bibfield  {author} {\bibinfo {author} {\bibfnamefont {C.}~\bibnamefont {Simon}}\ and\ \bibinfo {author} {\bibfnamefont {J.-P.}\ \bibnamefont {Poizat}},\ }\bibfield  {title} {\bibinfo {title} {Creating single time-bin-entangled photon pairs},\ }\href {https://doi.org/https://doi.org/10.1103/PhysRevLett.94.030502} {\bibfield  {journal} {\bibinfo  {journal} {Phys. Rev. Lett.}\ }\textbf {\bibinfo {volume} {94}},\ \bibinfo {pages} {030502} (\bibinfo {year} {2005})}\BibitemShut {NoStop}%
\bibitem [{\citenamefont {Jayakumar}\ \emph {et~al.}(2014)\citenamefont {Jayakumar}, \citenamefont {Predojevi{\'c}}, \citenamefont {Kauten}, \citenamefont {Huber}, \citenamefont {Solomon},\ and\ \citenamefont {Weihs}}]{jayakumar2014time}%
  \BibitemOpen
  \bibfield  {author} {\bibinfo {author} {\bibfnamefont {H.}~\bibnamefont {Jayakumar}}, \bibinfo {author} {\bibfnamefont {A.}~\bibnamefont {Predojevi{\'c}}}, \bibinfo {author} {\bibfnamefont {T.}~\bibnamefont {Kauten}}, \bibinfo {author} {\bibfnamefont {T.}~\bibnamefont {Huber}}, \bibinfo {author} {\bibfnamefont {G.~S.}\ \bibnamefont {Solomon}},\ and\ \bibinfo {author} {\bibfnamefont {G.}~\bibnamefont {Weihs}},\ }\bibfield  {title} {\bibinfo {title} {Time-bin entangled photons from a quantum dot},\ }\href {https://doi.org/https://doi.org/10.1038/ncomms5251} {\bibfield  {journal} {\bibinfo  {journal} {Nat. Commun.}\ }\textbf {\bibinfo {volume} {5}},\ \bibinfo {pages} {1} (\bibinfo {year} {2014})}\BibitemShut {NoStop}%
\bibitem [{\citenamefont {Carnio}\ \emph {et~al.}(2021)\citenamefont {Carnio}, \citenamefont {Buchleitner},\ and\ \citenamefont {Schlawin}}]{Carnio2021}%
  \BibitemOpen
  \bibfield  {author} {\bibinfo {author} {\bibfnamefont {E.}~\bibnamefont {Carnio}}, \bibinfo {author} {\bibfnamefont {A.}~\bibnamefont {Buchleitner}},\ and\ \bibinfo {author} {\bibfnamefont {F.}~\bibnamefont {Schlawin}},\ }\bibfield  {title} {\bibinfo {title} {How to optimize the absorption of two entangled photons},\ }\href {https://doi.org/10.21468/SciPostPhysCore.4.4.028} {\bibfield  {journal} {\bibinfo  {journal} {SciPost Phys. Core}\ }\textbf {\bibinfo {volume} {4}},\ \bibinfo {pages} {028} (\bibinfo {year} {2021})}\BibitemShut {NoStop}%
\bibitem [{\citenamefont {Arzani}\ \emph {et~al.}(2018)\citenamefont {Arzani}, \citenamefont {Fabre},\ and\ \citenamefont {Treps}}]{Arzani2018}%
  \BibitemOpen
  \bibfield  {author} {\bibinfo {author} {\bibfnamefont {F.}~\bibnamefont {Arzani}}, \bibinfo {author} {\bibfnamefont {C.}~\bibnamefont {Fabre}},\ and\ \bibinfo {author} {\bibfnamefont {N.}~\bibnamefont {Treps}},\ }\bibfield  {title} {\bibinfo {title} {Versatile engineering of multimode squeezed states by optimizing the pump spectral profile in spontaneous parametric down-conversion},\ }\href {https://doi.org/10.1103/PhysRevA.97.033808} {\bibfield  {journal} {\bibinfo  {journal} {Phys. Rev. A}\ }\textbf {\bibinfo {volume} {97}},\ \bibinfo {pages} {033808} (\bibinfo {year} {2018})}\BibitemShut {NoStop}%
\bibitem [{\citenamefont {Parker}\ \emph {et~al.}(2000)\citenamefont {Parker}, \citenamefont {Bose},\ and\ \citenamefont {Plenio}}]{Parker2000}%
  \BibitemOpen
  \bibfield  {author} {\bibinfo {author} {\bibfnamefont {S.}~\bibnamefont {Parker}}, \bibinfo {author} {\bibfnamefont {S.}~\bibnamefont {Bose}},\ and\ \bibinfo {author} {\bibfnamefont {M.~B.}\ \bibnamefont {Plenio}},\ }\bibfield  {title} {\bibinfo {title} {Entanglement quantification and purification in continuous-variable systems},\ }\href {https://doi.org/10.1103/PhysRevA.61.032305} {\bibfield  {journal} {\bibinfo  {journal} {Phys. Rev. A}\ }\textbf {\bibinfo {volume} {61}},\ \bibinfo {pages} {032305} (\bibinfo {year} {2000})}\BibitemShut {NoStop}%
\bibitem [{\citenamefont {Lamata}\ and\ \citenamefont {Le{\'o}n}(2005)}]{Lamata2005}%
  \BibitemOpen
  \bibfield  {author} {\bibinfo {author} {\bibfnamefont {L.}~\bibnamefont {Lamata}}\ and\ \bibinfo {author} {\bibfnamefont {J.}~\bibnamefont {Le{\'o}n}},\ }\bibfield  {title} {\bibinfo {title} {Dealing with entanglement of continuous variables: {{Schmidt}} decomposition with discrete sets of orthogonal functions},\ }\href {https://doi.org/10.1088/1464-4266/7/8/004} {\bibfield  {journal} {\bibinfo  {journal} {J. Opt. B: Quantum Semiclass. Opt.}\ }\textbf {\bibinfo {volume} {7}},\ \bibinfo {pages} {224} (\bibinfo {year} {2005})}\BibitemShut {NoStop}%
\bibitem [{\citenamefont {Law}\ \emph {et~al.}(2000)\citenamefont {Law}, \citenamefont {Walmsley},\ and\ \citenamefont {Eberly}}]{law2000}%
  \BibitemOpen
  \bibfield  {author} {\bibinfo {author} {\bibfnamefont {C.~K.}\ \bibnamefont {Law}}, \bibinfo {author} {\bibfnamefont {I.~A.}\ \bibnamefont {Walmsley}},\ and\ \bibinfo {author} {\bibfnamefont {J.~H.}\ \bibnamefont {Eberly}},\ }\bibfield  {title} {\bibinfo {title} {Continuous {{Frequency Entanglement}}: {{Effective Finite Hilbert Space}} and {{Entropy Control}}},\ }\href {https://doi.org/10.1103/PhysRevLett.84.5304} {\bibfield  {journal} {\bibinfo  {journal} {Phys. Rev. Lett.}\ }\textbf {\bibinfo {volume} {84}},\ \bibinfo {pages} {5304} (\bibinfo {year} {2000})}\BibitemShut {NoStop}%
\bibitem [{\citenamefont {Steck}(2019)}]{Steck2019}%
  \BibitemOpen
  \bibfield  {author} {\bibinfo {author} {\bibfnamefont {D.~A.}\ \bibnamefont {Steck}},\ }\href {https://steck.us/alkalidata/} {\bibinfo {title} {Sodium {{D Line Data}}}} (\bibinfo {year} {2019})\BibitemShut {NoStop}%
\bibitem [{\citenamefont {Szczykulska}\ \emph {et~al.}(2016)\citenamefont {Szczykulska}, \citenamefont {Baumgratz},\ and\ \citenamefont {Datta}}]{Szczykulska2016}%
  \BibitemOpen
  \bibfield  {author} {\bibinfo {author} {\bibfnamefont {M.}~\bibnamefont {Szczykulska}}, \bibinfo {author} {\bibfnamefont {T.}~\bibnamefont {Baumgratz}},\ and\ \bibinfo {author} {\bibfnamefont {A.}~\bibnamefont {Datta}},\ }\bibfield  {title} {\bibinfo {title} {Multi-parameter quantum metrology},\ }\href {https://doi.org/10.1080/23746149.2016.1230476} {\bibfield  {journal} {\bibinfo  {journal} {Adv. Phys. X}\ }\textbf {\bibinfo {volume} {1}},\ \bibinfo {pages} {621} (\bibinfo {year} {2016})}\BibitemShut {NoStop}%
\bibitem [{\citenamefont {Albarelli}\ \emph {et~al.}(2020)\citenamefont {Albarelli}, \citenamefont {Barbieri}, \citenamefont {Genoni},\ and\ \citenamefont {Gianani}}]{Albarelli2019c}%
  \BibitemOpen
  \bibfield  {author} {\bibinfo {author} {\bibfnamefont {F.}~\bibnamefont {Albarelli}}, \bibinfo {author} {\bibfnamefont {M.}~\bibnamefont {Barbieri}}, \bibinfo {author} {\bibfnamefont {M.~G.}\ \bibnamefont {Genoni}},\ and\ \bibinfo {author} {\bibfnamefont {I.}~\bibnamefont {Gianani}},\ }\bibfield  {title} {\bibinfo {title} {A perspective on multiparameter quantum metrology: {{From}} theoretical tools to applications in quantum imaging},\ }\href {https://doi.org/10.1016/j.physleta.2020.126311} {\bibfield  {journal} {\bibinfo  {journal} {Phys. Lett. A}\ }\textbf {\bibinfo {volume} {384}},\ \bibinfo {pages} {126311} (\bibinfo {year} {2020})}\BibitemShut {NoStop}%
\bibitem [{\citenamefont {Tsang}\ \emph {et~al.}(2020)\citenamefont {Tsang}, \citenamefont {Albarelli},\ and\ \citenamefont {Datta}}]{Tsang2019}%
  \BibitemOpen
  \bibfield  {author} {\bibinfo {author} {\bibfnamefont {M.}~\bibnamefont {Tsang}}, \bibinfo {author} {\bibfnamefont {F.}~\bibnamefont {Albarelli}},\ and\ \bibinfo {author} {\bibfnamefont {A.}~\bibnamefont {Datta}},\ }\bibfield  {title} {\bibinfo {title} {Quantum {{Semiparametric Estimation}}},\ }\href {https://doi.org/10.1103/PhysRevX.10.031023} {\bibfield  {journal} {\bibinfo  {journal} {Phys. Rev. X}\ }\textbf {\bibinfo {volume} {10}},\ \bibinfo {pages} {031023} (\bibinfo {year} {2020})}\BibitemShut {NoStop}%
\bibitem [{\citenamefont {Abramowitz}\ and\ \citenamefont {Stegun}(1964)}]{abramowitz1964handbook}%
  \BibitemOpen
  \bibfield  {author} {\bibinfo {author} {\bibfnamefont {M.}~\bibnamefont {Abramowitz}}\ and\ \bibinfo {author} {\bibfnamefont {I.~A.}\ \bibnamefont {Stegun}},\ }\href@noop {} {\emph {\bibinfo {title} {Handbook of mathematical functions with formulas, graphs, and mathematical tables}}},\ Vol.~\bibinfo {volume} {55}\ (\bibinfo  {publisher} {US Government printing office},\ \bibinfo {year} {1964})\BibitemShut {NoStop}%
\end{thebibliography}%


\onecolumngrid
\appendix
\section{Equivalence between spontaneous emission into many modes or a single mode for the reduced dynamics}\label{app:effectiveGammmaPerp}
In this Appendix we show that a photonic environment composed by an infinity of (initially empty) field modes that are distinct from the travelling pulse, i.e. different spatial and polarization degrees of freedom, can be effectively described as a single collective bosonic mode that interacts with the atom (the E subsystem in the main text), as mentioned after Eq.~\eqref{eq:Hint_PE} in Sec.~\ref{subsec:model}.

A two-level atom in free space can be described as interacting with a discrete set of infinitely many modes of the electromagnetic field.
With the usual dipole and Markovian approximations (explained in the main text in Sec.~\ref{subsec:model}) the interaction-picture Hamiltonian is
\begin{equation}
    \begin{split}
	H_{\mathrm{I}}(t) =& -\I \sqrt{\Gamma_{\mathrm{tot}} \eta_{P}} \left( \sigmin a(t)^\dag - \sigplus a^\dag(t)  \right) \\
    & -\I  \sum_j \sqrt{\Gamma_{\mathrm{tot}} \eta_j } \left( \sigmin a_j(t) - \sigplus a_j^\dag(t) \right). 
    \end{split}
\end{equation}
In this expression $\Gamma_{\mathrm{tot}}$ is the standard Wigner-Weisskopf spontaneous-emission rate in free space, which could be suitably modified to model emission of radiation in a different propagating medium, while the parameters $\eta_l > 0 $ are geometric factors that determine the coupling of the atom with the mode $l$, see for instance Ref.~\cite{Ko2022} for a more in-depth discussion.
In particular, we have separated the term corresponding to the interaction with the travelling pulse mode, which we assume to be the only experimentally accessible one.
The others modes are initially in the vacuum and we treat them as an inaccessible, i.e. environmental degrees of freedom.
For this reason, it is more convenient to treat them as a single collective mode, defined as
\begin{equation}
	b(t) =  \sum_j \sqrt{\frac{ \eta_j }{\sum_{j'} \eta_{j'} }} a_j(t),
\end{equation}
and satisfying $[b(t),b^\dag(t')]=\delta(t-t')$ so that we can rewrite the Hamiltonian~\eqref{eq:Hint_PE} used in the main text with $ \Gamma = \Gamma_{\mathrm{tot}} \eta_{P}$ and $\Gperp = \Gamma_{\mathrm{tot}} \sum_{j} \eta_{j} $.

\section{Explicit check of single photon states normalization}\label{app:normalization1ph}

In this Appendix we show explicitly that the state given by Eqs. \eqref{eq:1ph_pulse_psie}, \eqref{eq:1ph_pulse_unnorm_state} and \eqref{eq:1ph_env_state} of Sec.~\ref{subsubsec:1photonEvo} is normalized to unity.
We only consider the case $\Gperp = 0$ for simplicity.

The amplitude of the excited atomic state is
\begin{equation}
\psi_e(t) = - \sqrt{ \Gamma} \int_{-\infty}^t dt' e^{-\frac{\Gamma}{2}(t-t')} \xi(t')
\end{equation}
and the corresponding probability is
\begin{equation}
    | \psi_e(t) | ^2 =  \Gamma \left\vert \int_{-\infty}^t dt' e^{-\frac{\Gamma}{2}(t-t')} \xi(t') \right\vert^2 = \Gamma  e^{-\Gamma t} \left\vert \int_{-\infty}^t dt' e^{\frac{\Gamma}{2}t'} \xi(t') \right\vert^2.
\end{equation}
The pulse component is
\begin{equation}
    \ket*{\widetilde{\psi}_{g}^\mathrm{P}(t)} = \int_{-\infty}^\infty d\tau \left( \xi(\tau) + \sqrt{\Gamma} \Theta(t-\tau) \psi_e(\tau) \right) a^\dag(\tau) \ket{0^\mathrm{P}}
\end{equation}
with modulus squared
\begin{equation}
    \label{eq:psigpsig}
    \begin{split}
    \braket*{\widetilde{\psi}^\mathrm{P}_{g}(t)}{\widetilde{\psi}^\mathrm{P}_{g}(t)} &= \int_{-\infty}^\infty d\tau \left\vert \xi(\tau) + \sqrt{\Gamma} \Theta(t-\tau) \psi_e(\tau) \right\vert^2 \\
    &= \int_{-\infty}^\infty d\tau | \xi(\tau) |^2 + \int_{-\infty}^t d\tau  \left( \Gamma | \psi_e(\tau) |^2 + 2 \sqrt{\Gamma} \Re \left[ \psi_e(\tau)\xi^*(\tau) \right]  \right) \\
    &=1 + \Gamma^2 \int_{-\infty}^t d\tau e^{-\Gamma \tau} \left\vert \int_{-\infty}^\tau dt' e^{\frac{\Gamma t'}{2}} \xi(t') \right\vert^2 - 2 \Gamma \int_{-\infty}^t d\tau e^{-\frac{\Gamma \tau}{2}} \Re \left[ \int_{-\infty}^\tau dt' e^{\frac{\Gamma t'}{2}} \xi(t')\xi^*(\tau)\right].
    \end{split}
\end{equation}
Since we are assuming that the atom is initially in the ground state, i.e. $| \psi_e(-\infty) | ^2 = 0$ we can rewrite the excitation probability as
\begin{equation}
    \begin{split}
        | \psi_e(t) | ^2 = & \Gamma e^{-\Gamma t } \left\vert \int_{-\infty}^t dt' e^{ \frac{\Gamma}{2} t'}    \xi(t')  \right\vert^2 =  \Gamma \int_{-\infty}^t d\tau \frac{d}{d\tau} \left[ e^{-\Gamma \tau }\left\vert \int_{-\infty}^\tau dt' e^{ \frac{\Gamma}{2} t'}    \xi(t')  \right\vert^2  \right], 
    \end{split}
\end{equation}
explicitly computing the derivative inside the integral one can recognize the last two terms of Eq.~\eqref{eq:psigpsig} with an overall opposite sign and verify that $| \psi_e(t) | ^2 + \braket*{\widetilde{\psi}^\mathrm{P}_{g}(t)}{\widetilde{\psi}^\mathrm{P}_{g}(t)}  =1$.
The reasoning when $\Gperp > 0$ is analogous.

\section{Single-photon QFI for real-valued wavepackets}\label{app:1photonQFI}

For $\xi(t) \in \mathbb{R}$ all the temporal amplitudes remain real and we can rewrite the QFI in terms of the unnormalized state $\ket*{ \widetilde{\psi}_\Gamma} = \sqrt{1-p_\Gamma} \ket*{\psi_\Gamma}  $, satisfying $\braket*{\widetilde{\psi}_\Gamma}{\widetilde{\psi}_\Gamma} = 1-p_\Gamma$ and $\ket{\partial_\Gamma \psi_\Gamma} =  \frac{1}{\sqrt{1-p_\Gamma}} \ket*{\partial_\Gamma \widetilde{\psi}_\Gamma} + \frac{\partial_\Gamma p_\Gamma}{2(1-p_\Gamma)^{3/2}} \ket*{\widetilde{\psi}_\Gamma}$ and substituting this expression in the second term of Eq.~\eqref{eq:QFI_1ph_normalized} we obtain an alternative expression for the QFI
\begin{equation}
    \label{eq:QFI_single_photon_mixed_unnorm}
    \mathcal{Q}(\rho_{\Gamma}) = (\partial_\Gamma p_\Gamma)^2/p_\Gamma + 4 \braket*{\partial_{\Gamma} \widetilde{\psi}_{\Gamma}}{\partial_\Gamma \widetilde{\psi}_{\Gamma}}.
\end{equation}
This is because
\begin{equation}
    \braket*{\partial_{\Gamma} \widetilde{\psi}_{\Gamma}}{\partial_\Gamma \widetilde{\psi}_{\Gamma}} 
    = (1-p_\Gamma) \braket{\partial_{\Gamma} \psi_{\Gamma}}{\partial_\Gamma \psi_{\Gamma}} + \frac{ (\partial_\Gamma p_\Gamma)^2 }{4 (1-p_\Gamma) }.
\end{equation}
The form in Eq.~\eqref{eq:QFI_single_photon_mixed_unnorm} is particularly convenient, since we can immediately use the unnormalized state in Eq.~\eqref{eq:1ph_pulse_unnorm_state} without renormalizing it first.
These identities hold because we have $\braket{\psi}{\partial_\Gamma \psi} = 0$, in accordance with Eq.~\eqref{eq:rank2ortho}, where the second terms in the summation vanish.

The terms appearing in~\eqref{eq:QFI_single_photon_mixed_unnorm} can be evaluated more explicitly as follows (denoting $\left\Vert v \right\Vert^2 =  
\braket{v}{v} $)
\begin{align}
    p_\Gamma(t) &= \psi_e(t)^2 + \left \Vert \psi_{g}^\mathrm{E}(t) \right\Vert^2 \\
    \psi_e(t)^2 & =  \Gamma \left(\int_{-\infty}^t dt' e^{-\frac{\Gamma + \Gperp}{2}(t-t')} \xi(t') \right)^2 \\
    \left \Vert \widetilde{\psi}_{g}^\mathrm{E}(t) \right\Vert^2 & = \Gperp \Gamma \int_{-\infty}^{t} d\tau \left( \int_{-\infty}^\tau dt' e^{-\frac{\Gamma + \Gperp}{2}(\tau-t')} \xi(t') \right)^2 \\
    \partial_\Gamma p_\Gamma(t) & = \left(\int_{-\infty}^t dt' e^{-\frac{\Gamma + \Gperp}{2}(t-t')} \xi(t') \right)^2 + 2 \Gamma \left(\int_{-\infty}^t dt' e^{-\frac{\Gamma + \Gperp}{2}(t-t')} \xi(t') \right) \left(\int_{-\infty}^t dt' \frac{(t'-t)}{2}e^{-\frac{\Gamma + \Gperp}{2}(t-t')} \xi(t') \right) \nonumber \\
    &+ \Gperp \int_{-\infty}^{t} d\tau \left( \int_{-\infty}^\tau dt' e^{-\frac{\Gamma + \Gperp}{2}(\tau-t')} \xi(t') \right)^2 \nonumber \\
    &+ 2 \Gperp \Gamma \int_{-\infty}^{t} d\tau \left( \int_{-\infty}^\tau dt' e^{-\frac{\Gamma + \Gperp}{2}(\tau-t')} \xi(t') \right)\left( \int_{-\infty}^\tau dt' \frac{t'-\tau}{2} e^{-\frac{\Gamma + \Gperp}{2}(\tau-t')} \xi(t') \right)  \\ 
    \left\Vert \partial_\Gamma \widetilde{\psi}_{g}^\mathrm{P} \right\Vert^2  &
    = \int_{-\infty}^t  d \tau \left( \int_{-\infty}^\tau dt' e^{-\frac{\Gamma + \Gperp}{2}(\tau-t')} \xi(t') \right)^2 + \Gamma^2 \int_{-\infty}^t  d \tau \left( \int_{-\infty}^\tau dt'\frac{(t'-\tau)}{2} e^{-\frac{\Gamma + \Gperp}{2}(\tau-t')} \xi(t') \right)^2 \nonumber \\ 
    & + 2 \Gamma \int_{-\infty}^t d \tau \left( \int_{-\infty}^\tau dt' e^{-\frac{\Gamma + \Gperp}{2}(\tau-t')} \xi(t') \right)\left( \int_{-\infty}^\tau dt'\frac{(t'-\tau)}{2} e^{-\frac{\Gamma + \Gperp}{2}(\tau-t')} \xi(t') \right).
\end{align}
We also report the probability of the asymptotic single-photon component to be in the pulse temporal mode
\begin{align}
    p_{\mathrm{orig}}(t) & = \braket*{\widetilde{\psi}_g^\mathrm{P}(t)}{\xi}^2 \\
    \braket*{ \widetilde{\psi}_g^\mathrm{P}(t)}{\xi} & = 1 + \sqrt{\Gamma} \int_{-\infty}^t d\tau \psi_e(\tau) \xi(\tau) = 1-\Gamma \int_{-\infty}^t d\tau \left( e^{-\frac{(\Gamma+\Gperp)\tau}{2}} \xi(\tau) \int_{-\infty}^\tau dt' e^{\frac{(\Gamma+\Gperp)t'}{2}} \xi(t') \right).
\end{align}

\section{Single-photon wavepackets details}
\label{app:1phShapes}

In Table~\ref{tab:SinglePhotonShapes} we report details for all the pulse shapes mentioned in Fig.~\ref{fig:QFIshapes} of Sec.~\ref{subsec:1phPerfCouplRes}, including their definition, the excitation probability, the (dimensionless) QFI and the (dimensionless) CFI corresponding to a measurement in the original temporal mode; as mentioned in the main text the only relevant parameter for these quantities is the dimensionless product $\Gamma T$.
For the Gaussian pulse analytical expressions for the QFI and CFI are not available.
The arrival time of all the pulses corresponds to their peak, except for the rectangular pulse for which it corresponds to the beginning of the region with a nonzero photon density.
\begin{table}[h!]
    \caption{\label{tab:SinglePhotonShapes} Pulse shapes used in the main text. The arrival of all the pulses is at $t=0$. $\Theta$ is the Heaviside step function.}
    \begin{ruledtabular}
        \begin{tabular}{cccccc}
            Shape & $\xi(t)$ & $T_\sigma$  &  $p_e(t)$ & $ \Gamma^2 \mathcal{Q}(\rho_\Gamma^\infty)$ & $\Gamma^2 \mathcal{C}(p_\mathrm{orig})$ \\
            \hline
            Rectangular & $\frac{\Theta (t) \Theta (T -t)}{\sqrt{T }}$ & $ \frac{T}{\sqrt{12}}$ & $\begin{cases} 0 \qquad & t \leq 0 \\ \frac{4 e^{-\Gamma t } \left(e^{\frac{\Gamma  t}{2}}-1\right)^2}{\Gamma  T} \quad &0 < t < T \\ \frac{4 e^{-\Gamma t } \left(e^{\frac{\Gamma  T}{2}}-1\right)^2}{\Gamma  T} \quad  & t \geq T  \end{cases}$ & $\frac{8 \left(2-e^{-\frac{\Gamma  T}{2} } (\Gamma  T+2)\right)}{\Gamma T} $ & $\frac{2 \left(\Gamma  T-2 e^{\frac{\Gamma  T}{2}}+2\right)^2}{\left(e^{\frac{\Gamma  T}{2}}-1\right) \left(e^{\frac{\Gamma  T}{2}} (\Gamma  T-2)+2\right)}$ \\
            \hline
            Rising Exp &  $ \frac{1}{\sqrt{T}} e^{\frac{t}{2 T} } \Theta(-t)$ & $T$ & $\begin{cases}  \frac{4 \Gamma  T e^{ t / T }}{(\Gamma  T+1)^2} \quad &t\leq 0 \\ \frac{4 \Gamma  T e^{-\Gamma  t}}{(\Gamma  T+1)^2}  \quad &t > 0 \end{cases}$ & $\frac{8 \Gamma T}{  (\Gamma  T+1)^2} $ & $\frac{4 \Gamma T}{ (\Gamma  T+1)^2}$ \\
            \hline
            Decaying Exp & $\frac{1}{\sqrt{T}} e^{-\frac{t}{2T} } \Theta(t)$ & $T$ & $\frac{4 \Gamma  T e^{- \Gamma t}}{(\Gamma  T-1)^2}\left(e^{\frac{t (\Gamma  T-1)}{2 T}}-1\right)^2 \Theta (t)$ & $\frac{8 \Gamma T}{  (\Gamma  T+1)^2}$ & $\frac{4 \Gamma T}{  (\Gamma  T+1)^2}$ \\
            \hline
            Symmetric Exp & $\frac{1}{\sqrt{T} } e^{- \frac{|t|}{T} }$ & $\frac{T}{\sqrt{2}}$ & $ \begin{cases}  \frac{4 \Gamma  T e^{\frac{2 t}{T}}}{(\Gamma  T+2)^2} \quad &t\leq 0\\ 
                \frac{4 \Gamma  T e^{-\Gamma t}}{\left[ (\Gamma T)^2-4\right]^2} \left[(\Gamma  T+2) e^{\frac{1}{2} t \left(\Gamma -\frac{2}{T}\right)}-4\right]^2 \quad &t > 0 \end{cases}$  & $\frac{64 \Gamma T}{ (\Gamma  T+2)^3}$  & $\frac{64 \Gamma  T}{(\Gamma  T+2)^2 (\Gamma  T+4)}$ \\
            \hline
            Gaussian & $ \frac{1}{\sqrt{T} (2\pi)^{1/4}} e^{-\frac{t^2}{4 T^2} } $ & $T$ & $\sqrt{\frac{\pi }{2}} \Gamma  T e^{\frac{ (\Gamma T) ^2-2 \Gamma t}{2}} \left[\text{erf}\left(\frac{t}{2 T} - \frac{\Gamma T}{2} \right)+1\right]^2$ & n.a. & n.a. \\
        \end{tabular}
    \end{ruledtabular}
\end{table}

\section{Numerical evidence for the validity of the time-dependent Jaynes-Cummings model in the short-time regime}
\label{app:Molmer}

\begin{figure}
    \includegraphics{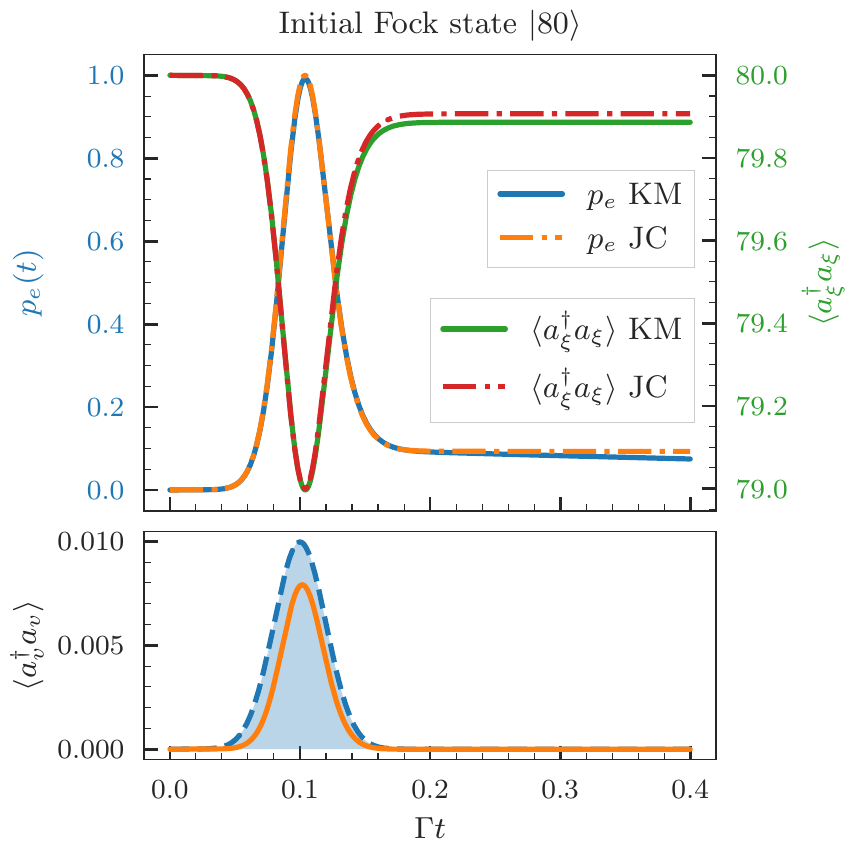}
    \caption{
    Populations and probabilities for an initial Fock state with 80 photons, prepared in a Gaussian pulse of duration $\Gamma T= 1/50 $ for $ \Gperp =0$.
    Top panel: Comparison between the atom excitation probability $p_e(t)$ and the average number of excitations in the pulse mode $\xi$ for the state obtained with the time-dependent Jaynes-Cummings (JC) approximation and for the state obtained in the Kiilerich and Mølmer (KM) formalism.
    Bottom panel: Average number of excitations in the auxiliary orthogonal mode $v$ appearing in the KM description, the shaded blue region shows the Gaussian temporal distribution $|\xi(t)|^2$ of the pulse as a guide for the eye (not to scale on the vertical axis).     }
    \label{fig:KMvsJC_vs_time}
\end{figure}

In this Appendix we use the formalism of Ref.~\cite{Christiansen2022} to present numerical evidence for the approximation explained in Sec.~\ref{sec:short_pulses}, which is also employed in the results of Sec.~\ref{sec:sodium}.

When $\Gperp = 0$ the Schrödinger equation for the joint pulse-atom system can be formally solved for input pulses containing a finite number of photons, as shown in Ref.~\cite{Konyk2016}, but the integrals rapidly become intractable as the number of photons increase.
To the best of our knowledge there is no general approach to obtain the field state analytically for arbitrary $\Gamma$ and $\Gperp$.
The main difficulty is that the interaction does not only transform the input quantum state, initially defined in a single temporal model only through the operators $A_{\xi}^\dag$ introduced in Eq.~\eqref{eq:temporalmode_annihil}, but also changes the temporal mode structure due to the spontaneous emission of the atom, as schematically depicted in Fig.~\ref{fig:scheme_interaction}.

A more tractable problem is to obtain the state of the light after the interaction only for a particular temporal mode, i.e., the reduced state obtained by tracing out all the other field temporal modes.
A general formalism to solve this problem was introduced by Kiilerich and Mølmer~\cite{Kiilerich2019,Kiilerich2020} (KM), based on the use of virtual cavities and cascaded master equations.
More recently this approach has been improved in Ref.~\cite{Christiansen2022} taking advantage of an appropriate interaction representation for the virtual cavities.
This more recent method is particularly suited to study the dynamics of the quantum state of light in a fixed single temporal mode that interacts with a quantum system, exactly what we need to validate the approximation to a time-dependent JC model.

We will now briefly summarize the formalism we have employed for the numerical validation, referring the interested reader to Ref.~\cite{Christiansen2022} for further details.
We consider the dynamics of a two-level atom interacting with an incident quantized radiation field governed by the Hamiltonian~\eqref{eq:interaction_hamiltonian_Gamma}.
When the incident radiation is a pulse in a single temporal mode, the interaction can be described using an effective cascaded-system master equation~\cite{Kiilerich2019,Kiilerich2020}.
In this formalism, the quantum pulse is represented by the radiation leaking from an upstream virtual cavity, while the component of outgoing radiation that eventually occupies an arbitrary fixed temporal mode is represented by the radiation picked up by a virtual downstream cavity.
All other temporal modes of the outgoing radiation are reflected by the downstream cavity, and in this formalism are essentially described as Markovian loss.
The coupling between the atom and the virtual cavities is time-dependent and it is a function of the chosen temporal modes~\cite{Kiilerich2019,Kiilerich2020}. 
In this approach excitations travel in a preferred direction; the initial state of the upstream cavity contains photons but at the end of the evolution the upstream cavity is left empty, on the contrary the downstream cavity starts empty and is eventually populated with the state of the radiation in the chosen temporal mode.

The idea of Ref.~\cite{Christiansen2022} is to use an interaction picture with respect to the interaction Hamiltonian of the two virtual cavities, responsible for the propagation of the radiation from the incident temporal mode to the chosen outgoing temporal mode, even in the absence of interaction with the atom.
The picture of an upstream and a downstream cavity is no longer valid in this frame, but the overall dynamics is still described by a tripartite system composed by the atom and two bosonic modes.
This approach is particularly illuminating if one focuses on the same temporal mode for the incident and outgoing radiation, which we assume to be described by a real-valued wavepacket $\xi(t)$ as in the main text.
In this situation we have a ``main'' (initially populated) bosonic mode $a_\xi$, describing the temporal evolution of the state of pulse (P) in the fixed temporal mode $\xi(t)$, and an auxiliary orthogonal (O) mode $a_v$.
The mode $v$ is needed to fully capture the transiet dynamics during the atom-pulse interaction, since it is initialized in the vacuum but eventually decays into the vacuum again after the interaction.
For conceptual clarity we keep the bosonic operator $a_\xi$ employed in the KM formalism (formally obtained from the interaction picture applied to the virtual cavities) distinct from the physical photon-wavepackeet operator $A_\xi$ defined in Eq.~\eqref{eq:temporalmode_annihil}.
The dynamics of the tripartite system composed by the atom, the pulse mode and the auxiliary orthogonal mode obeys a time-dependent Lindblad-like master equation: 
\begin{equation}
	\label{eq:KM_ME}
	\dot{\rho}^{\mathrm{APO}}(t) = - \frac{\I}{\hbar} [ H(t) , \rho^{\mathrm{APO}}(t) ] + \Gperp \mathcal{D}[ \sigmin]\rho^{\mathrm{APO}}(t) + \mathcal{D}[ L(t) ] \rho^{\mathrm{APO}}(t),
\end{equation}
where the Hamiltonian has the suggestive form
\begin{equation}
H(t) = \I \hbar \sqrt{\Gamma} \xi(t) \bigl( a_\xi^\dag \sigmin - \sigplus a_\xi \bigr) + \frac{\I \hbar}{2} \sqrt{\Gamma} f_1(t) \bigl( a_v^\dag \sigmin - \sigplus a_v \bigr),
\end{equation}
showing exactly the time-dependent JC interaction in Eq.~\eqref{eq:tdepJC} that we are seeking, plus an additional time-dependent interaction with the mode $v$ (no free atom Hamiltonian appears because we are assuming no detuning, as in the rest of the paper).
The time-dependent collapse operator reads 
\begin{equation}
	L(t) = \sqrt{\Gamma} \sigmin -  f_2(t) \xi(t) a_v.
\end{equation}
We have introduced two functions that depend on the wavefunction $\xi(t)$ and on the integral of its modulus squared
\begin{equation}
    f_1(t) = \frac{\left[ 1 - 2 I_{\xi}(t) \right] \xi(t)}{\sqrt{\left[ 1-I_{\xi}(t) \right] I_{\xi}(t)}} \qquad f_2(t) = \frac{\xi(t)}{\sqrt{\left[ 1-I_{\xi}(t) \right] I_{\xi}(t)}} \qquad I_{\xi}(t) = \int_{t_0}^t ds |\xi(s)|^2, 
\end{equation}
where as usual the starting time $t_0$ of the experiment is assumed to be far before the arrival (e.g. the main peak) of the pulse so that $I_\xi(t) \approx 1$ for times much greater than the pulse duration $t \gg T$.

In Fig.~\ref{fig:KMvsJC_vs_time} we show results for a Gaussian pulse of duration (i.e. standard deviation of $|\xi(t)|^2$) $\Gamma T= 0.02$ centered around $ \Gamma t_0 = 0.1 $, for times up to $\Gamma t = 0.4$.
In particular, we choose an initial Fock state with 80 photons, to show the expected coherent Rabi oscillations between the atom and the collective mode, similarly to what predicted in~\cite{Fischer2018a} using quantum stochastic calculus techniques.
As mentioned in Sec.~\ref{sec:short_pulses} the number of photons needs to be relatively high so that even if $\Gamma T$ is a small number we are beyond the linear absorption regime to witness a coherent exchange of excitations.
In this regime we already see a pretty good agreement between the population predicted by the two models and we see that the orthogonal mode $v$ contains less then $0.01$ photons on average during the whole dynamics.
Nonetheless, we also start to see a deviation from the JC model in the fact that the true $p_e(t)$ obtained from the KM method starts to slowly decay after the interaction (and it will eventually go to zero for $ t \gg \Gamma$), and correspondingly the photon number of the state in the pulse temporal mode $\xi$ after the interaction contains slightly less photons than predicted by the JC approximation.
Thus while $p_e(t) + \langle a_\xi^\dag a_\xi \rangle = 80$ in the JC model we see that this holds only approximately for the results simulated with the KM method.

\begin{figure}[t]
    \includegraphics{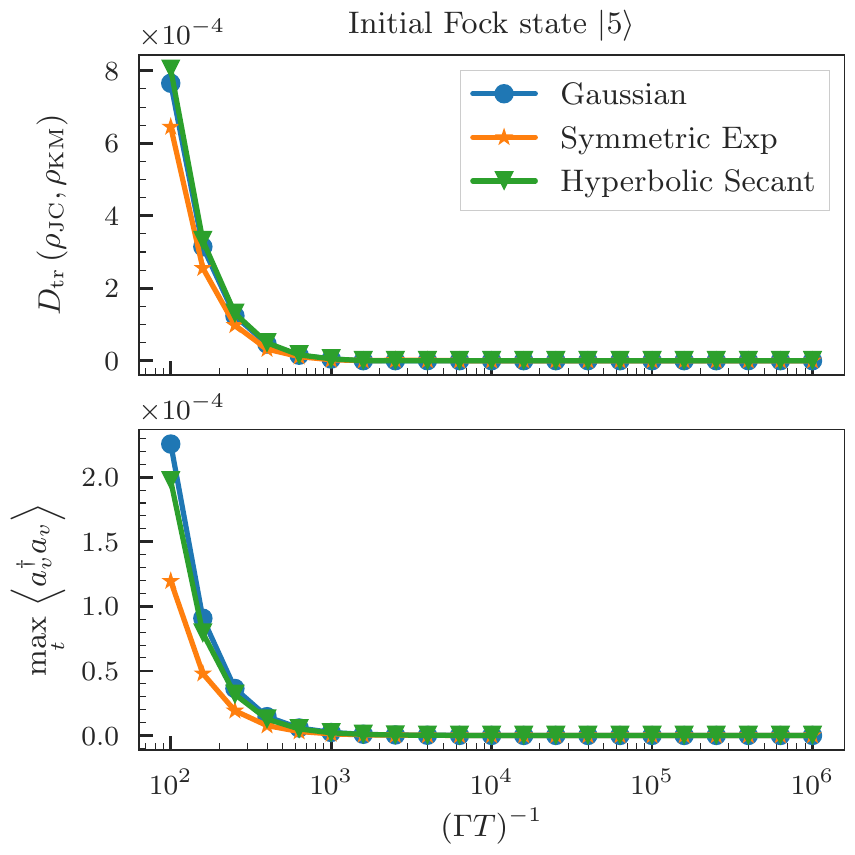}
    \includegraphics{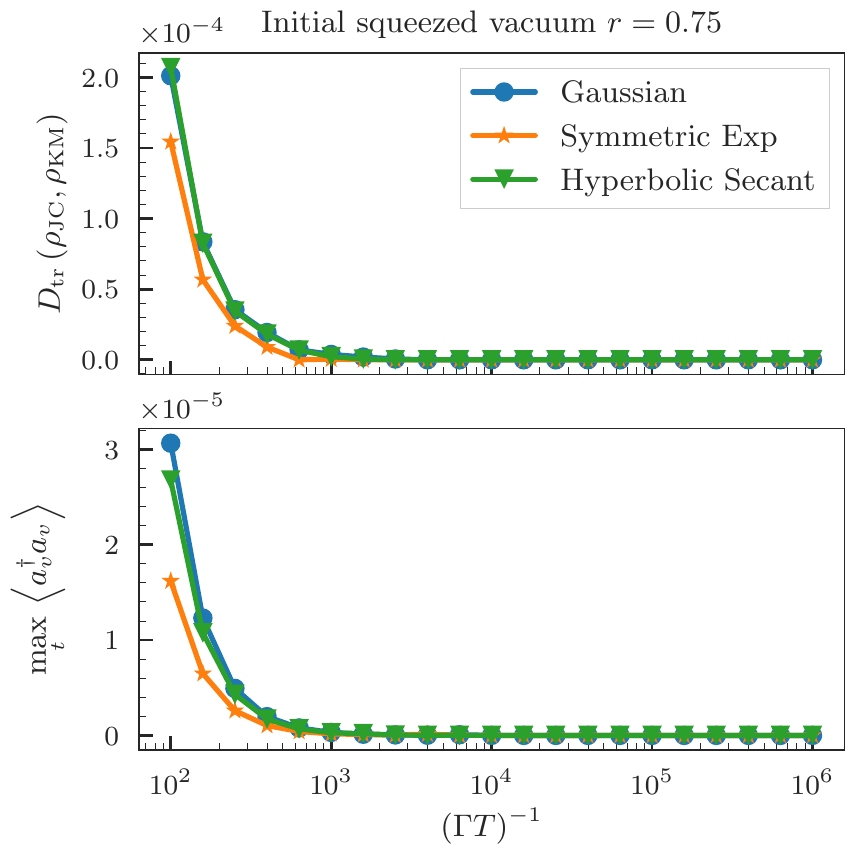}
    \caption{Top panels: Trace distance between the reduced state in the pulse temporal mode $\rho_{\mathrm{KM}}$ obtained from the Kiilerich and Mølmer (KM) formalism and the state $\rho_{\mathrm{JC}}$ obtained using the Jaynes-Cummings (JC) approximation. 
    The trace distance is evaluated at $t=10T$, i.e. after the pulse has fully interacted with the atom.
    Bottom panels: Maximum value reached by the average number of excitations in the auxiliary orthogonal mode $v$ appearing in the KM formalism during the evolution.
    }
    \label{fig:KMvsJC_vs_T}
\end{figure}

If we consider shorter pulses we can see these discrepancies disappear.
In particular, since in a spectroscopy setting we can only measure the output light scattered by the two-level atom, we want to test the goodness of the approximation at the level of the reduced state of the pulse, i.e. $\rho_{\mathrm{KL}}(t) \equiv \Tr_{\mathrm{A,O}} \rho^{\mathrm{APO}}(t)$.
Such a state is compared with the state $\rho_{\mathrm{JC}}(t) \equiv \Tr_{\mathrm{A}} \rho^{\mathrm{AP}}(t)$ obtained by unitarily evolving the atom-pulse bipartite system with the time-dependent JC Hamiltonian in Eq.~\eqref{eq:tdepJC} and then tracing out the atom subsystem.
In Fig.~\eqref{fig:KMvsJC_vs_T} we plot the trace distance between these two states $D_\mathrm{tr}(\rho,\sigma)= \frac{1}{2} \Tr | \rho-\sigma | $ evaluated at a time $t=10T$, such that pulse-atom interaction is complete but we are still in the short-time regime $\Gamma t \ll 1$.
We see that the trace distance vanishes as the pulse and evolution time are made shorter, similarly we see that the maximal population of the auxiliary mode $v$ during the evolution also decreases as the pulses get shorter.
This holds true not only for an initial Fock state $\ket{5}$ but also for a squeezed vacuum state which has coherences in the Fock basis.

From Figs.~\ref{fig:KMvsJC_vs_time} and~\ref{fig:KMvsJC_vs_T} it is also apparent why the method of Ref.~\cite{Christiansen2022} is more numerically efficient than the original upstream and downstream cavities approach of Refs.~\cite{Kiilerich2019,Kiilerich2020}.
As a matter of fact, even if the initial state of the pulse contains many photons and requires a high dimensional Hilbert space, the orthogonal mode $v$ often absorbs only a small portion of the initial photons, thus requiring a much smaller Fock space cutoff for the numerical simulation than the cascaded cavities.

\section{Short-time regime for short single photon pulses}
\label{app:appShortSinglePh}

In this Appendix we show that the QFI in Eq.~\eqref{eq:QFI_tdepJCFock} derived for the approximated time-dependent JC model of Sec.~\ref{sec:short_pulses} for the case a single-photon pulse corresponds to the result obtained from the exact expressions derived in Sec.~\ref{sec:1photon}.

We consider times much shorter than the lifetime, i.e., $t-t_0 \ll 1 / \Gamma_\mathrm{tot}$.
While $t_0$ is the ``start of the experiment'' and is assumed to be in the past, i.e., $t_0 < t$ for the final $t$ at which detection happens, we can actually think that the only relevant $t_0$ in the experiment is related to the region where the pulse $\xi(t)$ is non-zero~(justifying then the substitution $t_0 \to -\infty$ in the integrals in the main text).
For simplicity and without loss of generality we always assume the peak (or the ``arrival time'') of the pulse to be at $\bar{t}=0$, so that $t_0$ is always negative.
It follows that for short pulses with $\Gamma_\mathrm{tot} T \ll 1$, $t_0$ can be assumed to be $| t_0 |  \ll 1 / \Gamma_\mathrm{tot}$ so that we can neglect the exponentials also for $t<0$.

In this regime, spontaneous emission is negligible and the corresponding exponentials factors can be omitted from Eq.~\eqref{eq:1ph_pulse_psie}, obtaining
\begin{align}
    & \psi_e(t) = - \sqrt{\Gamma} \int_{t_0}^t dt' \xi(t') = - \sqrt{\Gamma T} \int_{t_0}^{t/T} dx f(x), \\
    & \braket*{\widetilde{\psi}_g^\mathrm{P}(t)}{\xi} =  1-\Gamma \int_{t_0}^t d\tau \left( \xi(\tau) \int_{t_0}^\tau dt' \xi^*(t')\right) =  1- \Gamma T \int_{t_0}^{t/T} d x \left( f(x) \int_{t_0}^x dx' f^*(x') \right),
\end{align}
where we have used the scale-invariant pulse shape $f(x)$ defined as $\xi(t) = f(t/T)/\sqrt{T}$, satisfying $\int_{-\infty}^\infty |f(x)|^2 dx = 1$.
For a short pulse satisfying $\Gamma T \ll 1 $ we can expand $p_\mathrm{orig}$ to first order in $\Gamma T$ and obtain
\begin{equation}
    \begin{split}
    p_\mathrm{orig} = \left | \braket*{ \widetilde\psi_g^\mathrm{P}(t)}{\xi} \right |^2 
    			& \approx 1 - \Gamma T \left[ \int_{t_0}^{t/T} dx \left( \xi(x) \int_{t_0}^x dx' f^*(x') \right) + \int_{t_0}^{t/T} dx \left( f^*(x) \int_{t_0}^x dx' f(x') \right) \right] \\
    & = 1 - \Gamma T \left| \int_{t_0}^{t/T} d x  f(x) \right|^2 = 1 - |\psi_e(t)|^2, 
    \end{split}
\end{equation}
where the equality between the first and second line can be obtained as simply an integration by parts $\int_{t_0}^{t/T} d x' \left( f(x') \int_{t_0}^{x'} d x'' f ^*(x'') \right) = \left| \int_{t_0}^{t/T} d x f(x) \right|^2  - \int_{t_0}^{t/T} d x \left( f^*(x) \int_{t_0}^{x} d x' f(x') \right) $.

This calculation shows that, to first order in $\Gamma T $, the small probability of not detecting a photon in the pulse temporal mode after the interaction with the atom is only due to the atom absorption and not because other temporal modes become populated (recall that $\ket*{\widetilde\psi_g^\mathrm{P}(t)}$ is not a normalized state).
This, in turn, means that we only have the classical contribution to the QFI, which becomes
\begin{equation}
    \label{eq:peTermCFIsingle}
    \frac{ \left[ \partial_\Gamma p_e(t) \right]^2 }{ p_e(t)[1-p_e(t)]} \approx \frac{ \left[ \partial_\Gamma p_e(t) \right]^2 }{ p_e(t)} = \frac{p_e(t)}{\Gamma^2}   = \frac{T}{\Gamma} \left[ \int_{t_0}^{t/T} f(x) dx \right]^2 \equiv \frac{T F_t^2}{\Gamma}, 
\end{equation}
where the first approximate equality holds for $\Gamma T \ll 1$ because $p_e(t)=|\psi_e(t)|^2 = \Gamma T \left| \int_{t_0}^{t/T} d x  f(x) \right|^2$ is also a very small quantity.
We have also used that $\partial_\Gamma p_e(t) = p_e(t) / \Gamma$.
This expression in Eq.~\eqref{eq:peTermCFIsingle} corresponds to the Fock state QFI in Eq.~\eqref{eq:QFI_tdepJCFock} obtained from the approximate time-dependent JC model, for $n=1$.

At the same time, we can show that the quantum contribution $\tilde{\mathcal{Q}}(\ket{\psi_{\Gamma}})$ vanishes faster than the $\mathcal{C}(p_{\mathrm{orig}})$ in the limit of short times $t-t_0 \ll 1 / \Gamma_\mathrm{tot}$.
In order to show this, we first note the following expression for QFI of the normalized state $\ket{\psi}_{\Gamma}$ in terms of the overlaps of unnormalized wavepacket $\ket*{\widetilde\psi_g^\mathrm{P}}$:
\begin{equation}\label{eq:quantumQFIshorttime}
    \mathcal{Q}(\ket{\psi_{\Gamma}}) = 4\frac{\langle\partial_{\Gamma} \widetilde\psi_g^\mathrm{P}|\partial_{\Gamma}\widetilde{\psi}_g^\mathrm{P}\rangle}{\langle \widetilde{\psi}_g^\mathrm{P}| \widetilde\psi_g^\mathrm{P}\rangle} - 4\left(\frac{\langle \widetilde\psi_g^\mathrm{P}|\partial_{\Gamma} \widetilde\psi_g^\mathrm{P}\rangle}{\langle \widetilde\psi_g^\mathrm{P}|\widetilde\psi_g^\mathrm{P}\rangle}   \right)^2
\end{equation}
where we have employed the earlier assumption that $\psi_{\Gamma},\partial_{\Gamma}\psi_{\Gamma}\in\mathbb{R}$.
The first term in the above expression can be expressed as the following power series of pulse duration $T$ from which we can extract the leading term~(in keeping with the assumption of short pulses so that $\Gamma T \ll 1$):
\begin{align}
    \frac{\langle\partial_{\Gamma}\widetilde\psi_g^\mathrm{P}|\partial_{\Gamma}\widetilde\psi_g^\mathrm{P}\rangle}{\langle\widetilde\psi_g^\mathrm{P}|\widetilde\psi_g^\mathrm{P}\rangle} &= \frac{T\,\int_{t_0}^{t}\,d\tau\,F_{\tau}^2}{1-\Gamma T F_t^2} \nonumber\noindent \\
    &= T\,\int_{t_0}^{t}\,d\tau\,F_{\tau}^2 + \Gamma T^2 F_t^2\,\int_{t_0}^{t}\,d\tau\,F_{\tau}^2 + \dots \approx T\,\int_{t_0}^{t}\,d\tau\,F_{\tau}^2
\end{align}
Similarly, the (square root of) the second term is, to leading power in $T$:
\begin{equation}
    \frac{\langle \widetilde \psi_g^\mathrm{P}|\partial_{\Gamma} \widetilde \psi_g^\mathrm{P} \rangle}{\langle \widetilde \psi_g^\mathrm{P}|\widetilde \psi_g^\mathrm{P}\rangle}  \approx  T\left[ \int_{t_0}^{t/T}\!dx\,\left( \int_{t_0/T}^x\,dx'\,f(x') \right)\,\left(\Gamma T\,  \int_{t_0/T}^x\!dx'\,f(x') - f(x) \right)  \right].
\end{equation}
Using the mean value theorem for definite integrals to approximate $F_t= \frac{t-t_0}{T}\,f\left(\frac{t_1-t_0}{T}\right)\,\,\mathrm{s.t.}\,\,t_0<t_1<t$, the quantum contribution $\mathcal{Q}(\ket{\psi}_{\Gamma})$ obeys the following bound:
\begin{equation}
    \mathcal{Q}(\ket{\psi_{\Gamma}}) \leq \frac{4}{3}\,T^2\,\left( \frac{t-t_0}{T}\right)^3\,f\left( \frac{t_1-t_0}{T} \right)^2 - 4T^2\,\left( \frac{t-t_0}{T}   \right)^4\,\,\left[ \frac{\Gamma T}{3}\,  \frac{t-t_0}{T} f\left( \frac{t_2-t_0}{T} \right)^2  - \frac{1}{2} f\left( \frac{t_3-t_0}{T} \right)     \right]^2, \,\, t_0<t_1,t_2,t_3<t
\end{equation}
meaning that $\mathcal{Q}(\ket{\psi_{\Gamma}})$ scales as $O([t-t_0]^3)$, whereas the classical contribution in Eq.~(\ref{eq:peTermCFIsingle}) scales as $O([t-t_0]^2)$. Therefore, as $\Gamma_{\mathrm{tot}}(t-t_0)\rightarrow 0$, the quantum contribution vanishes faster, leaving only $\mathcal{C}(p_{\mathrm{orig}})$.

\section{Derivation of useful quantum estimation results}

\subsection{Optimality of projecting on the probe state for pure-state local quantum estimation}\label{app:OptProjSameState}

In this Appendix we show explicitly that a projection on the state itself saturates the pure state QFI as stated in Sec.~\ref{subsubsec:1phOptMeas}.

The state of the system is $\ket{\psi_{\Gamma}}$ and we consider a projective measurement $ \Pi_1 = \ket{\psi_{\Gamma'}}\bra{\psi_{\Gamma'}}$ and $\Pi_0 = \id - \Pi_1$, so that $p_1= | \braket{\psi_{\Gamma'} }{ \psi_{\Gamma} } |^2$, $p_0 = 1-p_1$ and $ \partial_\Gamma p_1 = - \partial_\Gamma p_0 = 2 \Re \left(\braket{\psi_{\Gamma} }{ \psi_{\Gamma'} } \braket{\psi_{\Gamma'} }{ \partial_\Gamma \psi_{\Gamma} }   \right)$. 
We aim to take the limit $\Gamma' \to \Gamma$ for which $\lim_{\Gamma' \to \Gamma} \braket{ \psi_{\Gamma'}} { \psi_\Gamma } = \braket{ \psi_{\Gamma}} { \psi_\Gamma }  = 1$ and $\lim_{\Gamma' \to \Gamma} \Re \braket{ \psi_{\Gamma'}} { \partial_\Gamma \psi_\Gamma } = \Re \braket{ \psi_{\Gamma}} { \partial_\Gamma \psi_\Gamma } = 0 $, thus  $\lim_{\Gamma' \to \Gamma} p_1 = 1$ and $\lim_{\Gamma' \to \Gamma} \partial_\Gamma p_1 = 0$.
The CFI of such a two-outcome measurement is $\mathcal{C}(p_1) = \frac{(\partial_\Gamma p_1 )^2}{p_1 (1-p_1)}$ and becomes a $0/0$ indeterminate form in the limit $\Gamma'\to\Gamma$.
Using L'Hôpital's rule we obtain
\begin{equation}
    \label{eq:lhopitalpure}
    \lim_{\Gamma'\to \Gamma} \mathcal{C}(p_1) = \lim_{\Gamma'\to \Gamma} \frac{(\partial_\Gamma p_1 )^2}{p_1 (1-p_1)} = \lim_{\Gamma'\to \Gamma} \frac{2 \partial_\Gamma p_1 \, \partial^2_\Gamma p_1 }{ \partial_\Gamma p_1 (1- 2 p_1)} = - 2 \partial^2_\Gamma p_1 \vert_{\Gamma'=\Gamma} = -4 \left(  \Re \braket{\psi_{\Gamma} }{ \partial^2_\Gamma \psi_{\Gamma} } + \left| \braket{ \partial_\Gamma \psi_{\Gamma} }{ \psi_{\Gamma} } \right|^2      \right).
\end{equation}
Differentiating the equality $\braket{ \psi_{\Gamma}} { \partial_\Gamma \psi_\Gamma } + \braket{  \partial_\Gamma \psi_{\Gamma}} { \psi_\Gamma } = 0$ we obtain $\Re \left(  \braket{ \psi_{\Gamma}}{ \partial^2_\Gamma \psi_\Gamma } \right) = - \braket{ \partial_\Gamma \psi_{\Gamma}}{ \partial_\Gamma \psi_\Gamma }$ and thus $\lim_{\Gamma'\to \Gamma} \mathcal{C}(p_1) = \mathcal{Q}(\ket{\psi_\Gamma}) $ according to Eq.~\eqref{eq:qfi_pure}.

\subsection{QFI of a rank-2 state}
\label{app:QFI_rank2}

In this Appendix we evaluate the QFI of a rank-2 density matrix, written as a mixture of two non-orthogonal pure states.
This is employed in the main paper for the state in Eq.~\eqref{eq:JCreducedFieldState} obtained in the approximate time-dependent JC model introduced in Sec.~\ref{sec:short_pulses}, in particular it is applied to coherent and squeezed states in Sec.~\ref{sec:sodium}.
However, more generally, the reduced state of the field is described by rank-2 density matrix whenever $\Gperp=0$ and $t< \infty$, i.e., when the quantum state of the pulse is mixed only for being entangled with the two-level atom.

We consider the rank-2 density matrix
\begin{equation}
    \label{eq:rhorank2app}
	\rho_\Gamma = \ket*{\widetilde{\psi}_e}\bra*{ \widetilde{\psi}_e} + \ket*{ \widetilde{\psi}_g} \bra*{ \widetilde{\psi}_g},
\end{equation} 
such as the one in Eq.~\eqref{eq:JCreducedFieldState} for the reduced state of the field.
We denote with $\mathcal{B}$ the (generally nonorthogonal) basis formed by these two vectors and their derivatives with respect to the parameter of interest $\Gamma$
\begin{equation}
	\mathcal{B} = \left\{ \ket*{\widetilde{\psi}_e } , \ket*{\widetilde{\psi}_g } , \ket*{\partial_\Gamma \widetilde{\psi}_e }, \ket*{\partial_\Gamma \widetilde{\psi}_g }  \right\}.
\end{equation}
with the Gramiam matrix
\begin{equation}
	G^\mathcal{B} = \begin{bmatrix} 
		\braket*{ \widetilde{\psi}_e }{ \widetilde{\psi}_e} & \braket*{ \widetilde{\psi}_e }{ \widetilde{\psi}_g } & \braket*{ \widetilde{\psi}_e }{ \partial_\Gamma \widetilde{\psi}_e } & \braket*{ \widetilde{\psi}_e }{ \partial_\Gamma \widetilde{\psi}_g } \\
		\braket*{ \widetilde{\psi}_g}{ \widetilde{\psi}_e} & \braket*{ \widetilde{\psi}_g }{ \widetilde{\psi}_g } & \braket*{ \widetilde{\psi}_g }{ \partial_\Gamma \widetilde{\psi}_e } & \braket*{ \widetilde{\psi}_g }{ \partial_\Gamma \widetilde{\psi}_g } \\ 
		\braket*{ \partial_\Gamma \widetilde{\psi}_e }{ \widetilde{\psi}_e} & \braket*{ \partial_\Gamma \widetilde{\psi}_e }{ \widetilde{\psi}_g } & \braket*{ \partial_\Gamma \widetilde{\psi}_e }{ \partial_\Gamma \widetilde{\psi}_e } & \braket*{ \partial_\Gamma \widetilde{\psi}_e }{ \partial_\Gamma \widetilde{\psi}_g } \\
		\braket*{ \partial_\Gamma \widetilde{\psi}_g }{ \widetilde{\psi}_e} & \braket*{ \partial_\Gamma \widetilde{\psi}_g }{ \widetilde{\psi}_g } & \braket*{ \partial_\Gamma \widetilde{\psi}_g }{ \partial_\Gamma \widetilde{\psi}_e } & \braket*{ \partial_\Gamma \widetilde{\psi}_g }{ \partial_\Gamma \widetilde{\psi}_g }
	\end{bmatrix}.
\end{equation}
Assuming that $\mathcal{B}$ is a basis means that the vectors must be linearly independent and thus $G^\mathcal{B}$ invertible.
While the linear independence of $\ket{\psi_e}$ and $\ket{\psi_g}$ is implied by the assumption that $\rho_\Gamma$ is rank-2, the linear independence of the whole basis $\mathcal{B}$ is as an extra assumption in this derivation, but it is valid for the applications considered in this paper.

Using the notation of Ref.~\cite{Fiderer2021a} we can represent operators as matrices expressed on the basis $\mathcal{B}$ and Eq.~\eqref{eq:Lyap_def} becomes
\begin{equation}
	2 \partial_\Gamma \rho^\mathcal{B} = L_\Gamma^\mathcal{B} G_\Gamma^\mathcal{B} \rho_\Gamma^\mathcal{B} + \rho_\Gamma^\mathcal{B} G_\Gamma^\mathcal{B} L_\Gamma^\mathcal{B}.
\end{equation}
This equation can be solved efficiently by using block vectorization~\cite{Bisketzi2019,Fiderer2021a}.
Once a solution is found, the QFI can be evaluated as 
\begin{equation}
	\mathcal{Q}(\rho_\Gamma) = \Tr \left[ L_\Gamma^\mathcal{B} G^\mathcal{B} \partial_\Gamma \rho_\Gamma^\mathcal{B} G^\mathcal{B} \right]
\end{equation}

For the rank-2 model in~\eqref{eq:rhorank2app} the density matrix and its derivative have a very simple form in the basis $\mathcal{B}$:
\begin{equation}\rho^\mathcal{B} = \begin{bmatrix} 1 & 0 & 0 & 0 \\ 
	0 & 1 & 0 & 0 \\ 
	0  & 0 & 0 & 0 \\
	 0 & 0 & 0 & 0 
\end{bmatrix}  \qquad
\partial_\Gamma \rho^\mathcal{B} = \begin{bmatrix} 0 & 0 & 1 & 0 \\ 
	0 & 0 & 0 & 1 \\ 
	1  & 0 & 0 & 0 \\
	 0 & 1 & 0 & 0 
\end{bmatrix}
\end{equation}
and the Lyapunov equation can be solved analytically to obtain
an explicit, albeit complicated, expression for the QFI that depends only on the matrix elements of $G^{\mathcal{B}}$.
\begin{equation}
    \label{eq:QFIrank2}
\begin{split}
	\mathcal{Q}(\rho_\Gamma) = &\frac{-4}{\Delta (G_{11}+G_{22})} \Biggl\{ 
	\Delta \left[(\Im G_{13}-\Im G_{24})^2+(\Im G_{14}+\Im G_{23})^2\right]
	\\ 
&+4 \Delta G_{11} (\Im G_{33}+\Im G_{44})+4 \Delta G_{22} (\Im G_{33}+\Im G_{44})\\
&	-4 (\Im G_{13})^2 G_{22}^2+8 \Im G_{13} \Re G_{12} G_{22} (\Im G_{14}+\Im G_{23})+8 \Im G_{13} \Im G_{12} G_{22} (\Re G_{23}-\Re G_{14})\\
& -4 G_{11} G_{22} \left[ 2 \Im G_{13} \Im G_{24}+(\Re G_{14}-\Re G_{23})^2 \right]+8 \Im G_{12} \Re G_{12} (\Im G_{14}+\Im G_{23}) (\Re G_{14}-\Re G_{23})\\
& +8 \Im G_{24} \Re G_{12} G_{11} (\Im G_{14}+\Im G_{23})\\
& -4 (\Re G_{12})^2 (\Im G_{14}+\Im G_{23}+\Re G_{14}-\Re G_{23}) (\Im G_{14}+\Im G_{23}-\Re G_{14}+\Re G_{23}) \\
& +8 \Im G_{12} \Im G_{24} G_{11} (\Re G_{23}-\Re G_{14})-4 (\Im G_{24})^2 G_{11}^2 \Biggr\},
\end{split}
\end{equation}
where $\Delta = G_{11}G_{22}- |G_{12}|^2 > 0$ is the determinant of the first diagonal block of $G^\mathcal{B}$ and the superscript $\mathcal{B}$ has been suppressed for compactness.

A much simpler expression can be obtained when the two parameter-dependent rank-1 states live in orthogonal subspaces, i.e. $\braket*{ \widetilde{\psi}_e }{ \widetilde{\psi}_g }= 0 $, $\braket*{ \widetilde{\psi}_e }{ \partial_\Gamma \widetilde{\psi}_g } = 0$ and $\braket*{ \widetilde{\psi}_g }{ \partial_\Gamma \widetilde{\psi}_e } = 0$:
\begin{equation}
    \label{eq:rank2ortho}
    \mathcal{Q}(\rho_\Gamma) = \sum_{x=e,g} 4 \braket*{ \partial_\Gamma \widetilde{\psi}_x }{ \partial_\Gamma \widetilde{\psi}_x }  + \frac{ \Im \left[ \braket*{ \widetilde{\psi}_x}{ \partial_\Gamma \widetilde{\psi_x } } \right]^2 }{ \braket*{ \widetilde{\psi_x}}{\widetilde{\psi_x } }},
\end{equation}
which makes the computation easier by avoiding a renormalization of the two orthogonal states.
However, since the two states in the mixture are orthogonal, this QFI can also be obtained from the standard formulas based on the eigendecomposition of the density matrix~\cite{Paris2009,Liu2014a}.

\section{Description of Parametric Down Converted (PDC) State}
\label{app:PDC}
The biphoton state generated at the end of low-gain type-II PDC interaction in birefringent crystals~(such as BBO or KTP cystals) that converts the classical pump photon into (signal and idler) daughter photons is obtained as the first order perturbation term,
\begin{equation}\label{eq:PDCstate}
    \ket{\Phi_{\mathrm{PDC}}} = \frac{1}{\sqrt{N_{\mathrm{PDC}}}}\,\left( \ket{0} + \int d\omega_{\mathrm{S}}\int d\omega_{\mathrm{I}}\, \tilde{\Phi}_{\mathrm{PDC}}(\omega_{\mathrm{S}},\omega_{\mathrm{I}})\,a^{\dag}_{\mathrm{S}}(\omega_{\mathrm{S}})a^{\dag}_{\mathrm{I}}(\omega_{\mathrm{I}})\ket{0^{\mathrm{S}}}\ket{0^{\mathrm{I}}}     \right),
\end{equation}
where $N_{\mathrm{PDC}}$ is the normalization factor which ensures that $\ket{\Phi_{\mathrm{PDC}}}$ is well normalized. For a more complete description of the PDC process, including in the high-gain regime, see Ref.~\cite{Christ2013}.
The bivariate joint spectral amplitude~(JSA) $\tilde{\Phi}_{\mathrm{PDC}}(\omega_{\mathrm{S}},\omega_{\mathrm{I}})$ for PDC states is the following product of the classical pump pulse envelope~(which is assumed to be Gaussian with spectral width given by $\sigma_{\mathrm{p}}$), and the sinc phase-matching function for collinear setups,
\begin{equation}\label{eq:JSA1}
     \tilde{\Phi}_{\mathrm{PDC}}(\omega_{\mathrm{S}},\omega_{\mathrm{I}}) = -\frac{i\alpha_{\mathrm{pump}}}{\hbar}\,\mathrm{sinc}\left( \frac{\Delta k(\omega_{\mathrm{S}},\omega_{\mathrm{I}})L}{2} \right)\,\frac{1}{\sqrt{2\pi\sigma_{\mathrm{p}}^2}}\,e^{-(\omega_{\mathrm{S}}+\omega_{\mathrm{I}}-\omega_{\mathrm{p}})^2/2\sigma_{\mathrm{p}}^2},
\end{equation}
where $\alpha_{\mathrm{pump}}/\hbar$ depends on the crystal properties~(such as crystal length $L$, and the second-order -- as PDC is a three-wave mixing process -- non-linear susceptibility $\chi^{(2)}$), as well as beam properties~(chief amongst them being the beam width that fixes the area of quantization in the paraxial description).
For simplicity, we bunch these experimental parameters together into the efficiency of the downconversion process~\cite{Schlawin2017a}.

The phase-matching function $\Delta k(\omega_{\mathrm{S}},\omega_{\mathrm{I}})$ can be related to the different group velocities and times of arrival of the two photons, by Taylor expanding the signal/idler wavevectors around their respective central frequencies~(for which conservation of energy dictates $\bar{\omega}_{\mathrm{S}} + \bar{\omega}_{\mathrm{I}} = \omega_{\mathrm{p}}$),
\begin{equation}
    k(\omega_{\mathrm{X}}) = \bar{k}_{\mathrm{X}} + \frac{\partial k}{\partial \omega_{\mathrm{X}}} \biggr\vert_{\omega_{\mathrm{X}}=\bar{\omega}_{\mathrm{X}}}(\omega_{\mathrm{X}}-\bar{\omega}_{\mathrm{X}}) + \dots,~~\mathrm{X}\,=\,\mathrm{S,I}.
\end{equation}
The first-order coefficient can be identified as the inverse of the wavepacket group velocity $1/v_{\mathrm{X}} = \partial k/\partial \omega_{\mathrm{X}}\vert_{\omega_{\mathrm{X}}=\bar{\omega}_{\mathrm{X}}} $.
Keeping then only the linear terms in the Taylor expansion, the phase-matching function is
\begin{equation}
    \Delta k(\omega_{\mathrm{S}},\omega_{\mathrm{I}})L = \left( \frac{1}{v_{\mathrm{p}}} - \frac{1}{v_{\mathrm{S}}} \right)L\,(\omega_{\mathrm{S}}-\bar{\omega}_{\mathrm{S}}) +  \left( \frac{1}{v_{\mathrm{p}}} - \frac{1}{v_{\mathrm{I}}} \right)L\,(\omega_{\mathrm{I}}-\bar{\omega}_{\mathrm{I}}) = T_{\mathrm{S}}\,(\omega_{\mathrm{S}}-\bar{\omega}_{\mathrm{S}}) + T_{\mathrm{I}}\,(\omega_{\mathrm{I}}-\bar{\omega}_{\mathrm{I}}),
\end{equation}
where $T_{\mathrm{S}} = (1/v_{\mathrm{p}}\,-\,1/v_{\mathrm{S}})L$ is the time difference between the arrival of the wavepacket travelling at the group velocity of the pump versus that of the signal photon, and similarly for $T_{\mathrm{I}}$.
The time delay between the arrival of the two photons is captured by the quantity $T_{\mathrm{qent}} = T_{\mathrm{S}} - T_{\mathrm{I}}$, henceforth referred to as the entanglement time.
In the main text
we only study two-photon states with frequency anti-correlations~($T_{\mathrm{S}}>0,T_{\mathrm{I}}>0$), with the specific choice of $T_{\mathrm{S}} = 0.12 \,T_{\mathrm{qent}}$ and $T_{\mathrm{I}} = 1.12 \,T_{\mathrm{qent}}$.
The entanglement time $T_{\mathrm{qent}}$ itself is varied for the purposes of the calculation of the metrological quantities between $50\,\mathrm{fs}$ and $3.0\,\mathrm{ps}$.


Finally, the $\mathrm{sinc}$ function can be approximates as a Gaussian~\cite{Grice2001a,kuzucu2008joint,Christ2013} ignoring their minor maxima, as
\begin{equation}
    \mathrm{sinc}\left( \frac{\Delta k(\omega_{\mathrm{S}},\omega_{\mathrm{I}})L}{2} \right) \approx \mathrm{exp}\left( -\gamma(\Delta k(\omega_{\mathrm{S}},\omega_{\mathrm{I}})L)^2\right), ~~\gamma = 0.04822.
\end{equation}
yielding a JSA that is now proportional to a two-dimensional Gaussian function,
\begin{equation}\label{eq:JSAgaussian}
    \tilde{\Phi}_{\mathrm{PDC}}(\omega_{\mathrm{S}},\omega_{\mathrm{I}}) \approx -\frac{i\alpha_{\mathrm{pump}}}{\hbar}\,\frac{1}{\sqrt{2\pi\sigma_{\mathrm{p}}^2}}\mathrm{exp}\left(-a(\omega_{\mathrm{S}}-\bar{\omega}_{\mathrm{S}})^2 + 2b\,(\omega_{\mathrm{S}}-\bar{\omega}_{\mathrm{S}})(\omega_{\mathrm{I}}-\bar{\omega}_{\mathrm{I}}) -c(\omega_{\mathrm{I}}-\bar{\omega}_{\mathrm{I}})^2 \right)
\end{equation}
where 
\begin{align}
    a = \frac{1}{2\sigma_{\mathrm{p}}^2} + \gamma T_{\mathrm{S}}^2 , ~b = \frac{1}{2\sigma_{\mathrm{p}}^2} + \gamma T_{\mathrm{S}} T_{\mathrm{I}}, ~ c = \frac{1}{2\sigma_{\mathrm{p}}^2} + \gamma T_{\mathrm{I}}^2. 
\end{align}
While it is always possible to (numerically) construct a Schmidt decomposition for arbitrary bivariate JSAs $\tilde{\Phi}(\omega_{\mathrm{S}},\omega_{\mathrm{I}})$~\citep{Lamata2005}, the approximate double Gaussian JSA in Eq.~(\ref{eq:JSAgaussian}) admits an analytical Schmidt decomposition in terms of the Hermite-Gaussian~(HG) mode functions, defined as
\begin{equation}\label{eq:hgmodedef}
    h_n(x) = \frac{1}{\sqrt{2^n n! \sqrt{\pi}}}~e^{-x^2/2}\,H_n(x)~\forall\,n\in\{0,1,\dots\}.
\end{equation}
where $H_n(x)$ is the $n$-th order Hermite polynomial. Then, using Mehler's Hermite polynomial formula~\citep{abramowitz1964handbook},
\begin{equation}\label{eq:mehler}
    \sum_{n=0}^{\infty}~\frac{w^n H_n(x)H_n(y) }{2^n n!} = \frac{1}{\sqrt{1-w^2}}\,\mathrm{exp}\left[ \frac{2w xy - w^2(x^2+y^2)}{1-w^2}   \right],
\end{equation}
we can express the two-dimensional Gaussian JSA as the following sum of products of univariate functions,
\begin{equation}\label{eq:schmidtJSA}
    \tilde{\Phi}_{\mathrm{PDC}}(\omega_{\mathrm{S}},\omega_{\mathrm{I}}) \approx \sum_{n=0}^{\infty}\,r_{n,\mathrm{PDC}}\, h_n(k_{\mathrm{S}}(\omega_{\mathrm{S}}-\bar{\omega}_{\mathrm{S}}))\,h_n(k_{\mathrm{I}}(\omega_{\mathrm{I}}-\bar{\omega}_{\mathrm{I}})),\, r_{n,\mathrm{PDC}} = -\frac{i\alpha_{\mathrm{pump}}}{\hbar}\,\sqrt{\frac{1+w^2}{4\sqrt{ac}\sigma_{\mathrm{p}}^2}}\,w^n,
\end{equation}
where $k_{\mathrm{S}}$ and $k_{\mathrm{I}}$ are the projections of the elliptical JSA onto the $\omega_{\mathrm{S}}$- and $\omega_{\mathrm{I}}$-axes respectively,
\begin{equation}
    \label{eq:kappa_Schmidt}
    k_{\mathrm{S}} = \sqrt{\frac{2a(1-w^2)}{(1+w^2)}}, ~ k_{\mathrm{I}} = \sqrt{\frac{2c(1-w^2)}{(1+w^2)}},
\end{equation}
the Schmidt weight factor $w$ is obtained using the quadratic formula,
\begin{equation}
    w = \frac{-\sqrt{ac} + \sqrt{ac-b^2}}{b}.
\end{equation}
Defining mode creation operators for signal and idler modes as
\begin{equation}
    a_{n,\mathrm{S}}^{\dag} = \int d\omega_{\mathrm{S}}\,h_n(k_{\mathrm{S}}(\omega_{\mathrm{S}}-\bar{\omega}_{\mathrm{S}}))\,a_{\mathrm{S}}^{\dag}(\omega_{\mathrm{S}}),~~a_{n,\mathrm{I}}^{\dag} = \int d\omega_{\mathrm{I}}\,h_n(k_{\mathrm{I}}(\omega_{\mathrm{I}}-\bar{\omega}_{\mathrm{I}}))\,a^{\dag}_{\mathrm{I}}(\omega_{\mathrm{I}}),
\end{equation}
so the bosonic commutation relations $[a_{m,\mathrm{S}}^{},a_{n,\mathrm{S}}^{\dag}] = \delta_{mn}$,  $[a_{m,\mathrm{I}}^{},a_{n,\mathrm{I}}^{\dag}] = \delta_{mn}$ hold, the approximate PDC state then has the following Schmidt form
\begin{equation}
    \ket{\Phi_{\mathrm{PDC}}} \approx \frac{1}{\sqrt{N_{\mathrm{PDC}}}}~\left(\ket{0} + \sum_{n=0}^{\infty}\,r_{n,\mathrm{PDC}}\,\,a^{\dag}_{n,\mathrm{S}}a^{\dag}_{n,\mathrm{I}}\ket{0^{\mathrm{S}}}\ket{0^{\mathrm{I}}} \right) =   \frac{1}{\sqrt{N_{\mathrm{PDC}}}} \left( \ket{0} + \sum_{n=0}^{\infty}\,r_{n,\mathrm{PDC}}\,\,\ket{\xi_n^{\mathrm{S}}}\ket{\xi_n^{\mathrm{I}}}    \right)
\end{equation}
where $\ket{\xi_n^{\mathrm{S}}} = a^{\dag}_{n,\mathrm{S}}\ket{0^{\mathrm{S}}}$~($\ket{\xi_n^{\mathrm{I}}} = a^{\dag}_{n,\mathrm{I}}\ket{0^{\mathrm{I}}}$) are $n$-mode Schmidt basis kets for the signal~(idler) photons.
Finally, if we post-select for only successful detections of the two-photon state, the biphoton PDC state becomes
\begin{equation}
   \ket{\psi_{\mathrm{biph,PDC}}} = \sum_{n=0}^{\infty}\,\tilde{r}_{\mathrm{n,PDC}}\,\,\ket{\xi_n^S}\ket{\xi_n^I}, ~~\tilde{r}_{\mathrm{n,PDC}} = \frac{r_{\mathrm{n,PDC}}}{\sqrt{\sum_{n=0}^{\infty}\,|r_{\mathrm{n,PDC}}|^2}} = -\I w^n \sqrt{1-w^2},
\end{equation}
which has the same form as Eq.~(\ref{eq:two_photon _entangled_schmidt_expression}).
Notice that this is equivalent to renormalizing $f_{\mathrm{PDC}}(\omega_S,\omega_I)$ to be treated as a proper wavefunction, thus the efficiency parameter $\alpha_{\mathrm{pump}}$ does not enter explicitly in the description of the post-selected state.

\end{document}